\documentclass{amsart}
\usepackage{graphicx} 

\title{Bethe roots for periodic TASEP and algebraic curve}

\author{Shinsuke Iwao}
\address{Faculty of Business and Commerce, Keio University, Hiyosi 4–1–1, Kohoku-ku, Yokohama-si, Kanagawa 223-8521, Japan.}
\email{iwao-s@keio.jp}

\author{Kohei Motegi}
\address{Faculty of Marine Technology, Tokyo University of Marine Science and Technology, Etchujima 2-1-6, Koto-Ku, Tokyo, 135-8533, Japan.}
\email{kmoteg0@kaiyodai.ac.jp}

\date{\today}

\usepackage[twoside,lmargin=9em,rmargin=9em,top=4cm,bottom=4cm]{geometry}
\usepackage{graphicx,xcolor}

\usepackage{amsmath}
\usepackage{amssymb}
\usepackage{amsthm}
\usepackage[all]{xy}
\usepackage[centertableaux]{ytableau}
\usepackage{tikz}
\usetikzlibrary{positioning, arrows.meta}

\newtheorem{thm}{Theorem}[section]
\newtheorem{prop}[thm]{Proposition}
\newtheorem{lemma}[thm]{Lemma}

\newtheorem{example}[thm]{Example}
\newtheorem{rem}[thm]{Remark}

\newtheorem{cor}[thm]{Corollary}

\def\CC{\mathord{\mathbb{C}}}

\def\RR{\mathord{\mathbb{R}}}
\def\ZZ{\mathord{\mathbb{Z}}}
\def\PP{\mathord{\mathbb{P}}}

\newcommand\ee{\mathord{\mathrm{e}}} 

\begin{document}

\begin{abstract}
We present an algebraic method for solving the Bethe ansatz equations for the periodic totally asymmetric exclusion process (TASEP) with an arbitrary number of sites and particles.
The Bethe ansatz equations are realized as an algebraic equation on a certain Riemann surface.
While our Riemann surface is essentially the same as the one introduced by Prolhac, we focus on its algebraic realization as a (singular) plane curve.
Through a counting argument on the Riemann surface, we establish a rigorous proof that the Bethe ansatz equation has the expected number of solutions when counted with multiplicity. Consequently, under appropriate generic conditions, the completeness of the Bethe ansatz follows.
The decomposition of the Riemann surface into connected components determines how often each value of the product of Bethe roots appears.
We classify the connected components and their multiplicities using a similar argument to the spectral degeneracy of the Markov matrix discussed by Golinelli-Mallick.
As a result, we give an algebro-geometric characterization of Golinelli-Mallick-type spectral degeneracy of the Markov matrix.
We also give explicit formulas for the number of connected components, the number of ramification points, and the total genus of the Riemann surface.
These formulas recover the table of examples presented by Prolhac.
Moreover, we explore applications of the special type of Bethe roots that appear in this case to partition functions of the five-vertex model.
We introduce a version of the free energy and evaluate the thermodynamic limit to find the explicit form in terms of the Riemann zeta function.
\end{abstract}

\maketitle

\section{Introduction}


The Bethe ansatz \cite{Bethe} is one of the most traditional and powerful methods in the field of quantum integrable models \cite{Baxterbook}.
Integrable probability is another field of integrable models for stochastic processes, which has also attracted attention in recent years, and some of the models in both fields have relations.
One of the most notable correspondences is the relation between
the XXZ quantum spin chain and the asymmetric simple exclusion process (ASEP) \cite{Derrida,Liggett}.
The Hamiltonian of the former and the Markov matrix of the latter are related by a gauge transformation.
The eigenvalues of the Hamiltonian of the periodic XXZ spin chain can be computed by the Bethe ansatz,
and likewise, the eigenvalues of the Markov matrix of the periodic ASEP can be described by the Bethe ansatz.

The Bethe ansatz in the most naive sense is the statement that the eigenvalues of the Hamiltonians or the Markov matrices of the integrable many-body systems can be described by equations called the Bethe ansatz equations.
Deriving the Bethe ansatz equations for the periodic case for various systems was one of the extensive research areas on the Bethe ansatz in the last century.
However, solving or investigating Bethe ansatz equations is not easy with the exception of the free-fermion point, where all the Bethe ansatz solutions can be described explicitly. There has been extensive research for the XXX and XXZ spin chains which led to discoveries of new mathematical notions and structures, and still deserve to be studied.
See, for example, \cite{Ki,KR,LSA,TV,Baxter,MTV,Tarasov} for the periodic quantum spin chains.
See also \cite{BCPS} for investigations of the completeness
on the infinite line.

As for the ASEP, a multiparticle extension of random walks with exclusion constraints, a special case in which particles move in one direction is called the totally asymmetric simple exclusion process (TASEP).
The investigation of the Bethe ansatz equations for the ASEP and its related asymmetric XXZ chain was pioneered by 
Gwa-Spohn and Kim
\cite{GSone,GStwo,Kimone,Kimtwo}, and later extended to open boundaries in \cite{deGE}.
For the TASEP case, the structure of the Bethe ansatz equations becomes simpler than the ASEP case or the XXZ chain. However, it is still rather nontrivial.

In a series of their works, Golinelli-Mallick \cite{GMone,GMtwo,GMthree,GMspecdeg,GMfour} proposed an ansatz on the distribution and labeling of quantum numbers of the roots of the TASEP Bethe ansatz equations on the `Cassini oval', and investigated an application to the spectral gaps.
They also studied the spectral degeneracy of the eigenvalues of the Markov matrix.
Prolhac further pursued this approach extensively
and investigated important classes of excited states and asymptotics \cite{Prolhacone,Prolhactwo}
with successful applications to current fluctuations and large deviations \cite{Prolhacthree,Prolhacfour,Prolhacfive}
related to the KPZ universality class \cite{KPZ, Corwin,Spohn,Takeuchi}, extending the mathematical analysis which was mainly restricted to the infinite line at that time \cite{Johone,Johtwo,BFPS,TW,SS,ACQ}.

Moreover, in recent years, Prolhac proposed a mathematical formulation of the Bethe ansatz equations of the TASEP using some Riemann surface and expressed probabilities as integrals over curves on the Riemann surface, which are valid under the assumption that the total number of sites $L$ and the total number of particles $N$ are coprime \cite{ProlhacRSone,ProlhacRStwo,ProlhacRSthree,ProlhacSciPost}.
See also de Gier-Kenyon-Watson \cite{deGKW} for recent works on the TASEP Bethe ansatz equations from a related but different perspective with applications to the limit shape of the five-vertex model and \cite{NPH} for investigations on the spectral boundary.
As for related works, see \cite{BDS} for another mathematical approach to the Bethe ansatz equations, \cite{ISN} for an investigation of certain classes of excited states of the ASEP, \cite{BLone,BLtwo} for multipoint distributions of the TASEP under periodic boundaries using different techniques from \cite{Prolhacthree,Prolhacfour,Prolhacfive}, and  \cite{AKSS} for the spectrum of the Markov matrix for the multispecies ASEP.

The main idea of the approach to the TASEP Bethe ansatz equations proposed by Golinelli-Mallick
and further pursued by Prolhac and also by de Gier-Kenyon-Watson
is to first start from investigations on the `Cassini oval'
and imposing some ansatz on the behavior of the Bethe roots,
partially based on numerical experimentations.
This approach is important and also effective for practical computations.
For example, classes of `low-energy excited' Bethe roots which give dominant contributions
to the asymptotics were investigated in detail in \cite{Prolhacone} with applications to the KPZ universality in subsequent works.
For a mathematical justification, one of the problems is that the investigation starts from some ansatz on the behavior of the Bethe roots, which is likely to be true when the fugacity parameter is turned on and the total number of sites $L$ and the total number of particles $N$ are coprime, which seems to be hard to prove in general.
Note that the Bethe roots are points on the `Cassini oval' and are not the `Cassini oval' itself.
When there is no fugacity parameter, the `Cassini oval' for the steady state Bethe roots collapses to a point, and also counterexamples to the behavior of the Bethe roots can be found when $L$ and $N$ are large (there are two solutions when $L=10$, $N=5$ which do not satisfy the ansatz on the behavior).
The Riemann surface construction of the Bethe ansatz equations for the periodic TASEP proposed by Prolhac \cite{ProlhacRSone} includes cases when $L$ and $N$ are not coprime, and depends on some assumptions mentioned above and applies to the case when the fugacity parameter is turned on.

In this paper, we study an algebro-geometric formulation of the Bethe ansatz equations on a Riemann surface.
While our Riemann surface is essentially the same as the one introduced by Prolhac \cite{ProlhacRSone}, we focus on its algebraic realization as a plane curve.
With an appropriate choice of coordinate functions, the Riemann surface can be mapped to a (possibly singular, non-irreducible, and non-reduced) plane curve that admits a computable defining equation.

We emphasize that our argument does not rely on experimental observations or ansatz regarding the behavior of the Bethe roots on the  `Cassini oval', as such approaches are not valid for the above-mentioned case.

Based on our algebro-geometric construction, we will provide:
\begin{enumerate}
\item a rigorous proof that the Bethe anzats equation has the expected number of solutions counted with multiplicity (Theorem \ref{completenessequation}), which implies that, under appropriate genericness conditions, the completeness of the Bethe anzats follows (Corollary \ref{cor:generic_gamma}. See also Remark \ref{rem:note}),
\item explicit formulas for the number of connected components (Lemma \ref{lemma:formula_for_N} and Eq.~\eqref{eq:N_partition}), the genus of a connected component (Eq.~\eqref{eq:sub_formula_for_total_genus}), and the total genus of the Riemann surface (Eq.~\eqref{eq:total_genus}), reproducing examples presented in \cite{ProlhacRSone}; and
\item a detailed investigation of the irreducible decomposition of the plane curve (Proposition \ref{prop:decomp_of_defining_eq_of_C0}), which
determines the degeneracy of eigenvalues (see Proposition \ref{prop:the_same_point}).
\end{enumerate}

We also give several applications of our construction to the case when $L$ and $N$ are not coprime.
Golinelli-Mallick \cite{GMspecdeg} discovered and explored a type of spectral degeneracy of the Markov matrix, which appears in this case.
We apply the same type of argument to the plane curves and discuss a criterion for multiplicity.
Golinelli-Mallick's work also implies the existence of a special type of Bethe ansatz roots, which are both nontrivial and simple. 
We insert these special Bethe roots for the case of half-filling $L=2N$ into the factorized forms of the partition functions of the five-vertex model or equivalently the norm of the on-shell Bethe vector for the TASEP \cite{Bo,MSS} obtained by the algebraic Bethe ansatz or the quantum inverse scattering method, and evaluate the partition functions and express in terms of roots of unity.
Introducing the free energy of the partition functions and evaluating the thermodynamic limit, we find it is given by the Riemann zeta function $\zeta(3)$.
We also discuss a vanishing of the overlap between a particular type of state and the on-shell Bethe vector corresponding to the special type of Bethe roots for 1/3-filling, generalizing the observation of the vanishing for the case of half-filling in \cite{MSS} by using some results from \cite{MS}.

This paper is organized as follows.
In Section 2, we reformulate the Bethe ansatz equations for the periodic TASEP in the standard form
and give a Riemann surface construction of the Bethe ansatz equations.
By a counting argument on the Riemann surface, we show the completeness.
In Sections 3 and 4, we give a detailed investigation of the irreducible decompositions and give the explicit forms of the genus of the algebraic curve, each of its roots giving the product of the Bethe roots.
In Section 5, we give an argument on the multiplicities of the irreducible components of the algebraic curve
in the same way as the Golinelli-Mallick type degeneration for the spectrum of the Markov matrix.
In Section 6, we discuss the half-filling case and evaluate a certain type of partition function of the five-vertex model, and evaluate the thermodynamic limit.
We discuss several other applications in Section 7.

\subsection{Notations}
Throughout the paper, we assume $1\leq N<L$.
Let $e=\mathrm{gcd}(N,L)$ be the greatest common divisor of $L$ and $N$.
Define $L'$ and $N'$ by
\begin{equation}\label{eq:def_of_L'_N'}
L=eL',\quad N=eN'.
\end{equation}

We also use the following standard notations:
\begin{itemize}
    \item For a positive integer $n$,
    the factorial $n!$ denotes the product of all positive integers up to $n$, and $0!:=1$.
    \item The binomial coefficient $\binom{n}{k}$ is defined by
    \[
    \binom{n}{k} = \frac{n!}{k!(n-k)!}
    \quad \text{for integers } n \geq k \geq 0.
    \]
    \item The M\"obius function $\mu(n)$ is defined by
    \[
    \mu(n) = 
    \begin{cases}
        1 & \text{if } n = 1, \\
        (-1)^k & \text{if } n \text{ is a product of } k \text{ distinct primes}, \\
        0 & \text{if } n \text{ has a squared prime factor}.
    \end{cases}
    \]
    \item The Euler totient function $\varphi(n)$ denotes the number of integers between $1$ and $n$ that are coprime to $n$.
\end{itemize}

\section{Bethe equations and Riemann surface}

\subsection{Riemann surface}

The \textit{Bethe ansatz equations} for the periodic TASEP \cite{GSone,GMthree,GMspecdeg} are given as
\begin{equation}\label{eq:Bethe_in_one_line}
\frac{z_i^N}{(1-z_i)^L}=(-1)^{N+1}
\ee^{L\gamma}
\prod_{j=1}^Nz_j,\quad (i=1,2,\dots,N),
\end{equation}
where $\gamma$ is a fugacity parameter.
From a solution $(z_1,z_2,\dots,z_N)$ to \eqref{eq:Bethe_in_one_line}, one can construct an eigenvector of the ($\gamma$-deformation of the) Markov matrix for the periodic TASEP with the exception of the singular solution $z_1=z_2=\dots=z_N=0$.
If $\ee^{L\gamma}\neq 1$, the singular solution is called an \textit{unphysical solution}, which does not correspond to any eigenstate.
For a physical solution, the corresponding eigenvalue is given by
\begin{equation}\label{eq:eigenvalue}
E=\sum_{k=1}^N\frac{z_i}{1-z_i}.
\end{equation}
For detailed construction of the eigenvectors, see~\cite{Derrida,ProlhacRStwo} for example.

Introducing the auxiliary variables $v$ and $w$ to \eqref{eq:Bethe_in_one_line}, we obtain the equations
\begin{equation}\label{eq:Bethe_original}
\begin{cases}
    w=\dfrac{z_i^N}{(1-z_i)^L},\quad(i=1,2,\dots,N)\\
    v=\prod\limits_{i=1}^Nz_i,\\
    w=(-1)^{N+1}\ee^{L\gamma}v.
\end{cases}    
\end{equation}
To analyze the structure of the system \eqref{eq:Bethe_original}, we introduce the rational map
\begin{equation}\label{eq:def_of_varphi}
\varphi:\PP\to \PP;\qquad t\mapsto \frac{t^N}{(1-t)^L},
\end{equation}
where $\PP=\mathbb{C} \cup \{ \infty \}$.
The map $\varphi$ has four critical points $t=0,-\frac{N}{L-N},1,\infty$, with multiplicities $N$, $2$, $L$, and $L-N$, respectively.
The critical values of $\varphi$ are given as
\[
\varphi(0)=\varphi(\infty)=0,\quad \varphi(1)=\infty,\quad \varphi(-\tfrac{N}{L-N})=(-1)^N\tfrac{N^N(L-N)^{L-N}}{L^L}.
\]
Let $c^\ast:=(-1)^N\tfrac{N^N(L-N)^{L-N}}{L^L}$.

The map $\varphi:\PP\to \PP$ is an $ L$-to-$1$ covering of the projective line.
Let $K:=\{0,\infty,c^\ast\}\subset \PP$ be the set of branching points of $\varphi$.
Define the open domain $U:=\PP\setminus K$ and its subset $D\subset U$ by
\[
D:=\begin{cases}
U\setminus \{w\;;\;w\in \RR_{\geq 0}\}, & (N:\mbox{even})\\
U\setminus \{w\;;\;w\in \RR_{\leq c^\ast}\}\cup \{w\;;\;w\in \RR_{\geq 0}\}, & (N:\mbox{odd}).
\end{cases}
\]
Since $D$ is simply-connected and contains no branching point of $\varphi$, there exist $L$ independent inverse functions $t_1,t_2,\dots,t_L:D\to \CC$ of $\varphi$, satisfying
\begin{equation}\label{eq:w_to_t}
w=\frac{t_i(w)^N}{(1-t_i(w))^L}.    
\end{equation}

These inverse functions are labeled according to the local behavior around $w=0$;
For $w=re^{\phi\sqrt{-1}}$ ($r>0$, $0<\phi\leq 2\pi$), we let
\begin{equation}\label{eq:local_behave_1}
t_i(w)\stackrel{w\to 0}{=}r^{1/N}\exp\left(\frac{\phi+2\pi (i-1)}{N}\sqrt{-1}\right)\times (1+O(w)), \quad(i=1,2,\dots,N)
\end{equation}
and
\begin{equation}\label{eq:local_behave_2}
t_{N+j}(w)\stackrel{w\to 0}{=}
-r^{-1/(L-N)}
\exp\left(
-\frac{\phi+2\pi (j-1)}{L-N}\sqrt{-1}
\right)\times (1+O(w)),\quad (j=1,2,\dots,L-N).
\end{equation}


Let $[L]:=\{1,2,\dots,L\}$ and $\Omega:=\{I\subset [L]\;;\;\sharp I=N\}$.
{
Fix an arbitrary element $I_0=\{i_1,i_2,\dots,i_N\}$ of $\Omega$ and consider the $j$-th elementary symmetric polynomial $$\phi_j(w):=e_j(t_{i_1}(w),t_{i_2}(w),\dots,t_{i_N}(w)).$$
Then, we define the holomorphic map $\Phi_0:D\to \CC^N$ by
\begin{equation}\label{eq:def_of_v0}
w\mapsto (\phi_1(w),\phi_2(w),\dots,\phi_N(w)).
\end{equation}
By analytically continuing $\Phi_0$, we obtain a multivalued map $\Phi$ defined on $U$.}
Standard arguments in complex analysis guarantee the existence of a compact Riemann surface on which $\Phi$ is regarded as a single-valued map.
However, this Riemann surface may not realize all $\binom{L}{N}$ possible choices of the branch of \eqref{eq:def_of_v0}.
In such a case, we choose a different initial $I_0$ and repeat the procedure to obtain a new Riemann surface.
By continuing this process until all possible branches appear, we eventually obtain a collection of several Riemann surfaces.
Let $X$ denote their disjoint union.

{
In what follows, we particularly focus on the $N$-th coordinate function $\phi_N(w)$ of $\Phi$.
Throughout, we write $v(w):=\phi_N(w)$.
Then, $X$ possesses the following properties.
}

\begin{enumerate}
    \item $v$ is regarded as a single-valued function $v:X\to \PP$.
    \item there exists a projection $w : X\to \PP$ satisfying the following (2-a), (2-b):
    \begin{itemize}
       \item[(2-a)] The map $w:X\to \PP$ is a $\binom{L}{N}$-to-$1$ covering of $\PP$, with branch points at $0,\infty,c^\ast$.
       \item[(2-b)] The inverse image $w^{-1}(D)$ is a disjoint union of $\binom{L}{N}$ domains $D_I$, which is isomorphic to $D$, labeled by an element $I$ of $\Omega$:
       \[
       w^{-1}(D)=\bigsqcup_{I\in \Omega}D_I\qquad (D_I\simeq D).
       \]
       We will refer to $D_I$ as `the $I$-th sheet of $X$.'
    \end{itemize}
    \item At a point $p$ on the $I$-th sheet $D_I$, $v(p)$ is expressed as
    \[
    v(p)=\prod_{i\in I}t_i(w(p)).
    \]
    We will simply write this relation as `$v=\prod_{i\in I}t_i(w)$ on $D_I$.'
\end{enumerate}

We refer to the locus $X_\infty:=v^{-1}(\infty)\cup w^{-1}(\infty)\subset X$ as the \textit{infinite part} of $X$, and $X_0:=X\setminus X_\infty$ as the \textit{affine part} of $X$.
Throughout the paper, we focus on the realization of $X_0$ as a plane curve via the map
\begin{equation}\label{eq:map_to_the_plane}
X_0\to \CC^2;\quad p\mapsto (v(p),w(p))
\end{equation}
with coordinate functions $v$ and $w$.
Let $C_0\subset \CC^2$ denote the image.
As is well known (see Weyl's textbook \cite{WM} on Riemann surfaces), the defining equation of the plane curve $C_0$ can be given by using elementary symmetric polynomials in all possible choices of branches of the multivalued function $v(w)$.

We derive the defining equation of $C_0$ as follows.
From \eqref{eq:w_to_t}, we have the algebraic equation
\begin{equation}\label{eq:alg_eq_of_t}
t^L-e_1t^{L-1}+e_2t^{L-2}-\dots+(-1)^Le_L=0,    
\end{equation}
where
\[
e_i=\binom{L}{i}-(-1)^Nw^{-1}\delta_{i,N}, \quad (i=1,2,\dots,L).
\]
Let $t=t_1(w),\dots,t_L(w)$ be the roots of \eqref{eq:alg_eq_of_t}.
By Vieta's formula, we have $e_i=e_i(t_1,t_2,\dots,t_L)$, where $e_i(t_1,t_2,\dots,t_L)$ denotes the $i$-th elementary symmetric polynomial in $t_1,t_2,\dots,t_L$.

For $I\in \Omega$, set $t_I:=\prod_{i\in I} t_i(w)$.
We define the \textit{$\Omega$-elementary symmetric polynomial} $E_i$ by
\[
E_i:=\sum_{\{I_1,I_2,\dots,I_i\}}t_{I_1}t_{I_2}\dots t_{I_i},\quad
\Bigg(i=1,2,\dots,\binom{L}{N} \Bigg),
\]
where the summation is taken over all sets $\{I_1,\dots,I_i\}$ of distinct $i$ elements of $\Omega$.
Equivalently, the polynomials $E_1,\dots,E_{\binom{L}{N}}$ are defined by the generating function
\[
\prod_{I\in \Omega}(1+t_I T)=1+E_1T+E_2T^2+\cdots+E_{\binom{L}{N}}T^{\binom{L}{N}}.
\]
Then, each $E_i$ is an element of $\ZZ[w^{-1}]$ because $E_i$ is a symmetric polynomial in $t_1,\dots,t_L$.

By using the $\Omega$-elementary symmetric polynomials $E_1,E_2,\dots,E_{\binom{L}{N}}$, the defining equation of the plane curve $C_0$ is given as
\[
v^{\binom{L}{N}}-E_1v^{\binom{L}{N}-1}+E_2v^{\binom{L}{N}-2}-\cdots+(-1)^{\binom{L}{N}}E_{\binom{L}{N}}=0.
\]
As mentioned below, $C_0$ is generally a singular, non-irreducible, and non-reduced algebraic curve.

In most cases, a point on $X_0$ can be specified by the image $(v,w)\in \CC^2$ under the map \eqref{eq:map_to_the_plane}.
However, if $(v,w)$ is a singular point of $C_0$, more than one point on $X_0$ can be mapped to $(v,w)$.
In such a case, we specify this point by referring to the sheet containing it; For example, `$(v,w)$ is on $D_I$ ($I\in \Omega$).'

On the Riemann surface $X$, the Bethe equations \eqref{eq:Bethe_original} are realized as one simple algebraic equation.
A tuple of complex numbers $(z_1,\dots,z_n,v,w)$ satisfies the Bethe equations \eqref{eq:Bethe_original} if and only if 
\begin{enumerate}
\item[(i)] $(v,w)\in D_I$,
\item[(ii)] $w=(-1)^{N+1}\ee^{L\gamma}v$, and
\item[(iii)] $t=z_1,z_2,\dots,z_n$ is a subset of solutions of $w=\frac{t^N}{(1-t)^L}$ satisfying $v=\prod_{i=1}^Nz_i$.
\end{enumerate}
In particular, a Bethe root corresponds to a point of intersection between the plane curve $C_0$ and the line $\{w=(-1)^{N+1}\ee^{L\gamma}v\}$.
In other words, to find Bethe roots, we have to solve the algebraic equation $w=(-1)^{N+1}\ee^{L\gamma}v$ on the Riemann surface $X$.

In simple cases, all Bethe roots can be determined directly.
See the two examples below.
\begin{example}[Bethe roots for $(L,N)=(3,2)$]\label{ex:L3N2_one}
Let $(L,N)=(3,2)$.
From the equation $w=\frac{t^2}{(1-t)^3}$, we obtain the cubic equation
\[
t^3-(3-w^{-1})t^2+3t-1=0.
\]
By Vieta's formula, we have
\[
t_1+t_2+t_3=3-w^{-1},\quad
t_{1}t_2+t_{1}t_3+t_{2}t_3=3,\quad
t_{1}t_2t_3=1.
\]
Since $\Omega=\{\{1,2\},\{1,3\},\{2,3\}\}$, the $\Omega$-elementary symmetric polynomials $E_1,E_2,E_3$ are given by
\[
\begin{gathered}
E_1=t_{12}+t_{13}+t_{23}=3,\quad
E_2=t_{12}t_{13}+t_{12}t_{23}+t_{13}t_{23}=3-w^{-1},\quad
E_3=t_{12}t_{13}t_{23}=1.
\end{gathered}
\]
Then the defining equation of $C_0$ is 
\[
v^3-3v^2+(3-w^{-1})v-1=0.
\]
By substituting $w=-\ee^{L\gamma} v$, we have the cubic equation $v^3-3v^2+3v-1+\ee^{-L\gamma}=0$. 

If the fugacity parameter $\gamma$ is $0$, the cubic equation $v^3-3v^2+3v=0$ has $3$ roots:
\[
v=0,\quad \frac{3+\sqrt{-3}}{2},\quad \frac{3-\sqrt{-3}}{2}.
\]
Substituting these values to the cubic equation $t^3-(3-w^{-1})t^2+3t-1=0$ and solving them, we get the Bethe roots
\[
\{z_1,z_2\}=\textstyle
\{0,0\},\quad 
\left\{
\frac{3+\sqrt{3-2\sqrt{-3}}}{2},
\frac{3-\sqrt{3-2\sqrt{-3}}}{2}
\right\},\quad
\left\{
\frac{3+\sqrt{3+2\sqrt{-3}}}{2},
\frac{3-\sqrt{3+2\sqrt{-3}}}{2}
\right\}.
\]
(Solving the cubic equations is not difficult because we know that the product of two of the three roots is $v$.)
One can verify that these values satisfy \eqref{eq:Bethe_in_one_line}.
\end{example}

\begin{example}[Bethe roots for $(L,N)=(4,2)$]
\label{ex:L4N2_one}
Let $(L,N)=(4,2)$.
Then, we have the quartic equation
\begin{equation}\label{eq:quartic_42}
t^4-4t^3+(6-w^{-1})t^2-4t+1=0.
\end{equation}
From this, we have $e_1=4$, $e_2=6-w^{-1}$, $e_3=4$, and $e_4=1$.
The $\Omega$-elementary symmetric polynomial $E_1$ is given by $E_1=
t_{12}+
t_{13}+
t_{14}+
t_{23}+
t_{24}+
t_{34}=e_2=6-w^{-1}$.
With standard techniques on symmetric polynomials, we can deduce
\begin{equation}\label{eq:Omega-elem-42}
\begin{gathered}
E_2=e_1e_3-e_4=15,\quad
E_3=e_1^2e_4-2e_2e_4+e_3^2=20+2w^{-1},\\
E_4=e_1e_3e_4-e_4^2=15,\quad
E_5=e_4^2e_2=6-w^{-1},\quad
E_6=e_4^3=1.
\end{gathered}
\end{equation}
(See Appendix \S \ref{sec:appA} for a derivation of \eqref{eq:Omega-elem-42}.)
Then, the defining equation of $C_0$ is
\[
v^6 - (6 - w^{-1})v^5 + 15v^4 - (20 + 2w^{-1})v^3 + 15v^2 - (6 - w^{-1})v + 1 = 0,
\]
which factors as
\begin{equation}\label{eq:def_of_C0_42}
(v - 1)^2(v^4 - (4 - w^{-1})v^3 + (6 + 2w^{-1})v^2 - (4 - w^{-1})v + 1) = 0.    
\end{equation}
Hence, $C_0$ is non-irreducible and non-reduced.
As discussed later (Proposition \ref{prop:decomp_of_defining_eq_of_C0}), this factorization can be understood by examining the monodromy of the covering $w: X \to \mathbb{P}$.

Let the fugacity parameter $\gamma$ be $0$. 
By substituting $w = -v$ into \eqref{eq:def_of_C0_42}, we obtain the equation $v(v - 1)^2(v - 3)(v^2 - v + 2) = 0$, which has six roots:
\[
v = 0,\ 1,\ 1,\ 3,\ \frac{1+ \sqrt{-7}}{2},\ \frac{1- \sqrt{-7}}{2}.
\]
Substituting them into \eqref{eq:quartic_42} and solving the quartic equations, we obtain the six Bethe roots
\[
\begin{gathered}
\{z_1, z_2\} = \{0, 0\}, \quad
\left\{
\textstyle
\frac{2 + \sqrt{-1} + \sqrt{-1 + 4\sqrt{-1}}}{2},\ 
\frac{2 + \sqrt{-1} - \sqrt{-1 + 4\sqrt{-1}}}{2}
\right\},\\
\left\{
\textstyle
\frac{2 - \sqrt{-1} + \sqrt{-1 - 4\sqrt{-1}}}{2},\ 
\frac{2 - \sqrt{-1} - \sqrt{-1 - 4\sqrt{-1}}}{2}
\right\},\\
\left\{
\textstyle
\frac{6 - \sqrt{-3} + \sqrt{-3 - 12\sqrt{-3}}}{6},\ 
\frac{6 + \sqrt{-3} + \sqrt{-3 + 12\sqrt{-3}}}{6}
\right\},\\
\left\{
\textstyle
\frac{2 + \zeta + \sqrt{\zeta^2 + 4\zeta}}{2},\ 
\frac{2 - \zeta - \sqrt{\zeta^2 - 4\zeta}}{2}
\right\},\quad
\left\{
\textstyle
\frac{2 + \eta + \sqrt{\eta^2 + 4\eta}}{2},\ 
\frac{2 - \eta - \sqrt{\eta^2 - 4\eta}}{2}
\right\}.
\end{gathered}
\]
Here, $\zeta$ is one of the two complex numbers satisfying $\zeta^2 = \frac{-1 + \sqrt{-7}}{4}$, and $\eta$ is one of the two complex numbers satisfying $\eta^2 = \frac{-1 - \sqrt{-7}}{4}$.  
Here, the square root symbol always denotes the complex number whose argument $\theta$ satisfies $-\frac{\pi}{2} < \theta \leq \frac{\pi}{2}$.

\end{example}

\begin{rem}
The Riemann surface $X$ is isomorphic to $\mathcal{R}_N$ in \cite{ProlhacRStwo}.
\end{rem}

\begin{rem}\label{rem:origin}
From \eqref{eq:local_behave_1}, we have $t_1(0) = \cdots = t_N(0) = 0$, so that $\prod_{i=1}^N t_i(0) = 0$.
This means that on the $\{1,2,\dots,N\}$-th sheet of $X$, the condition $w = 0$ implies $v = 0$.
In particular, the point $(0,0)$ lies on the curve $C_0$.
Therefore, the plane curve $C_0$ intersects the line $\{w = (-1)^{N+1}\ee^{L\gamma}v\}$ at $(0,0)$, regardless of the value of $\gamma$.
\end{rem}

\subsection{Completeness of the Bethe ansatz for periodic TASEP}\label{sec:completeness}

The expected number of independent eigenvectors for the periodic TASEP with $N$ particles and $L$ positions is $\binom{L}{N}$.  
In this subsection, we will show that the Bethe equations \eqref{eq:Bethe_original} have $\binom{L}{N}$ \textit{physical} roots, counted with multiplicities.
If all the roots are simple, the Bethe equations capture all the eigenvectors of the (deformed) Markov matrix.
(We emphasize that the multiplicities of Bethe roots and those of eigenvalues of the (deformed) Markov matrix are not the same. 
Different Bethe roots may correspond to a common eigenvalue. 
See Section \ref{sec:degeneracy} for details on the multiplicities of values of $\displaystyle v=\prod_{i=1}^N z_i $, and \cite{GMspecdeg}
for eigenvalues of the Markov matrix.)

To count solutions of \eqref{eq:Bethe_original}, we need to find all zeros of the meromorphic function $w-(-1)^{N+1} \ee^{L \gamma} v$ on $X$.
Let $h=v/w$.
On the open set $X\setminus \{(0,0)\}$, the zeros of $w-(-1)^{N+1}\ee^{L \gamma} v$ coincides with those of $h-(-1)^{N+1}\ee^{-L\gamma}$.
However, it is not the case for the origin $(0,0)$.

To explain this, let us consider the Taylor expansion of $v$ at $(0,0)\in X$.
Recall that $(0,0)$ lies on the $\{1,2,\dots,N\}$-th sheet
(Remark \ref{rem:origin}).
Then, from \eqref{eq:local_behave_1}, the function $v$ near $(0,0)$ expands as
\[
v = t_1t_2\cdots t_N = w\cdot \prod_{i=1}^N\exp\left(\frac{2\pi(i-1)}{N}\sqrt{-1}\right)(1+O(w))=(-1)^{N+1}w+\cdots.
\]
Therefore, $h=w/v$ has the expansion $h=(-1)^{N+1}+\cdots$ at $(0,0)$.
Hence, the meromorphic function $h-(-1)^{N+1}\ee^{-L\gamma}$ has a zero at $(0,0)$ if and only if $\ee^{L\gamma}=1$.
This implies that the set of zero locus of $h-(-1)^{N+1}\ee^{-L\gamma}$ on $X$ corresponds precisely to the set of all \textit{physical} Bethe roots.

Since the number of zeros and that of poles of a meromorphic function are equal, it suffices to count the poles of $h$. 
For any point $(v, w)$ on the $I$-th sheet $D_I$, we have
\[
v = \infty 
\quad\Rightarrow\quad 
t_i(w) = \infty \quad (\exists i \in I)
\quad\Rightarrow\quad 
w = 0
\]
since $v=\prod_{i\in I}t_i(w)$.
Then, it follows that
\[
h = \infty
\quad\Rightarrow\quad 
\text{($w = 0$ or $v = \infty$)}
\quad\Rightarrow\quad 
w = 0.
\]
Therefore, to count poles of $h$, it suffices to examine the locus $\Gamma_0 := \{(v, w) \in X \;;\; w = 0\}$.

As mentioned previously by Prolhac \cite[Section 3]{ProlhacRStwo}, the monodromy of the covering $w:X\to \PP$ around $w=0$ acts on the set of branches $\{t_1,\dots,t_{L}\}$ as
\[
t_1\mapsto t_2\mapsto \dots \mapsto t_N\mapsto t_1,\qquad
t_{N+1}\mapsto t_{N+2}\mapsto \dots \mapsto t_{L}\mapsto t_{N+1}.
\]
According to this, we define the one-to-one map $\sigma : [L] \to [L]$ by
\[
1 \mapsto 2 \mapsto \dots \mapsto N \mapsto 1,\qquad 
N+1 \mapsto N+2 \mapsto \dots \mapsto L \mapsto N+1,
\]
so that the monodromy action is written as $t_i \mapsto t_{\sigma(i)}$.

Let $\widehat{\sigma} : \Omega \to \Omega$ be the one-to-one map induced from $\sigma$, which sends $I=\{i_1,i_2,\dots,i_N\}$ to $\widehat{\sigma}(I)=\{\sigma(i_1),\sigma(i_2),\dots,\sigma(i_N)\}$.
Then, starting from a point on the $I$-th sheet of $X$, a counterclockwise loop around $w = 0$ on the $w$-plane takes us to the $\widehat{\sigma}(I)$-th sheet.  

The subgroup $G_1 = \langle \widehat{\sigma} \rangle \subset \mathrm{End}(\Omega)$ generated by $\widehat{\sigma}$ is a cyclic group of order $\mathrm{lcm}(N, L-N)$.
A point in $\Gamma_0$ is in one-to-one correspondence with an element in the set $\Omega / G_1$ of all $G_1$-orbits.
For a $G_1$-orbit $o \in \Omega / G_1$, let $P_o$ denote the point in $\Gamma_0$ corresponding to $o$.

\begin{example}[$\Gamma_0$ for $(L,N)=(3,2)$]
When $(L,M)=(3,2)$, $\sigma$ is given by $\sigma(1) = 2$, $\sigma(2) = 1$, and $\sigma(3) = 3$. Then, the induced map $\widehat{\sigma}:\Omega\to \Omega$ is defined as
\[
\widehat{\sigma}(\{1,2\}) = \{1,2\},\quad
\widehat{\sigma}(\{1,3\}) = \{2,3\},\quad
\widehat{\sigma}(\{2,3\}) = \{1,3\}.
\]
Hence, $\Gamma_0=\{P_{o_1},P_{o_2}\}$, where $P_{o_1}$ and $P_{o_2}$ corresponds to the $G_1$-orbits
\[
o_1 = \{\{1,2\}\},\qquad
o_2 = \{\{1,3\}, \{2,3\}\},
\]
respectively.
In the notation of Example~\ref{ex:L3N2_one}, they are expressed as  
$P_{o_1}=(0, 0)$ and $P_{o_2}=(\infty, 0)$.
\end{example}

\begin{example}[$\Gamma_0$ for $(L,N)=(4,2)$]
When $(L,N)=(4,2)$, $\sigma$ is given by  
$\sigma(1) = 2$, $\sigma(2) = 1$, $\sigma(3) = 4$, and $\sigma(4) = 3$.  
Then, there exist four $G_1$-orbits:
\[
\begin{gathered}
o_1=\{\{1,2\}\},\quad
o_2=\{\{3,4\}\},\quad
o_3=\{\{1,3\},\{2,4\}\},\quad
o_4=\{\{1,4\},\{2,3\}\},
\end{gathered}
\]
in $\Omega$.
Hence, $\Gamma_0=\{P_{o_1},P_{o_2},P_{o_3},P_{o_4}\}$.  
In the notation of Example \ref{ex:L4N2_one}, these points are expressed as  
$P_{o_1}=(0, 0)$, $P_{o_2}=(\infty, 0)$,  
$P_{o_3}=(1, 0)$, and $P_{o_4} = (-1, 0)$.
\end{example}

For a $G_1$-orbit $o$, let $d_o:=\# o$ be the cardinality of $o$.  
Given $I \in \Omega$, define $N_1 = N_1(I)$ as the number of elements of $I$ that are $\leq N$, and $N_2 = N_2(I)$ as the number of elements $> N$.  
Since $N_1$ and $N_2$ are constants on the orbit $o$, we can write them as $N_1(o)$ and $N_2(o)$.

The point $P_o$ is at the intersection of all $d_o$ sheets $D_I$ such that $I\in o$.
This means that, around $P_o$, one can choose a local coordinate $k$ so that
\[
w = k^{d_o} \qquad (|k| \ll 1).
\]

\begin{thm}
\label{completenessequation}
The Bethe equations have $\binom{L}{N}$ physical roots
counted with multiplicities.  
\end{thm}
\begin{proof}
It suffices to show that the total number of the order of poles of $h$ is $\binom{L}{N}$.
Recall that all poles of $h$ are contained in $\Gamma_0$.
From \eqref{eq:local_behave_1} and \eqref{eq:local_behave_2}, the function $v$ near $P_o$ expands as
\[
v = (\text{const.} \neq 0) \cdot k^{N_1\frac{d}{N} - N_2\frac{d}{L-N}} + \cdots, \qquad 
(d = d_o,\ N_1 = N_1(o),\ N_2 = N_2(o)).
\]
(This expression follows from the fact that $t_1, \dots, t_N$ contribute to $v$ as $k^{1/N}$, and $t_{N+1}, \dots, t_L$ as $k^{-1/(L-N)}$.)  
Hence, $h$ is expanded as
\begin{equation}\label{eq:expansion_of_h_at_0}
h = \frac{v}{w} = (\text{const.} \neq 0) \cdot k^{N_1\frac{d}{N} - N_2\frac{d}{L-N} - d} + \cdots.
\end{equation}
Since $N_1 \leq N$, the exponent $N_1\frac{d}{N} - N_2\frac{d}{L-N} - d$ is always non-positive.  
Therefore, the total order of the poles of $h$ is given by 
\[
\sum_{o \in \Omega/G_1} d_o\left(1 - \frac{N_1(o)}{N} + \frac{N_2(o)}{L-N} \right).
\]
Since $N_1(o)$ and $N_2(o)$ are constants on $o$, this sum can be rewritten as
\[
\begin{aligned}
\sum_{I \in \Omega} \left( 1 - \frac{N_1(I)}{N} + \frac{N - N_1(I)}{L-N} \right)
&=
\frac{L}{L-N} \binom{L}{N}
- \left( \frac{1}{N} + \frac{1}{L-N} \right)
\sum_{I \in \Omega} N_1(I).
\end{aligned}
\]
The number of $I\in \Omega$ satisfying $n=\sharp(I\cap \{i\;;\;i\leq N\})$ is given by $\binom{N}{n}\binom{L-N}{N-n}$ and hence we get 
\[
\sum_{I \in \Omega} N_1(I)
= \sum_{n \geq 0} n \cdot \binom{N}{n} \binom{L-N}{N-n}
= N \binom{L-1}{N-1}.
\]
Therefore, the sum of the orders of poles of $h$ is
\[
\begin{aligned}
\frac{L}{L-N} \binom{L}{N}
- \left( \frac{1}{N} + \frac{1}{L-N} \right) \cdot N \binom{L-1}{N-1}
&=
\left[
\frac{L}{L-N}
- \frac{1}{N} \cdot \frac{N^2}{L}
- \frac{1}{L-N} \cdot \frac{N^2}{L}
\right] \cdot \binom{L}{N} \\
&= \binom{L}{N}.
\end{aligned}
\]
\end{proof}

\begin{cor}\label{cor:generic_gamma}
For a generic $\gamma$, the Bethe equations have exactly $\binom{L}{N}$ physical roots.
\end{cor}


\section{Irreducible Decomposition}

When \( L \) and \( N \) are not coprime, the plane curve \( C_0 \) decomposes into several irreducible components.  
This decomposition can be analyzed via the monodromy group of the covering $w: X\to \PP$.

\subsection{Monodromy group of $w:X\to \PP$}

Let \( \widehat{\sigma}:\Omega\to \Omega \) be the map defined in \S \ref{sec:completeness}, and define \( \widehat{\tau}:\Omega\to \Omega \) as the induced map from the permutation
\[
\tau:1\mapsto 2\mapsto \dots \mapsto L\mapsto 1.
\]
It is known \cite{ProlhacRStwo} that the monodromy action of the covering $w:X\to \PP$ around $\infty$ is given by $\widehat{\tau}$.
We also show that the monodromy action around $c^\ast$ is generated by $\widehat{\sigma}^{-1}\circ \widehat{\tau}$.
Consider the subgroup \( G_2 := \langle \widehat{\sigma}, \widehat{\tau} \rangle\subset \mathrm{End}(\Omega) \) generated by \( \widehat{\sigma} \) and \( \widehat{\tau} \). 
(Obviously, $G_1\subset G_2$.)
Then, the group $G_2$ is isomorphic to the monodromy group of $w:X\to \PP$.

From the action of the groups $G_1,G_2$ on $\Omega$, we can determine the topology of $X$.
Let $\Omega/G_i$ be the set of $G_i$-orbits in $\Omega$.
\begin{itemize}
    \item An element of \( \Omega/G_1 \) is in one-to-one correspondence with a point of \( \Gamma_0 \).
    \item An element of \( \Omega/G_2 \) is in one-to-one correspondence with a connected component of \( X \).
    \begin{itemize}
        \item     If \( C_0 \) is reduced, an element of \( \Omega/G_2 \) is in one-to-one correspondence with an irreducible component of \( C_0 \).
        \item     If \( C_0 \) is non-reduced, an element of \( \Omega/G_2 \) corresponds to a union of several copies of an irreducible component of \( C_0 \).        
    \end{itemize}
\end{itemize}

For \( e = \mathrm{gcd}(N, L) \), define $e$ subsets \( \mathcal{P}_1, \mathcal{P}_2, \dots, \mathcal{P}_e\) of \( [L] \) by
\begin{equation}\label{eq:def_of_Pk}
\mathcal{P}_k := \{ i \in [L] \,;\, i \equiv k \mod e \}.    
\end{equation}
(In Golinelli-Mallick's terminology~\cite{GMspecdeg}, $\mathcal{P}_k$ is called a \textit{package}.)
For $k\in \ZZ$, we let $\mathcal{P}_{k+e}\equiv \mathcal{P}_{k}$.

Let $S_L=\mathrm{End}([L])$ be the symmetric group.
Consider the subgroup $H\subset S_L$ defined by
\[
H = \{ f \in S_L \,;\, \text{there exists some }s\text{ such that }f(\mathcal{P}_k) = \mathcal{P}_{k+s}\text{ for }k=1,2,\dots,e\}.
\] 
Then, the monodromy group $G_2\subset \mathrm{End}(\Omega)$ is characterized as follows.

\begin{lemma}\label{lemma:str_of_G_2}
Let \( \widehat{f}:\Omega \to \Omega \) denote the one-to-one map induced by \( f \in H \).  
Then we have $G_2 = \{ \widehat{f} \,;\, f \in H \}$.
\end{lemma}

\begin{proof}
Let \( H' = \langle \sigma, \tau \rangle \). 
Since the group homomorphism \( \mathrm{End}([L]) \mapsto \mathrm{End}(\Omega);\; f \mapsto \widehat{f} \) is injective if $0<L<N$, it suffices to show \( H = H' \).  

Since \( \sigma, \tau \in H \), it is clear that \( H' \subset H \).  
We now show the reverse inclusion \( H \subset H' \).
Let \( K \) be the subgroup of \( S_L \) consisting of permutations that fix all elements of \( [L] \) except for those in the $e$-th package $\mathcal{P}_e$.
Then, the subgroup $K_k:=\tau^k K\tau^{-k}\subset S_L$ acts on the $k$-th package $\mathcal{P}_k$ as a permutation. 
For given $f\in H$, let $s$ be an integer such that $f\cdot \tau^{-s}(\mathcal{P}_k)=\mathcal{P}_k$ for $k=1,2,\dots,e$.
Then, we have the decomposition
\[
f\tau^{-s} = x_1x_2\dots x_e\text{ for some } x_1\in K_1,\, x_2\in K_2,\, \dots,\, x_e \in K_e,
\]
which implies
$
H=K_1\cdot K_2\cdot \dots \cdot K_e\cdot \langle\tau\rangle
$.
Thus, it suffices to show \( K_e \subset H' \).  
Let \( (i,j) \) denote the transposition exchanging \( i \) and \( j \).  
Then, we have \( \tau^{-1} \circ \sigma = (N, L) \), and
\[
\tau^{ke} \circ (N, L) \circ \tau^{-ke} = (\tau^{ke}(N), \tau^{ke}(L)) = (N + ke, L + ke)
\]
holds for all \( k \in \mathbb{Z} \), where $N + ke$ and $L + ke$ are interpreted in modulo $L$.
Since \( K_e=\langle (N + ke, L + ke) \,;\, k \in \mathbb{Z} \rangle \), we have \( K_e \subset H' \).
This concludes the proof.
\end{proof}

Note that $K_k\simeq S_{L'}$, where $L'=L/e$ is the natural number defined in \eqref{eq:def_of_L'_N'}.
From the proof of Lemma \ref{lemma:str_of_G_2}, we have
\begin{equation}\label{eq:decomp_of_G2}
G_2
\simeq K_1\cdot K_2\cdots K_e\cdot \langle \tau\rangle
\simeq 
(S_{L'})^e\cdot \ZZ/e\ZZ,
\end{equation}
where $\ZZ/e\ZZ$ is the cyclic group of order $e$ and
$(S_{L'})^e=\overbrace{S_{L'}\cdot S_{L'}\cdots S_{L'}}^e$.

\begin{cor}\label{cor:isu_G2}
The cardinality of \( G_2 \) is \( e \times \left\{\left(\frac{L}{e}\right)!\right\}^e \).
\end{cor}

\begin{cor}\label{cor:G_2_coprime}
If \( L \) and \( N \) are coprime, then \( G_2 \simeq S_L \).
\end{cor}

\subsection{Classification of the $G_2$-orbits in $\Omega$}

An orbit of the monodromy group $G_2$ of $w:X\to \PP$ corresponds to a connected component of $X$.
In this subsection, we give a combinatorial way to classify $G_2$-orbits in \(\Omega \).

Let
\[
Z := \{(A_1, \dots, A_e) \; ; \; A_1 + \dots + A_e = N,\ 0 \leq A_i \leq L' \}.
\]
There exists a surjection
\begin{equation}\label{eq:def_of_pi}
\pi: \Omega \to Z; \quad I \mapsto (I_1,I_2,\dots,I_e),
\qquad\text{where } I_k:=\# (I\cap \mathcal{P}_k).
\end{equation}
On $Z$, we define a cyclic group action of \( \mathbb{Z}/e\mathbb{Z} = \langle \theta \; ; \; \theta^e = \mathrm{id} \rangle \) by
\[
\theta \cdot (A_1, \dots, A_{e-1}, A_e) = (A_2, \dots, A_e, A_1).
\]
Then, we have
\begin{equation}\label{eq:pi_and_G2_action}
\pi(\widehat{\tau}\cdot I)=\theta \cdot \pi(I) 
\ \text{ and }\ 
\pi(X\cdot I)=\pi(I)\qquad (X\in K_1\cdot K_2\cdots K_e).
\end{equation}

\begin{prop}\label{prop:G_2_orbit_crit}
Two elements \( I \) and \( J \) in \( \Omega \) lie in the same \( G_2 \)-orbit if and only if \( \pi(I) \) and \( \pi(J) \) lie in the same \( \mathbb{Z}/e\mathbb{Z} \)-orbit.
\end{prop}

\begin{proof}
The necessity follows from \eqref{eq:pi_and_G2_action}.
We will show the sufficiency.
Assume $\theta^s\pi(I)=\pi(J)$ for some integer $s$.
We may assume without loss of generality that \( i_a + s \equiv j_a \mod e \) for each \( a \).  
Then, we have \( 
\#(([L] \setminus I) \cap \mathcal{P}_k) 
=
\#(([L] \setminus J) \cap \mathcal{P}_{k+s})
\) for all $k$.  
Hence, there exists a bijection \( f_0: ([L] \setminus I) \to ([L] \setminus J) \) such that
$
f_0(( [L] \setminus I ) \cap \mathcal{P}_k) = ([L] \setminus J) \cap \mathcal{P}_{k+s}
$
for all $k$.
The bijection $f_0$ can be extended to the permutation \( f: [L] \to [L] \) by
\[
f(x) :=
\begin{cases}
j_a, & \text{if } x = i_a \\
f_0(x), & \text{if } x \notin I.
\end{cases}
\]
Since $f\in H$, we have \( \widehat{f} \in G_2 \) and \( \widehat{f}(I) = J \) by Lemma \ref{lemma:str_of_G_2}.
This implies $I$ and $J$ lie in the same $G_2$-orbit.
\end{proof}

Let $\overline{Z}:=Z/\sim$ be the quotient set defined by the equivalence relation
\[
(A_1,A_2,\dots,A_e)\sim
(B_1,B_2,\dots,B_e)
\iff
\exists s\in \ZZ\mbox{ s.t. }
(A_1,A_2,\dots,A_e)=
(B_{1+s},B_{2+s},\dots,B_{e+s}),
\]
where $A_{k+e}\equiv A_k$.
Then, by Proposition \ref{prop:G_2_orbit_crit}, $G_2$-orbits in $\Omega$ are in one-to-one correspondence with elements in $\overline{Z}$.
We denote the equivalence class containing $(A_1,\dots,A_e)\in Z$ by $[A_1,\dots,A_e]\in \overline{Z}$.

\begin{example}
[Irreducible decomposition of $C_0$ for $(L,N)=(6,3)$]
\label{eq:63_remification}
When \( L = 6 \) and \( N = 3 \), the $\binom{6}{3}=20$ elements of \( \Omega \) are partitioned into the following $8$ subsets under the action of \( G_1 \):
\[
\begin{gathered}
P_1:\{\{1,2,3\}\},\quad
P_2:\{\{1,2,4\},\{2,3,5\},\{1,3,6\}\},\quad
P_3:\{\{1,2,5\},\{2,3,6\},\{1,3,4\}\},\\
P_4:\{\{1,2,6\},\{2,3,4\},\{1,3,5\}\},\quad
P_5:\{\{1,4,5\},\{2,5,6\},\{3,4,6\}\},\\
P_6:\{\{1,4,6\},\{2,4,5\},\{3,5,6\}\},\quad
P_7:\{\{1,5,6\},\{2,4,6\},\{3,4,5\}\},\quad
P_8:\{\{4,5,6\}\}.
\end{gathered}
\]
Each of them is in one-to-one correspondence with an element of $\Gamma_0$.
When these points are further classified into \( G_2 \)-orbits, we have:
\begin{itemize}
\item Points \( P_1, P_4, P_7, P_8 \) belong to the orbit corresponding to $[1,1,1]$,
\item Points \( P_2, P_5 \) belong to the orbit corresponding to $[2,1,0]$,
\item Points \( P_3, P_6 \) belong to the orbit corresponding to $[2,0,1]$.
\end{itemize}

Thus, the curve \( X \) decomposes into three connected components:
one containing \( \{P_1, P_4, P_7, P_8\} \),
one containing \( \{P_2, P_5\} \), and
one containing \( \{P_3, P_6\} \).
The first component contains $1+3+3+1=8$ elements of \(\Omega \), and the second and third ones contain $3+3=6$ elements.
Therefore, the defining equation of \( C_0 \) factors as
\[
(\text{a degree-8 polynomial in } v) \times
(\text{a degree-6 polynomial in } v) \times
(\text{a degree-6 polynomial in } v)=0.
\]
See \eqref{eq:def_poly_63} in Appendix \S \ref{sec:appB} for explicit numerical results.
\end{example}

\subsection{Ramification Points in \( \Gamma_0 \)}

To compute algebro-geometric invariants of $X$, such as the genus, we need to study the ramification behavior of the covering \( w: X \to \mathbb{P} \). 
However, elementary computations like those in Example~\ref{eq:63_remification} are quite tedious for general $L,N$.
In this subsection, focusing on the locus $\Gamma_0$, we present a representation theory-based technique to compute both the number of points on $\Gamma_0$ and their ramification indices.
Similar computations can also be performed for the loci \( \Gamma_{c^\ast}:=\{w = c^*\} \) and \( \Gamma_\infty:=\{w = \infty\} \).

Recall that the subgroup \( G_1=\langle \widehat{\sigma}\rangle \) is a cyclic group of order \( d = \mathrm{lcm}(N, L - N) \). As is well known, any irreducible representation of a cyclic group of order \( d \) is one-dimensional.
There are exactly \( d \) one-dimensional representations (up to isomorphism). 
We denote them by \( V_0, V_1, \dots, V_{d-1} \). The character \( \chi_i(g) := \mathrm{tr}_{V_i}(g) \) on $V_j$ is computed by
\[
\chi_i(\widehat{\sigma}^j) = \exp\left(2\pi \sqrt{-1} \cdot \frac{ij}{d}\right).
\]

Let $E(G_1)$ be the set of all class functions on $G_1$:
\[
E(G_1):=\{\chi:G_1\to \CC\;;\;\chi(gxg^{-1})=\chi(x),\ \forall x,g\in G_1\}.
\]
It is known that $E(G_1)$ is a finite dimensional $\CC$-vector space and the bilinear form
\[
(\chi,\mu)_{G_1}:=\frac{1}{\# G_1}\sum_{g\in G_1}\chi(g)\cdot \overline{\mu(g)}
\]
is symmetric and nondegenerate.
The set of characters of all (finite-dimensional) irreducible representations forms an orthonormal basis of $E(G_1)$.

Let $\mathbb{C}[\Omega]$ be the group algebra
\[
\mathbb{C}[\Omega] = \bigoplus_{x \in \Omega} \mathbb{C} \cdot x
\]
over the set $\Omega$.
Consider the decomposition of \( \mathbb{C}[\Omega] \) into irreducible \( G_1 \)-representations:
\begin{equation}\label{eq:decomp_of_Omega}
\mathbb{C}[\Omega] \simeq V_0^{\oplus e_0} \oplus V_1^{\oplus e_1} \oplus \dots \oplus V_{d-1}^{\oplus e_{d-1}}.    
\end{equation}
Then, the multiplicity \( e_i \) is computed as
\[
e_i = (\chi_{\mathbb{C}[\Omega]}, \chi_i)_{G_1} = \frac{1}{d} \sum_{j=0}^{d-1} \chi_{\mathbb{C}[\Omega]}(\widehat{\sigma}^j) \cdot \exp\left(-2\pi \sqrt{-1} \cdot \frac{ij}{d}\right).
\]
For $\ell\mid d$, a point in $\Gamma_0$ with ramification index $\ell$ contributes as $V_0\oplus V_{d/\ell}\oplus V_{2d/\ell}\oplus \dots \oplus V_{d-d/\ell}$ to the direct sum \eqref{eq:decomp_of_Omega}.
Then, we can examine how many points are contained in $\Gamma_0$ and compute their ramification indices from the numbers $e_0,e_1,\dots,e_{d-1}$.

Summarizing the above, we have the following proposition:
\begin{prop}
Suppose that the locus \( \Gamma_0:=\{w = 0\} \) in \( X \) contains \( m_i \) points with ramification index \( d/i \). Then, as a \( G_1 \)-module, $\CC[\Omega]$ has an irreducible decomposition as
\[
\mathbb{C}[\Omega] = \bigoplus_{i \mid d} (V_0 \oplus V_i \oplus V_{2i} \oplus \dots \oplus V_{d-i})^{\oplus m_i}.
\]
In particular, the cardinality of $\Gamma_0$ is \( \sum_{i \mid d} m_i = e_0 = \langle \chi_{\mathbb{C}[\Omega]}, \chi_0 \rangle_{G_1} \).
\end{prop}

\begin{example}
[Ramification points in $\Gamma_0$ for $(L,N)=(6,3)$]
\label{ex:64_decomp_of_Omega}
When \( (L,N)=(6,3) \), we have \( d = \mathrm{lcm}(3,6-3)=3 \). 
Since 
\[
\chi_{\mathbb{C}[\Omega]}(g)=\# \{\mbox{fixed points under the action of $g$ on the set $\Omega$}\},
\]
we find
\[
\chi_{\mathbb{C}[\Omega]}(\mathrm{id}) = 20, \quad
\chi_{\mathbb{C}[\Omega]}(\widehat{\sigma}) = \chi_{\mathbb{C}[\Omega]}(\widehat{\sigma}^2) = 2.
\]
Therefore,
\[
e_0 = \frac{20 + 2 + 2}{3} = 8, \quad e_1 = e_2 = \frac{20 + 2\omega_3 + 2\omega_3^2}{3} = 6,
\]
where $\omega_3$ is a primitive third root of unity.
Hence, we obtain the decomposition:
\begin{equation}\label{eq:decomp_in_a_example_63}
\mathbb{C}[\Omega] \simeq V_0^{\oplus 8} \oplus V_1^{\oplus 6} \oplus V_2^{\oplus 6}.
\end{equation}
The direct sum \eqref{eq:decomp_in_a_example_63} can be uniquely decomposed into finitely many copies of $V_0$ and $V_0\oplus V_1\oplus V_2$ as
\[
\mathbb{C}[\Omega] \simeq (V_0)^{\oplus 2} \oplus (V_0 \oplus V_1 \oplus V_2)^{\oplus 6}.
\]
The first two copies of \( V_0 \) correspond to the points \( P_1\) and \( P_8 \) in Example \ref{eq:63_remification}, and the six copies of \( V_0 \oplus V_1 \oplus V_2 \) correspond to the points \( P_2, P_3, \dots, P_7 \).
\end{example}

\begin{example}
[Ramification points in $\Gamma_0$ for $(L,N)=(8,4)$]
\label{ex:84_decomp_of_Omega}
When \( (L,N)=(8,4) \), we have \( d = \mathrm{lcm}(4
,8-4)=4 \). By counting fixed points of \( \widehat{\sigma}^j \), we obtain
\[
\chi_{\mathbb{C}[\Omega]}(\mathrm{id}) = 70, \quad
\chi_{\mathbb{C}[\Omega]}(\widehat{\sigma}) = \chi_{\mathbb{C}[\Omega]}(\widehat{\sigma}^3) = 2, \quad
\chi_{\mathbb{C}[\Omega]}(\widehat{\sigma}^2) = 6,
\]
and the multiplicities
\[
e_0=\frac{70+2+6+2}{4}=20,\quad e_1=e_3=\frac{70+2\sqrt{-1} -6-2\sqrt{-1}}{4}=16,\quad
e_2=\frac{70-2+6-2}{4}=18.
\]
Thus, $\CC[\Omega]$ is decomposed as
\[
\begin{aligned}
\mathbb{C}[\Omega] 
&\simeq V_0^{\oplus 20} \oplus V_1^{\oplus 16} \oplus V_2^{\oplus 18} \oplus V_3^{\oplus 16}\\
&=(V_0)^{\oplus 2} \oplus (V_0 \oplus V_2)^{\oplus 2} \oplus (V_0 \oplus V_1 \oplus V_2 \oplus V_3)^{\oplus 16}.
\end{aligned}
\]
This implies that $\Gamma_0$ consists of two unramified points,
two points with ramification index $2$,
and sixteen points with ramification index $4$.
\end{example}

The same type of computation can be performed connected component-wise ($=G_2$-orbit-wise). 
Let \( \mathcal{O} \subset \Omega \) be a \( G_2 \)-orbit, which corresponds to a connected component \( X_\mathcal{O} \) of \( X \).

\begin{prop}
Suppose the locus \( \Gamma_0\cap  X_\mathcal{O}\) contains \( m_i \) points with ramification index \( d/i \). Then, as a \( G_1 \)-module, $\CC[\mathcal{O}]$ is decomposed as
\[
\mathbb{C}[\mathcal{O}] = \bigoplus_{i \mid d} (V_0 \oplus V_i \oplus V_{2i} \oplus \dots \oplus V_{d-i})^{\oplus m_i}.
\]
In particular, the number of such points is \( \sum_{i \mid d} m_i = \langle \chi_{\mathbb{C}[\mathcal{O}]}, \chi_0 \rangle_{G_1} \).
\end{prop}

\begin{example}
[Ramification points in $\Gamma_0\cap X_{111}$ for $(L,N)=(6,3)$]
\label{ex:decompO_111_63}
Let \( (L,N)= (6,3) \) and \( d = 3 \). 
The \( G_2 \)-orbit \( \mathcal{O}_{111} \) corresponding to \( [1,1,1] \) contains the following $8$ elements:
\[
\{1,2,3\}, \{4,2,3\}, \{1,5,3\}, \{4,5,3\},
\{1,2,6\}, \{4,2,6\}, \{1,5,6\}, \{4,5,6\}.
\]
Let \( \mathbb{C}[\mathcal{O}_{111}] \) be the group algebra over $\mathcal{O}_{111}$. 
Through a similar calculation as in Examples \ref{ex:64_decomp_of_Omega} and \ref{ex:84_decomp_of_Omega}, we obtain the decomposition
\[
\mathbb{C}[\mathcal{O}_{111}] \simeq V_0^{\oplus 4} \oplus V_1^{\oplus 2} \oplus V_2^{\oplus 2} = (V_0)^{\oplus 2} \oplus (V_0 \oplus V_1 \oplus V_2)^{\oplus 2}.
\]
This implies that the locus \( \Gamma_0\cap X_{111} \) consists of two unramified points and two points with ramification index $3$.
\end{example}

\subsection{Counting connected components of \( X \)}

The number of connected components of $X$, equal to $\# \overline{Z}$, is also an important invariant.
We need some combinatorial technique to enumerate it.
For each tuple \( (a_0, a_1, \dots, a_{L'}) \) of natural numbers satisfying
\begin{equation}\label{eq:exp_of_N}
a_0+a_1+\dots+a_{L'}=e,\quad
a_1+2a_2+\dots+L'a_{L'}=N,
\end{equation}
we define the subset
\begin{equation}\label{eq:def_of_Z(a,a)}
Z(a_0, a_1, \dots, a_{L'}) := \{ (A_1, \dots, A_e)\in Z \; ; \;
\# (i \text{ in } \{A_1,\dots,A_e\})=a_i\}
\end{equation}
of $Z$.
Then, the set \( Z \) is decomposed as
\begin{equation}\label{eq:decomp_of_Z}
Z = \bigsqcup_{(a_0, \dots, a_{L'})} Z(a_0, \dots, a_{L'}).    
\end{equation}
Note that the cardinality of \( Z(a_0, \dots, a_{L'}) \) is expressed as
\[
\# Z(a_0, a_1, \dots, a_{L'}) = \binom{e}{a_0,a_1, \dots, a_{L'}},
\]
where $\binom{e}{a_0,a_1, \dots, a_{L'}}$ is the multinomial coefficient
\[
\binom{n}{m_1, m_2, \dots, m_p} := \frac{n!}{m_1! m_2! \dots m_p! (n - m_1 - m_2 - \dots - m_p)!}.
\]
Since each \( Z(a_0, \dots, a_{L'}) \) is invariant under the \( \mathbb{Z}/e\mathbb{Z} \)-action, the problem is reduced to counting the $\ZZ/e\ZZ$-orbits in \( Z(a_0, \dots, a_{L'}) \).

Recall the surjection $\pi:\Omega\to Z$ defined in \eqref{eq:def_of_pi}.
For any element \( (A_1, \dots, A_e) \in Z(a_0, \dots, a_{L'}) \), the cardinality of the preimage \( \pi^{-1}(A_1, \dots, A_e) \) is given by
\[
\binom{L'}{0}^{a_0} \binom{L'}{1}^{a_1} \dots \binom{L'}{L'}^{a_{L'}}.
\]
Therefore, if the length of the \( \mathbb{Z}/e\mathbb{Z} \)-orbit of \( \pi(I) \) is \( \ell \), the cardinality of a \( G_2 \)-orbit \( \mathcal{O} \ni I \) is
\begin{equation}\label{eq:card_of_O}
\# \mathcal{O} = \ell \cdot \binom{L'}{0}^{a_0} \binom{L'}{1}^{a_1} \dots \binom{L'}{L'}^{a_{L'}}
, \quad \text{where } \pi(I) \in Z(a_0, \dots, a_{L'}).
\end{equation}

\begin{rem}
For $\ell\mid e$, let $m_\ell(a_0,\dots,a_{L'})$ denote the number of $\ZZ/e\ZZ$-orbits in $Z(a_0,\dots,a_{L'})$ of length $\ell$.
The number $m_\ell(a_0,\dots,a_{L'})$ is given by the equation
\begin{equation}\label{eq:explicit_form_m}
m_{\ell}(a_0,\dots,a_{L'})=\frac{1}{\ell}
\sum_{\substack{\ell'\mid \ell\\
e\mid a_i\ell'\ (\forall i)
}}
\binom{\ell'}{\frac{a_0\ell'}{e},\dots,
\frac{a_{L'}\ell'}{e}}\cdot \mu(\ell/\ell'),
\end{equation}
where $\mu$ is the M\"{o}bius function.
(We will give a proof of \eqref{eq:explicit_form_m} in Appendix \S \ref{sec:appD}.)
Then, we have $\ZZ/e\ZZ$-orbit decomposition:
\begin{equation}\label{eq:decomp_of_multinomial}
\# Z(a_0,\dots,a_{L'})=
\binom{e}{a_0,a_1,\dots,a_{L'}}=\sum_{\ell\mid e}m_\ell(a_0,\dots,a_{L'})\cdot \ell.
\end{equation}    

On the other hand, taking the preimage of both sides of \eqref{eq:decomp_of_Z} under \( \pi \), we obtain the decomposition
\[
\Omega = \bigsqcup_{\substack{a_0 + a_1 + \dots + a_{L'} = e \\ a_1 + 2a_2 + \dots + L'a_{L'} = N}} \pi^{-1}(Z(a_0, \dots, a_{L'})),
\]
which yields the equation
\[
\binom{L}{N} = \sum_{\substack{a_0 + a_1 + \dots + a_{L'} = e \\ a_1 + 2a_2 + \dots + L'a_{L'} = N}} 
\# Z(a_0,\dots,a_{L'})
\binom{L'}{0}^{a_0} \binom{L'}{1}^{a_1} \dots \binom{L'}{L'}^{a_{L'}}.
\]
Furthermore, by substituting \eqref{eq:decomp_of_multinomial}, we obtain
\begin{equation}\label{eq:decomp_G2_Omega}
\binom{L}{N} = \sum_{\substack{a_0 + a_1 + \dots + a_{L'} = e \\ a_1 + 2a_2 + \dots + L'a_{L'} = N}} \sum_{\ell \mid e} m_\ell(a_0,\dots,a_{L'}) \cdot \ell \cdot \binom{L'}{0}^{a_0} \binom{L'}{1}^{a_1} \dots \binom{L'}{L'}^{a_{L'}}.
\end{equation}
Recall \eqref{eq:card_of_O}. Then we see that this equation represents the $G_2$-orbit decomposition of \( \Omega \).
\end{rem}

Let $\mathcal{N}$ be the number of $G_2$-orbits in $\Omega$.
The following lemma gives an explicit formula for $\mathcal{N}$.

\begin{lemma}\label{lemma:formula_for_N}
Let \( \varphi(n) \) denote Euler's totient function
(the number of the positive integers up to integer $n$ that are relatively prime to $n$). Then, the number \( \mathcal{N} \) of \( G_2 \)-orbits in \( \Omega \) is given by
\[
\mathcal{N} = \frac{1}{e} \sum_{\substack{a_0 + a_1 + \dots + a_{L'} = e \\ a_1 + 2a_2 + \dots + L'a_{L'} = N}} \sum_{\substack{d \mid e \\ e \mid a_i \cdot d\ (\forall i)}} \varphi(e/d) \binom{d}{\frac{a_0 d}{e}, \dots, \frac{a_{L'} d}{e}}.
\]
\end{lemma}
\begin{proof}
From \eqref{eq:decomp_G2_Omega}, the number $\mathcal{N}$ of $G_2$-orbits is given by
\[
\mathcal{N} = \frac{1}{e} \sum_{\substack{a_0 + a_1 + \dots + a_{L'} = e \\ a_1 + 2a_2 + \dots + L'a_{L'} = N}} \sum_{\ell\mid e} m_\ell(a_0,\dots,a_{L'}).
\]
Then, the lemma follows from the equation
\begin{equation}\label{eq:sum_of_m}
\begin{aligned}
\sum_{\ell\mid e}m_{\ell}(a_0,\dots,a_{L'})
=
\frac{1}{e}
\sum_{
\substack{
\ell'\mid e\\
e\mid a_i\ell'\ (\forall i)
}
}\binom{\ell'}{\frac{a_0\ell'}{e},\dots,
\frac{a_{L'}\ell'}{e}
}\cdot \varphi(e/\ell').
\end{aligned}
\end{equation}
We will give proof of this equation in Appendix \S \ref{sec:appD}.
\end{proof}

The explicit formula in Lemma \ref{lemma:formula_for_N} can be simplified by using \textit{partitions}.
A tuple $(a_0,\dots,a_{L'})$ under the constraints \eqref{eq:exp_of_N} can be identified with a partition of the natural number $N$, where the number of parts does not exceed $e$ and each part does not exceed $L'$:
\[
(a_0, a_1, \dots, a_{L'}) \leftrightarrow \lambda = (\underbrace{L', \dots, L'}_{a_{L'}}, \dots, \underbrace{1, \dots, 1}_{a_1}, \underbrace{0, \dots, 0}_{a_0}).
\]
Let \( \mathcal{P}_{L', e}(N) \) be the set of partitions
of $N$ with
length at most \( e \), each part not exceeding \( L' \).

For a partition \( \lambda \), let \( m_i(\lambda) \) denote the number of \( i \)'s contained in \( \lambda \).  
Rearranging the non-negative integers $m_0(\lambda),m_1(\lambda),\dots,m_{L'}(\lambda)$ in decreasing order, we again obtain a partition, denoted by \( m(\lambda) \).
Let \( \mathcal{P} \) denote the set of all partitions. 
Through the procedure explained above, we define the map
\[
m : \mathcal{P}_{L', e}(N) \to \mathcal{P}; \quad \lambda \mapsto m(\lambda).
\]
For a partition \( \mu \in \mathcal{P} \), let \( \mathcal{N}(\mu) \) denote the number of elements in \( m^{-1}(\mu) \). Then equation~\eqref{eq:exp_of_N} can be rewritten as
\begin{equation}\label{eq:N_partition}
\mathcal{N} = \frac{1}{e} \sum_{\mu \in \mathcal{P}} \mathcal{N}(\mu) \sum_{\substack{d \mid e \\ e \mid \mu_i \cdot d\,(\forall i)}} \varphi(e/d) \binom{d}{\frac{\mu_1 d}{e}, \dots, \frac{\mu_{\ell(\mu)} d}{e}},
\end{equation}
which is better suited for direct computation.
See the following examples.

\ytableausetup{smalltableaux}

\begin{example}[$\mathcal{N}$ for $(L,N)=(6,3)$]
When \( (L,N) = (6,3) \), we have $e=3$ and \( L' = 2 \). 
The set \( \mathcal{P}_{2,3}(3) \) consists of
\[ \ydiagram{2,1}\quad \ydiagram{1,1,1} \]
with images under \( m \) being \( (1,1) \) and \( (3) \), respectively. Thus, the total number of \( G_2 \)-orbits is
\[
\begin{aligned}
\mathcal{N}&=
\frac{1}{3}\left[
\mathcal{N}(1,1)\sum_{d}\varphi(3/d)\binom{d}{\frac{d}{3},\frac{d}{3}}
+
\mathcal{N}(3)\sum_{d}\varphi(3/d)\binom{d}{\frac{3d}{3}}
\right]\\
&=
\frac{1}{3}\left[
\varphi(1)\binom{3}{1,1}+\varphi(3)\binom{1}{1}+\varphi(1)\binom{3}{3}
\right]
=\frac{6+2+1}{3}=3.
\end{aligned}
\]
\end{example}

\begin{example}[$\mathcal{N}$ for $(L,N)=(9,3)$]\label{ex:num_of_orbit_93}
When \( (L,N)= (9,3) \), we have \( e = L'=3 \). The set \( \mathcal{P}_{3,3}(3) \) consists of
\[ \ydiagram{3}\quad \ydiagram{2,1}\quad \ydiagram{1,1,1} \]
with images under \( m \) being \( (1) \), \( (1,1) \), and \( (3) \), respectively. Thus,
\[
\begin{aligned}
\mathcal{N}&
=
\frac{1}{3}\left[
\mathcal{N}(1)\sum_{d}\varphi(3/d)\binom{d}{\frac{d}{3}}
+
\mathcal{N}(1,1)\sum_{d}\varphi(3/d)\binom{d}{\frac{d}{3},\frac{d}{3}}
+
\mathcal{N}(3)\sum_{d}\varphi(3/d)\binom{d}{\frac{3d}{3}}
\right]\\
&=
\frac{1}{3}\left[
\varphi(1)\binom{3}{1}+\varphi(1)\binom{3}{1,1}+\varphi(3)\binom{1}{1}
+\varphi(1)\binom{3}{3}
\right]
=\frac{3+6+2+1}{3}=4.
\end{aligned}
\]
\end{example}



\subsection{The defining equation of a specific connected component}



Recall \( e = \mathrm{gcd}(L, N) \), \( L = eL' \), and \( N = eN' \). If $e\neq 1$, the equation $
w = \frac{t^L}{(1 - t)^N}
$
(Eq.~\eqref{eq:w_to_t}) can be decomposed into \( e \) independent equations:
\begin{equation}\label{eq:decomposed_eq_w_to_t}
\omega_e^{k-1} w^{1/e} = \frac{t^{N'}}{(1 - t)^{L'}}, \qquad (k = 1, 2, \dots, e),    
\end{equation}
where \( \omega_e \) denotes the primitive \( e \)-th root of unity.
The branch of the \( e \)-th root \( w^{1/e} \) is chosen near \( w \to 0 \) so that
\[
w^{1/e} = \frac{t_1(w)^{N'}}{(1 - t_1(w))^{L'}}.
\]
Then, the equation \eqref{eq:decomposed_eq_w_to_t} has $L'$ roots $t=t_k, t_{e + k}, t_{2e + k}, \dots, t_{(L' - 1)e + k}$.

Let \( \mathcal{O} \) be a \( G_2 \)-orbit. 
Define the \textit{\( \mathcal{O} \)-elementary symmetric polynomial} \( E_n \) by
\[
\prod_{I \in \mathcal{O}} (1 + t_I T) = 1 + E_1 T + \dots + E_{|\mathcal{O}|} T^{|\mathcal{O}|},
\]
and the \textit{\( \mathcal{O} \)-power sum} \( P_n \) by
\[
P_n := \sum_{I \in \mathcal{O}} (t_I)^n.
\]
(Recall $t_I=\prod_{i\in I}t_i$.)
Since \( E_n, P_n \) are invariant under the transpositions \( t_k \leftrightarrow t_{k+e} \) for all \( k \), and under the substitution map \( w^{1/e} \mapsto \omega_e \cdot w^{1/e} \), we have \( E_n, P_n \in \mathbb{Z}[\omega_e][w^{-1}] \).

\begin{rem}
One cannot improve the statement \( E_n, P_n \in \mathbb{Z}[\omega_e][w^{-1}] \) to \( E_n, P_n \in\mathbb{Z}[w^{-1}] \).
See Appendix \S \ref{eq:appC}.
\end{rem}

The defining equation of $X_{\mathcal{O}}$ is given by
\[
v^{\# \mathcal{O}}-E_1v^{\# \mathcal{O}-1}+E_2v^{\# \mathcal{O}-2}-\dots+(-1)^{\# \mathcal{O}}E_{\# \mathcal{O}}=0,
\]
where $E_n\in \ZZ[\omega_e][w^{-1}]$ is the $\mathcal{O}$-elementary symmetric polynomial.


For small examples, we demonstrate how to calculate the defining equation of a specific connected component.
When \( (L,N) = (4,2) \), we have \( e = 2 \). 
In this case, $X$ has two connected components associated to  $[1,1]$, and $[2,0]$.
From \( (-1)^{k-1} w^{1/2} = \frac{t}{(1 - t)^2} \), we obtain the quadratic equation
\begin{equation}\label{eq:alg_equation_42}
t^2 - (2 + (-1)^{1-k} w^{-1/2})t + 1 = 0,\quad (k=1,2).
\end{equation}
The branches \( t=t_1, t_3 \) are the roots of \eqref{eq:alg_equation_42} with \( k = 0 \), and \( t=t_2, t_4 \) with \( k = 1 \). 
From Vieta's formula, we have
\[
t_1 + t_3 = 2 + w^{-1/2}, \quad t_2 + t_4 = 2 - w^{-1/2}, \quad t_1 t_3 = t_2 t_4 = 1.
\]

\begin{example}[The defining polynomial of $X_{11}$ for $(L,N)=(4,2)$]
Let \( \mathcal{O}_{11} \subset \Omega \) be the \( G_2 \)-orbit corresponding to \( [1,1] \). As previously seen, we have $ \mathcal{O}_{11} = \{\{1,2\}, \{3,2\}, \{1,4\}, \{3,4\}\}$.
The \( \mathcal{O}_{11} \)-power sums \( P_n \) are given by
\[
P_n := \sum_{\{i_1, i_2\} \in \mathcal{O}_{11}} (t_{i_1} t_{i_2})^n.
\]
Then, it follows that \( P_n = (t_1^n + t_3^n)(t_2^n + t_4^n) \), which yields
\[
P_1 = 4 - w^{-1}, \quad P_2 = 4 - 12 w^{-1} + w^{-2}.
\]
By Newton's formula (see Lemma \eqref{lemma:Newton} in Appendix \S \ref{sec:appA}), we obtain the $\mathcal{O}_{11}$-elementary symmetric polynomials $E_1,E_2,E_3,E_4$ by
\[
\begin{gathered}
E_1 = P_1 = 4 - w^{-1}, \quad E_2 = \frac{P_1^2 - P_2}{2} = 6 + 2w^{-1}, \\
E_3 = t_{12} t_{32} t_{14} + t_{12} t_{32} t_{34} + t_{12} t_{14} t_{34} + t_{32} t_{14} t_{34} = E_1 = 4 - w^{-1}, \\
E_4 = t_{12} t_{32} t_{14} t_{34} = 1.
\end{gathered}
\]
Hence, the connected component $X_{11}$ corresponding to \( [1,1] \) is defined by
\[
v^4 - (4 - w^{-1})v^3 + (6 + 2w^{-1})v^2 - (4 - w^{-1})v + 1 = 0.
\]    
\end{example}

\begin{example}[The defining polynomial of $X_{20}$ for $(L,N)=(4,2)$]
For the orbit \( \mathcal{O}_{20} = \{\{1,3\}, \{2,4\}\} \) corresponding to \( [2,0] \), the \( \mathcal{O}_{20} \)-symmetric functions are
\[
E_1 = t_{13} + t_{24} = 2, \quad E_2 = t_{13} t_{24} = 1.
\]
Thus, the connected component $X_{20}$ corresponding to \([2,0] \) is defined by
\[
v^2 - 2v + 1 = (v - 1)^2 = 0.
\]    
Compare this to \eqref{eq:def_of_C0_42}.
In Corollary \ref{GMparticulartype}, we will show that the defining polynomial corresponding to a certain type of $G_2$-orbit must be a power of $v-1$.
\end{example}

\section{Genus}
In this section, we give several explicit expressions
for the genus of $X$ for $L$,$N$ generic.
Some of the formulas reproduce examples studied by
Prolhac \cite{ProlhacRStwo}.

\subsection{Genus computation for a specific connected component}

For a $G_2$-orbit \( \mathcal{O} \subset \Omega \), let \( X_{\mathcal{O}} \) denote the connected component of \( X \) corresponding to \( \mathcal{O} \).
The ramification points of \( w : X_{\mathcal{O}} \to \mathbb{P}^1 \), which is a $\# \mathcal{O}$-to-$1$ covering of the projective line, are contained in the set \( K=\{0, c^*, \infty\} \).
Recall that the group $G_1:=\langle\widehat{\sigma}\rangle$ gives the monodromy action around $w=0$.
Similarly, the groups $Y_1:=\langle\widehat{\tau}\rangle$ and $Y_2:=\langle\widehat{\sigma}^{-1}\circ \widehat{\tau}\rangle$ give the monodromy actions around $w=\infty$ and $w=c^\ast$, respectively.

Let \( g(X_{\mathcal{O}}) \) be the genus of $X_\mathcal{O}$.
By applying the Riemann-Hurwitz formula to $w:X_\mathcal{O}\to \PP$, we obtain the formula
\[
g(X_{\mathcal{O}}) = 1 - \# \mathcal{O} + \frac{1}{2} \sum_p (\vartheta(p) - 1),
\]
where $\vartheta(p)$ is the ramification index of $p\in \PP$.

Consider the decomposition of the group algebra \( \mathbb{C}[\mathcal{O}] \) as a $G_1$-representation:
\[
\mathbb{C}[\mathcal{O}] = \bigoplus_{j=0}^{\# \mathcal{O} - 1} V_j^{\oplus e_j},
\]
where $V_0,V_1,\dots,V_{d}$ are the one-dimensional irreducible representations of $G_1$.
From the fiber $\Gamma_0=w^{-1}(0)$, we have
\[
\sum_{p \in w^{-1}(0)} \vartheta(p) = \# \mathcal{O}, \quad \sum_{p \in w^{-1}(0)} 1 = e_0 = (\chi_{\mathbb{C}[\mathcal{O}]}, \chi_{V_0})_{G_1}.
\]

Similarly, from the loci $w^{-1}(\infty)$ and $w^{-1}(c^\ast)$, we obtain 
\[
\sum_{p \in w^{-1}(\infty)} \vartheta(p) = 
\sum_{p \in w^{-1}(c^\ast)} \vartheta(p) = 
\# \mathcal{O}, \quad 
\sum_{p \in w^{-1}(0)} 1 =  (\chi_{\mathbb{C}[\mathcal{O}]}, \chi_{V_0})_{Y_1},\quad
\sum_{p \in w^{-1}(c^\ast)} 1 =  (\chi_{\mathbb{C}[\mathcal{O}]}, \chi_{V_0})_{Y_2}.
\]
Summing up these equalities, we obtain
\begin{equation}\label{eq:genus_indv}
\begin{aligned}
g(X_{\mathcal{O}})
&= 1 - \# \mathcal{O} + \frac{3}{2} \# \mathcal{O} - \frac{1}{2} \left[
(\chi_{\mathbb{C}[\mathcal{O}]}, \chi_{V_0})_{G_1} +
(\chi_{\mathbb{C}[\mathcal{O}]}, \chi_{V_0})_{Y_1} +
(\chi_{\mathbb{C}[\mathcal{O}]}, \chi_{V_0})_{Y_2}
\right] \\
&= 1 + \frac{\# \mathcal{O}}{2} - \frac{1}{2} \left[
(\chi_{\mathbb{C}[\mathcal{O}]}, \chi_{V_0})_{G_1} +
(\chi_{\mathbb{C}[\mathcal{O}]}, \chi_{V_0})_{Y_1} +
(\chi_{\mathbb{C}[\mathcal{O}]}, \chi_{V_0})_{Y_2}
\right].
\end{aligned}    
\end{equation}

\begin{example}[Genus of $X$ for $(L,N)=(3,2)$]
When \( (L,N)=(3,2) \), $\Omega$ itself is a \( G_2 \)-orbit.
On $\Omega$, the orders of $\widehat{\sigma}$ and $\widehat{\tau}$ are $2$ and $3$, respectively.
By counting fixed points, we have
\[
\chi_{\mathbb{C}[\Omega]}(\mathrm{id}) = \binom{3}{2} = 3, \quad
\chi_{\mathbb{C}[\Omega]}(\widehat{\sigma}) = 1, \quad
\chi_{\mathbb{C}[\Omega]}(\widehat{\tau}) =
\chi_{\mathbb{C}[\Omega]}(\widehat{\tau}^2) = 0, \quad
\chi_{\mathbb{C}[\Omega]}(\widehat{\tau}^{-1} \cdot \widehat{\sigma}) = 1.
\]
Thus, from \eqref{eq:genus_indv}, we have
\[
g(X) = 1 + \frac{1}{2} \cdot 3 - \frac{1}{2} \left[ \frac{3+1}{2} + \frac{3}{3} + \frac{3+1}{2} \right] = 0.
\]
\end{example}

\begin{example}[Genus of $X_{20}$ for $(L,N)=(6,2)$]
When \( (L,N)=(6,2)\), we have $e=2$.
There are two \( G_2 \)-orbits in \( \Omega \); One corresponds to \( [2,0] \) and the other to \( [1,1] \).
For the first orbit \( \mathcal{O}_{20} = \{\{1,3\}, \{1,5\}, \{3,5\}, \{2,4\}, \{2,6\}, \{4,6\}\} \), we have \( \# \mathcal{O}_{20} = 6 \).
The orders of $\widehat{\sigma}$ and $\widehat{\tau}$ on $\mathcal{O}_{20}$ are $4$ and $6$, respectively.
Counting fixed points gives
\[
\begin{aligned}
&\chi_{\mathbb{C}[\mathcal{O}_{11}]}(\mathrm{id}) = 6, \quad
\chi_{\mathbb{C}[\mathcal{O}_{11}]}(\widehat{\sigma}) = \chi_{\mathbb{C}[\mathcal{O}_{11}]}(\widehat{\sigma}^3) = 0, \quad
\chi_{\mathbb{C}[\mathcal{O}_{11}]}(\widehat{\sigma}^2) = 2, \\
&\chi_{\mathbb{C}[\mathcal{O}_{11}]}(\widehat{\tau}^k) = 0 \quad (1 \leq k \leq 5), \quad
\chi_{\mathbb{C}[\mathcal{O}_{11}]}(\widehat{\tau}^{-1} \cdot \widehat{\sigma}) = 4.
\end{aligned}
\]
Thus, from \eqref{eq:genus_indv}, we have
\[
g(X_{20}) = 1 + \frac{1}{2} \cdot 6 - \frac{1}{2} \left[ \frac{6+2}{4} + \frac{6}{6} + \frac{6+4}{2} \right] = 0.
\]
\end{example}

\subsection{Computation of total genus}

For a non-connected algebraic curve \( X \), the sum of the genera of its connected components is called the \textit{total genus}.
Let $g(X)$ be the total genus of $X$.
Summing up the genus formulas for individual connected components, we obtain the genus formula
\begin{equation}\label{eq:sub_formula_for_total_genus}
g(X) = \mathcal{N} + \frac{\# \Omega}{2}  - \frac{1}{2} \left[
(\chi_{\mathbb{C}[\Omega]}, \chi_{V_0})_{G_1} +
(\chi_{\mathbb{C}[\Omega]}, \chi_{V_0})_{Y_1} +
(\chi_{\mathbb{C}[\Omega]}, \chi_{V_0})_{Y_2}
\right].    
\end{equation}

Using techniques from the representation theory, we can derive an explicit formula for $g(X)$.
\begin{lemma}\label{lemma:character_formula_2}
The character $\chi_{\CC[\Omega]}$ satisfies the following equations:
\begin{align}
&\chi_{\mathbb{C}[\Omega]}(\mathrm{id}) = \binom{L}{N}, \label{subeq:to_prove_1}\\
&\chi_{\mathbb{C}[\Omega]}(\widehat{\sigma}^k) = \sum_{\substack{p \cdot \frac{N}{\mathrm{gcd}(N,k)} + q \cdot \frac{L - N}{\mathrm{gcd}(L - N,k)} = N}} \binom{\mathrm{gcd}(N,k)}{p} \binom{\mathrm{gcd}(L - N,k)}{q}, \label{subeq:to_prove_2}\\
&\chi_{\mathbb{C}[\Omega]}(\widehat{\tau}^k) =
\begin{cases}
\binom{\mathrm{gcd}(L,k)}{\mathrm{gcd}(L,k) \cdot N/L} & (L \mid \mathrm{gcd}(L,k) \cdot N) \\
0 & (\text{otherwise})
\end{cases}, \label{subeq:to_prove_3}\\
&\chi_{\mathbb{C}[\Omega]}(\widehat{\tau}^{-1} \cdot \widehat{\sigma}) = \binom{L}{N} - 2 \binom{L - 2}{N - 1}.\label{subeq:to_prove_4}
\end{align}
\end{lemma}
\begin{proof}
We will give a proof of this lemma in Appendix \S \ref{sec:appD}.
\end{proof}

Let
\[
F_{L,N}(k) := \sum_{\substack{p \cdot \frac{N}{\mathrm{gcd}(N,k)} + q \cdot \frac{L - N}{\mathrm{gcd}(L - N,k)} = N}} \binom{\mathrm{gcd}(N,k)}{p} \binom{\mathrm{gcd}(L - N,k)}{q}.
\]
Then, from Lemma \ref{lemma:character_formula_2}, we have
\[
(\chi_{\mathbb{C}[\Omega]}, \chi_{V_0})_{G_1}=\frac{\sum_{k=1}^{\mathrm{lcm}(N,L-N)} F_{L,N}(k)}{\mathrm{lcm}(N,L-N)},\quad
(\chi_{\mathbb{C}[\Omega]}, \chi_{V_0})_{Y_1}=\frac{\sum_{L\mid \mathrm{gcd}(L,k)\cdot N}\binom{\mathrm{gcd}(L,k)}{\mathrm{gcd}(L,k) \cdot N / L}}{L},
\]
and 
\[
(\chi_{\mathbb{C}[\Omega]}, \chi_{V_0})_{Y_2}=\frac{\# \Omega+\chi_{\CC[\Omega]}(\widehat{\tau}^{-1}\cdot\widehat{\sigma}) }{2}=\binom{L}{N}-\binom{L-2}{N-1}.
\]
By substituting them into \eqref{eq:sub_formula_for_total_genus}, we obtain
\begin{equation}\label{eq:total_genus}
g(X) = \mathcal{N} + \frac{1}{2} \binom{L - 2}{N - 1} - \frac{1}{2} \left[
\frac{\sum_{k = 1}^{\mathrm{lcm}(N, L - N)} F_{L,N}(k)}{\mathrm{lcm}(N, L - N)} +
\frac{\sum_{L \mid \mathrm{gcd}(L,k) \cdot N} \binom{\mathrm{gcd}(L,k)}{\mathrm{gcd}(L,k) \cdot N / L}}{L}
\right].
\end{equation}

For positive integers \( m, n \), define
\[
\varphi_{m,n}(i,j) := \#\{1 \leq k \leq \mathrm{lcm}(m,n) \; ; \; \mathrm{gcd}(m,k) = i, \ \mathrm{gcd}(n,k) = j\}.
\]
For genus computations, the following formulas are useful:
\[
\sum_{k = 1}^{\mathrm{lcm}(N, L - N)} F_{L,N}(k) = \sum_{\substack{i \mid N \\ j \mid L - N}} \varphi_{N, L - N}(i, j) \sum_{p \cdot \frac{N}{i} + q \cdot \frac{L - N}{j} = N} \binom{i}{p} \binom{j}{q},
\]
\[
\sum_{L \mid \mathrm{gcd}(L,k) \cdot N} \binom{\mathrm{gcd}(L,k)}{\mathrm{gcd}(L,k) \cdot N / L} = \sum_{i \mid L, L \mid iN} \varphi(L/i) \binom{i}{i \cdot N / L}.
\]

\begin{example}[Total genus of $X$ for $(L,N)=(16,4)$]
When \( (L,N) = (16,4) \), the formula \eqref{eq:N_partition} yields \( \mathcal{N} = 10 \).
The values of \( \varphi_{4,12}(i,j) \) are given by the following table:
\[
\begin{tabular}{c|cccccc}
\( i \backslash j \) & 1 & 2 & 3 & 4 & 6 & 12 \\\hline
1 & \( \varphi(3)\varphi(4) \) & 0 & \( \varphi(1)\varphi(4) \) & 0 & 0 & 0 \\
2 & 0 & \( \varphi(3)\varphi(2) \) & 0 & 0 & \( \varphi(1)\varphi(2) \) & 0 \\
4 & 0 & 0 & 0 & \( \varphi(3)\varphi(1) \) & 0 & \( \varphi(1)\varphi(1) \) \\
\end{tabular}
\]
Therefore, the number 
$\sum_{k=1}^{12} F_{16,4}(k)= \sum\limits_{\substack{i \mid 4 \\ j \mid 12}} \varphi_{4,12}(i,j) \sum\limits_{p\cdot \frac{4}{i} + q\cdot \frac{12}{j} = 4} \binom{i}{p} \binom{j}{q}$ is calculated as
{\footnotesize
\[
\begin{aligned}
&4 + 2\left\{
\binom{1}{1} \binom{3}{0} + \binom{1}{0} \binom{3}{1}
\right\}
+ 2 + 1\left\{
\binom{2}{2} \binom{6}{0} +
\binom{2}{1} \binom{6}{1} +
\binom{2}{0} \binom{6}{2}
\right\}+ 2\left\{
\binom{4}{4} \binom{4}{0} +
\binom{4}{1} \binom{4}{1}
\right\}
+ \binom{16}{4} \\\\
&= 1896.
\end{aligned}
\]
}
We also have
\[
\sum_{i \mid 16,\ 16 \mid 4i} \varphi(16/i) \binom{i}{i/4}
=
\varphi(4)\binom{4}{1} +
\varphi(2)\binom{8}{2} +
\varphi(1)\binom{16}{4} = 1856.
\]
Hence, the total genus of \( X \) is given by
\[
g(X) = 10 + \frac{1}{2} \binom{14}{3} - \frac{1}{2} \left[
\frac{1896}{12} + \frac{1856}{16}
\right] = 55.
\]
\end{example}

\section{Degeneracy}\label{sec:degeneracy}

If $C_0$ is non-irreducible, $X$ is non-connected.
Moreover, if $C_0$ is non-irreducible and non-reduced, more than two connected components of $X$ are mapped onto one irreducible component of $C_0$.
In this case, the defining polynomial of $C_0$ is factored as a product of irreducible polynomials
\[ 
\{f_1(v,w)\}^{\eta_1} \{f_2(v,w)\}^{\eta_2} \cdots \{f_p(v,w)\}^{\eta_p}=0,
\]
where the exponents \( \eta_i \) are not necessarily equal to 1.

Each Bethe root corresponds to a common point of $C_0$ and the line $\ell:=\{w=(-1)^{N+1}\ee^{L\gamma}v\}$.
Degeneration of Bethe roots occurs in either of the following cases: (i) $\ell$ intersects a curve defined by \( \{f_i(v,w)\}^{\eta_i}=0 \) with \( \eta_i > 1 \); or (ii) $\ell$ is tangent to a curve defined by some $f_i(v,w)=0$.

The case (i) generally occurs. 
For each intersection point of such type, one obtains \( \eta_i \) independent Bethe roots.
On the other hand, the case (ii) occurs only for special values of $\gamma$.
When degeneracy of type (ii) does not occur, the Bethe equations have exact \( \binom{L}{N} \) physical solutions (Corollary \ref{cor:generic_gamma}).

\subsection{Degeneracy of connected components}\label{sec:GM_type_degeneration}

Golinelli and Mallick \cite{GMspecdeg} studied a type of eigenvalue degeneracy that can occur when \( L \) and \( N \) are not coprime.
Following a similar approach, we will derive a necessary and sufficient condition for connected components of $X$ to be mapped onto a common irreducible component in $C_0$.

Suppose \( L \) and \( N \) are not coprime.
Let $L'=L/e$ and $N'=N/e$ be the numbers defined in \eqref{eq:def_of_L'_N'}.
Recall Equation \eqref{eq:decomposed_eq_w_to_t}:
\[ \omega_e^{k-1}w^{1/e} = \frac{t^{N'}}{(1 - t)^{L'}},\quad(k=1,2,\dots,e). \]
This equation has $L'$ roots $t=t_{k}, t_{k+e}, \dots, t_{k+(L'-1)e}$.
Then, by Vieta's formula, we have
\begin{equation}\label{eq:prod_of_package}
\prod_{i\in \mathcal{P}_1}t_i(w)
=
\prod_{i\in \mathcal{P}_2}t_i(w)
=\cdots
=
\prod_{i\in \mathcal{P}_e}t_i(w)
=1,
\end{equation}
where $\mathcal{P}_k=\{i\in [L]\;;\;i\equiv k\mod{e}\}$ be the \textit{$i$-th package} defined in \eqref{eq:def_of_Pk}.
In particular, if the branch set $I\in \Omega$ contains $\mathcal{P}_k$ for some $k$, the multi-valued function $v=\prod_{i\in I}t_i(w)$ is invariant under the replacement of $\mathcal{P}_k$ with $\mathcal{P}_{k'}$ ($k\neq k'$).

\begin{prop}\label{prop:multiple_component}
Let $X_{[A_1,\dots,A_e]}$ be the connected component of \( X \) corresponding to \([A_1,\dots,A_e] \). Suppose 
\begin{itemize}
  \item There exists some $k$ such that $A_k=L'$, and
  \item There exists some $k'$ such that $A_{k'}=0$.
\end{itemize}
Let $[B_1,\dots,B_e]$ be the element of $\overline{Z}$ obtained from $[A_1,\dots,A_e]$ by replacing \( k \) by \( k' \).
Then, the connected components $X_{[A_1,\dots,A_e]}$ and $X_{[B_1,\dots,B_e]}$ are mapped to a common irreducible component of $C_0\subset \CC^2$.
\end{prop}
\begin{proof}
Let $D_I$ ($I\in \Omega$) be a sheet contained in $X_{[A_1,\dots,A_e]}$.
By assumption, the two packages $\mathcal{P}_k$ and $\mathcal{P}_{k'}$ satisfy $\mathcal{P}_k\subset I$ and $\mathcal{P}_{k'}\cap I=\emptyset$.
Let $J:=(I\cap \mathcal{P}_\ell)\setminus \mathcal{P}_k$.
Then the $J$-th sheet $D_J$ is contained in $X_{[B_1,\dots,B_e]}$.
By \eqref{eq:prod_of_package}, we have $\prod_{i\in I}t_i(w)=\prod_{i\in J}t_i(w)$, which implies that the multi-valued function $v(w)$ has the same value on $X_{[A_1,\dots,A_e]}$ and $X_{[B_1,\dots,B_e]}$.
Hence, $X_{[A_1,\dots,A_e]}$ and $X_{[B_1,\dots,B_e]}$ are mapped onto a common subset of $\CC^2$.
\end{proof}

 If $[A_1,\dots,A_k]$ and $ [B_1,\dots,B_e]$ satisfy the assumption of Proposition \ref{prop:multiple_component}, we write
 \begin{equation}\label{eq:multisets_relation}
 [A_1,\dots,A_e]\approx
 [B_1,\dots,B_e].
 \end{equation}

{

The following proposition states that the converse of Proposition \ref{prop:multiple_component} also holds.
\begin{prop}\label{prop:two_components_ind}
Two connected components $X_{[A_1,\dots,A_e]}$ and $X_{[B_1,\dots,B_e]}$ are mapped to a common irreducible component of $C_0$ if and only if $ [A_1,\dots,A_e]\approx
 [B_1,\dots,B_e]$.
\end{prop}
\begin{proof}
If $X_{[A_1,\dots,A_e]}$ and $X_{[B_1,\dots,B_e]}$ are mapped to a common irreducible component, there exist some index sets $I,J\in \Omega$ such that (i) $\pi(I)$ is a representative of $[A_1,\dots,A_e]\in \overline{Z}$, (ii) $\pi(J)$ is a representative of $[B_1,\dots,B_e]\in \overline{Z}$, and (iii) $\prod_{i\in I}t_i(w)= \prod_{j\in J}t_j(w)$ for all $w$.
Let $\Delta:=I\cap J$, $I':=I\setminus \Delta$, and $J':=J\setminus \Delta$.
If $I',J'\neq \emptyset$, we have $\prod_{i\in I'}t_i(w)= \prod_{j\in J'}t_j(w)$.
By the identity theorem in complex analysis, we can replace $I',J'$ with $g(I'),g(J')$, where $g$ is an arbitrary element of the monodromy group $G_2$.
Recall that $G_2$ contains all transpositions of two elements within the same package (see Lemma~\ref{lemma:str_of_G_2}).
If there exists some package $\mathcal{P}_k$ such that $\mathcal{P}_k\cap I'\neq \emptyset$ and $\mathcal{P}_k\cap (I')^c\neq \emptyset$, we find some indexes $i\in \mathcal{P}_k\cap I'$ and $i'\in \mathcal{P}_k\cap (I')^c$, and $g\in G_2$ which permutes $i$ and $i'$.
By comparing $\prod_{i\in I'}t_i(w)= \prod_{j\in J'}t_j(w)$ and $\prod_{i\in g(I')}t_i(w)= \prod_{j\in g(J')}t_j(w)$, we deduce $\{t_i(w)\}^2=\{t_{i'}(w)\}^2$ if $i'\in J'$, and $t_i(w)=t_{i'}(w)$ if $i'\not\in J'$.
However, each equation contradicts the asymptotic behavior of $t_i(w)$ when $w\to \infty$.
Hence, the set $I$ must be of the form $I=\Delta\sqcup \mathcal{P}_{k_1} \sqcup \mathcal{P}_{k_2}\sqcup\cdots \sqcup \mathcal{P}_{k_p}$ for some $k_1,\dots,k_p$.
Similarly, we have the decomposition
$J=\Delta\sqcup \mathcal{P}_{k'_1} \sqcup \mathcal{P}_{k'_2}\sqcup\cdots \sqcup \mathcal{P}_{k'_p}$, which concludes the proposition.
\end{proof}

The following corollary refers to a `most degenerate component,' whose defining polynomial has a particularly simple form.
}

\begin{cor} \label{GMparticulartype}
Assume there exists some \( h \in \mathbb{Z}_{>0} \) such that \( hL' = eN' \).
Then, the connected component corresponding to an element of the form
\[
[A_1,\dots,A_e]\in \overline{Z}\qquad (\#\{i\;;\;A_i=L'\}=h,\ \#\{i\;;\;A_i=0\}=e-h)
\]
is defined by a power of \( v - 1 \).
\end{cor}
\begin{proof}
By Proposition~\ref{prop:multiple_component}, it suffices to prove when \( A_1=\dots=A_h=L' \) and $A_{h+1}=\dots=A_e=0$.
Then, there exist $e$ possible choices of $I\in \Omega$ representing $[L',\dots,L',0,\dots,0]$:
\[
I=\mathcal{P}_1\cup
\mathcal{P}_2\cup\dots\cup
\mathcal{P}_h,\quad
\mathcal{P}_2\cup
\mathcal{P}_3\cup\dots\cup
\mathcal{P}_{h+1},\quad\dots,\quad
\mathcal{P}_e\cup
\mathcal{P}_1\cup\dots\cup
\mathcal{P}_{h-1}.
\]
For each choice, we have $v=\prod_{i\in I}t_i(w) =1$ by \eqref{eq:prod_of_package}.
\end{proof}

{
\subsection{Degeneracy of eigenvalues}
Golinelli and Mallick \cite{GMspecdeg} have presented sufficient conditions for two Bethe roots to yield the same eigenvalue of the Markov matrix.
Using the terminology of algebraic curves, we briefly review their results.

The following lemma is equivalent to \cite[Eq.~(24)]{GMspecdeg}.
\begin{lemma} \label{lemforpartofsum}
If $\{t_{k},t_{k+e},\dots,t_{k+(L'-1)e}\}$ is the set of roots of the algebraic equation
\[
(1-t)^{L^\prime}=K t^{N^\prime},\quad (L'>1,N'\geq 1,K:\text{a constant})
\]
then we have
\begin{align}
\sum_{j=0}^{L^\prime-1} \frac{t_{k+je}}{1-t_{k+je}}=-L^\prime+N^\prime.
\label{partofsumformula}
\end{align}
\end{lemma}
\begin{proof}
Let $P(t)=(1-t)^{L'}-Kt^{N'}$.
Then, we have
\[
\sum_{j=0}^{L^\prime-1} \frac{t_{k+je}}{1-t_{k+je}}
=
-L'+\sum_{j=0}^{L^\prime-1} \frac{1}{1-t_{k+je}}
=
-L'+\frac{P'(1)}{P(1)}=
-L'+\frac{N'K1^{N'-1}}{K1^{N'}}=-L'+N'.
\]
\end{proof}

\begin{prop}\label{prop:the_same_point}
Consider two points on $X$ corresponding to two Bethe roots $\{z_i\}$ and $\{z_i'\}$, respectively.
If these two points are mapped to the same regular point on $C_0$, they yield the same eigenvalue $E$ (Eq.\eqref{eq:eigenvalue}).
\end{prop}
\begin{proof}
Let $p,q\in X$ be the points associated to $\{z_i\}$ and $\{z'_i\}$, respectively.
Consider the index sets $I,J\in \Omega$ with $p\in D_I$ and $q\in D_J$.
Since $p,q$ are mapped to the same regular point, there exists an open neighborhood $U\subset C_0$ such that $\prod_{i\in I}t_i(w)=\prod_{j\in J}t_j(w)$, ($w\in U$).
By arguments similar to those in the proof of Proposition~\ref{prop:two_components_ind}, we find that $J$ is obtained from $I$ by replacing several packages included in $I$ with other packages.
Because the contribution of any package to $E=\sum_{i}\frac{z_i}{1-z_i}$ equals $-L'+N'$ (see \eqref{eq:decomposed_eq_w_to_t} and Lemma \ref{lemforpartofsum}), $E$ is invariant under this replacement.
\end{proof}

In the setting of Corollary \ref{GMparticulartype}, the eigenvalue is given explicitly.

\begin{cor}
Let $X_{\mathcal{O}}$ be a connected component that is mapped onto the plane curve defined by $(v-1)^\eta=0$.
Then, the eigenvalue corresponding to a Bethe root associated with a point on $X_{\mathcal{O}}$ is $h(-L^\prime+N^\prime)$.
\end{cor}
}

\begin{example}[Degeneracy of eigenvalues for $(L,N)=(8,4)$]
Let \( (L,N) = (8,4) \). 
By Corollary \ref{GMparticulartype}, the irreducible components of \( X \) corresponding to \( [2,2,0,0] \) and \( [2,0,2,0] \) are both defined by a power of \( v - 1 \).
The intersection of the two lines $\{v=1\}$ and $\{w=-\ee^{8\gamma}v\}$ is $(1,-\ee^{8\gamma})\in C_0$.
Then, the associated Bethe root \( \{z_1, z_2, z_3, z_4\} \) must satisfy
\[
\begin{cases}
-\ee^{8\gamma}
=\dfrac{z_1^4}{(1-z_1)^8}
=\dfrac{z_2^4}{(1-z_2)^8}=\dfrac{z_3^4}{(1-z_3)^8}=\dfrac{z_4^4}{(1-z_4)^8},\\
1 = z_1 z_2 z_3 z_4.
\end{cases}
\]

The eight roots \( t_1, t_2, \dots, t_8 \) of the rational equation \( -\ee^{8\gamma} = \dfrac{t^4}{(1 - t)^8} \) are complex numbers satisfying:
\[
\begin{gathered}
t_1 + t_5 = 2 + \frac{\ee^{-2\gamma}}{\sqrt{2}} + \frac{\ee^{-2\gamma}}{\sqrt{2}}\sqrt{-1},\quad
t_2 + t_6 = 2 + \frac{\ee^{-2\gamma}}{\sqrt{2}} - \frac{\ee^{-2\gamma}}{\sqrt{2}}\sqrt{-1},\\
t_3 + t_7 = 2 - \frac{\ee^{-2\gamma}}{\sqrt{2}} - \frac{\ee^{-2\gamma}}{\sqrt{2}}\sqrt{-1},\quad
t_4 + t_8 = 2 - \frac{\ee^{-2\gamma}}{\sqrt{2}} + \frac{\ee^{-2\gamma}}{\sqrt{2}}\sqrt{-1},\\
t_1 t_5 = t_2 t_6 = t_3 t_7 = t_4 t_8 = 1.
\end{gathered}
\]
Then there are four Bethe roots 
\[
\{z_1,z_2,z_3,z_4\} =
\{t_1,t_5,t_2,t_6\},\
\{t_2,t_6,t_3,t_7\},\
\{t_3,t_7,t_4,t_8\},\
\{t_4,t_8,t_5,t_1\}
\]
corresponding to \( [2,2,0,0] \), and two Bethe roots \[
\{z_1,z_2,z_3,z_4\} =
\{t_1,t_5,t_3,t_7\},\
\{t_2,t_6,t_4,t_8\}
\]
corresponding to  \( [2,0,2,0] \).
The eigenvalue \( E \) (see Equation \eqref{eq:eigenvalue}) associated with each of these 6 solutions is the same and equal to \( E = -2 \).
\end{example}



\begin{prop}\label{prop:decomp_of_defining_eq_of_C0}
The defining polynomial of \( C_0 \) factors as
\[
\prod_{
\substack{\mathcal{O} \subset \Omega \\\mathcal{O} \text{ is a } G_2\text{-orbit}}
}
f_{\mathcal{O}}(v,w)=0,
\]
where $f_{\mathcal{O}}(v,w)\in \ZZ[\omega_e][v,w^{-1}]$.
As a polynomial in $v$, $f_{\mathcal{O}}(v,w)$ is of degree $\# \mathcal{O}$ (Eq.~\eqref{eq:card_of_O}).
Each $G_2$-orbit $\mathcal{O}$ is labeled by an element of $\overline{Z}$.
If there are two elements satisfying $[A_1,\dots,A_e]\approx [B_1,\dots,B_e]$, their defining polynomials of corresponding connected components are identical up to multiplicity.
\end{prop}

\begin{example}
When $(L,N)=(8,4)$, $X$ has six connected components corresponding to $[2,2,0,0]$, $[2,0,2,0]$, $[2,1,1,0]$, $[2,1,0,1]$, $[2,0,1,1]$, and $[1,1,1,1]$.
The cardinality of the six $G_2$-orbits in $\Omega$ is
\[
\#\mathcal{O}_{2200}=4,\quad
\#\mathcal{O}_{2020}=2,\quad
\#\mathcal{O}_{2110}=16,\quad
\#\mathcal{O}_{2101}=16,\quad
\#\mathcal{O}_{2011}=16,\quad
\#\mathcal{O}_{1111}=16.
\]
Since $[2,2,0,0]\approx [2,0,2,0]$, the defining equations of $X_{2200}$ and $X_{2020}$ coincide with each other up to multiplicity.
By the same reason, the defining equations of $X_{2110}$ and $X_{2011}$ coincide with each other.
Hence, the defining polynomial of $C_0$ is factored as a product of one polynomial of degree $2+4=6$, one polynomial of degree $16+16=32$, and two polynomials of degree $16$.
See \eqref{eq:def_poly_84} in Appendix for explicit numerical results.
\end{example}

\section{Half-filling case}

We present several results and applications to the half-filling case $L=2N$ in this section.

\subsection{A class of irreducible components of $C_0$}
When $L=2N$ and $N$ is odd,
one can write down explicitly a class of irredicuble components of $C_0$.

\begin{lemma}
For $L=2N$ and $N$ is odd, the irreducible component corresponding to
$ [ \underbrace{2,\dots,2}_{(N-1)/2},1,\underbrace{0,\dots,0}_{(N-1)/2}  ]
$
is given by
\begin{align}
\sum_{j=0}^{2N} (-1)^j \binom{L}{j} v^j-w^{-1} v^N=(v-1)^{2N}-w^{-1} v^N. \label{somesimplecomponent}
\end{align}
\end{lemma}
\begin{proof}
Let us denote the $G_2$-orbit corresponding to $ [ \underbrace{2,\dots,2}_{(N-1)/2},1,\underbrace{0,\dots,0}_{(N-1)/2}  ]
$ as $\mathcal{O}$. Using $t_j t_{N+j}=1$, we find the $\mathcal{O}$-power sums $P_n$ can be simplified as
\begin{align}
P_n=&(t_{1} t_{N+1})^n \cdots (t_{(N-1)/2} t_{(3N-1)/2})^n (t_{(N+1)/2}^n+t_{(3N+1)/2}^n)+\cdots \nonumber \\
+&(t_{(N+1)/2} t_{(3N+1)/2})^n \cdots (t_{N-1} t_{2N-1})^n (t_{N}^n+t_{2N}^n) \nonumber \\
+&(t_{(N+3)/2} t_{(3N+3)/2})^n \cdots (t_{N} t_{2N})^n (t_{1}^n+t_{N+1}^n)+\cdots \nonumber \\
+&(t_{N} t_{2N})^n \cdots (t_{(N-3)/2} t_{(3N-3)/2})^n (t_{(N-1)/2}^n+t_{(3N-1)/2}^n) \nonumber \\
=&(t_{(N+1)/2}^n+t_{(3N+1)/2}^n)+\cdots+(t_{N}^n+t_{2N}^n) 
+ (t_{1}^n+t_{N+1}^n)+\cdots + (t_{(N-1)/2}^n+t_{(3N-1)/2}^n) \nonumber \\
=&p_n(t_1,\dots,t_{2N}).
\end{align}
Since the $\mathcal{O}$-power sums $P_n$ are the same as the ordinary power sums, we can apply the same relation between the power sum and elementary symmetric polynomials and conclude that the $\mathcal{O}$-elementary symmetric polynomials are exactly the ordinary elementary symmetric polynomials. 
We get
\begin{align}
E_n=e_n(t_1,\dots,t_{2N})=\binom{2N}{N}+w^{-1} \delta_{n,N},
\end{align}
and hence, the corresponding irreducible component of $C_0$ can be expressed as
\begin{align}
\sum_{j=0}^{2N} (-1)^j \binom{L}{j} v^j-w^{-1} v^N=(v-1)^{2N}-w^{-1} v^N.
\end{align}
\end{proof}
For no fugacity case $\gamma=0$, setting $w=v$ in \eqref{somesimplecomponent}, we have
\begin{align}
(v-1)^{2N}- v^{N-1}=((v-1)^N-v^{(N-1)/2})((v-1)^N+v^{(N-1)/2}).
\end{align}

\subsection{Application to partition functions of five-vertex model}

We discuss an application of a special type of Bethe roots for half-filling to the norms of on-shell Bethe vectors, or equivalently, partition functions of the five-vertex model corresponding to TASEP.
Let us first review the results of the norms and the standard facts of the algebraic Bethe ansatz.
Note that we have performed changes of variables from the ones used in \cite{Bo,MSS,MS}, for example, to the ones more appropriate for and used in \cite{ProlhacSciPost}, for example, and in this paper.

Let $V=\mathbb{C}^2$ be the complex two-dimensional spanned by
$\binom{1}{0}=: | 0 \rangle $ and $\binom{0}{1}=:|1 \rangle$,
which can be regarded as the empty site and particle-occupied site.
We introduce operators $\sigma^+$, $\sigma^-$, $\sigma^z$, $s$, $n$ acting on elements of $V$ as
\begin{align}
\sigma^+=\begin{pmatrix}
0 & 1 \\
0 & 0 \\
\end{pmatrix},
\ \ \
\sigma^-=\begin{pmatrix}
0 & 0 \\
1 & 0 \\
\end{pmatrix},
\ \ \
\sigma^z=\begin{pmatrix}
1 & 0 \\
0 & -1 \\
\end{pmatrix},
\ \ \
s=\begin{pmatrix}
1 & 0 \\
0 & 0 \\
\end{pmatrix},
\ \ \
n=\begin{pmatrix}
0 & 0 \\
0 & 1 \\
\end{pmatrix}.
\end{align}
For the space $V_j$, we denote the corresponding vectors and operators as
$| 0 \rangle_j $, $|1 \rangle_j$, $\sigma_j^+$, $\sigma_j^-$, $\sigma_j^z$, $s_j$, $n_j$.

\begin{figure}[htbp]
\centering
\includegraphics[width=8truecm]{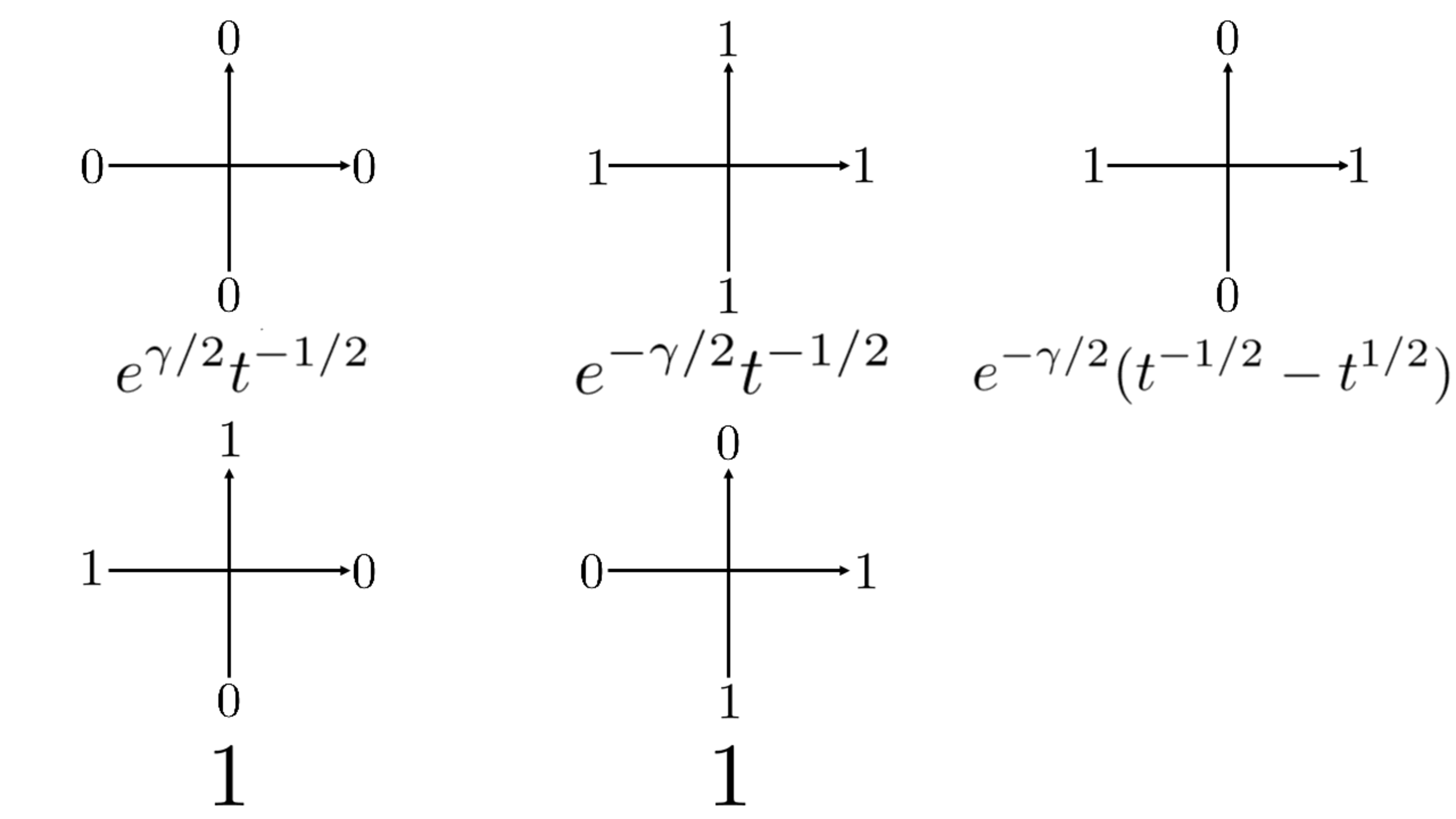}
\caption{The $L$-operator of the five vertex model
\eqref{fivevertexL}.
For each of the five non-zero local configurations (there are sixteen local configurations in total,
and eleven of them not displayed are all assigned weight zero), the corresponding weight is denoted below.
}
\label{fivevertexfigure}
\end{figure}

We introduce the $L$-operator for the five-vertex model (Figure \ref{fivevertexfigure})
\begin{align}
L_{\mu j}(t)=\ee^{\gamma/2} t^{-1/2} s_\mu s_j+\sigma_\mu^- \sigma_j^+
+\sigma_\mu^+ \sigma_j^-+\ee^{-\gamma/2} (t^{-1/2}-t^{1/2})n_\mu s_j+\ee^{-\gamma/2} t^{-1/2} n_\mu n_j,
\label{fivevertexL}
\end{align}
acting on $V_\mu \otimes V_j$.
We extend the definition to the tensor product of a large number of two-dimensional vector spaces, by defining the $L$-operator to act nontrivially on the components
$V_\mu$ and $V_j$ as \eqref{fivevertexL},
and as the identity on all other components.
$t$ is called the spectral parameter.

The $L$-operator satisfies the so-called $RLL$ relation
\begin{align}
R_{\mu \nu}(t,r)L_{\mu j}(t)L_{\nu j}(r)
=L_{\nu j}(r) L_{\mu j}(t)R_{\mu \nu}(t,r), \label{RLL}
\end{align}
acting on $V_\mu \otimes V_\nu \otimes V_j$.
Here, the $R$-matrix on the tensor product of
two $V$'s is given by
\begin{align}
R(t,r)=\begin{pmatrix}
\frac{r}{r-t} & 0 & 0 & 0 \\
0 & 0 & \frac{t^{1/2}r^{1/2}}{r-t} & 0 \\
0 & \frac{t^{1/2}r^{1/2}}{r-t} & 1 & 0 \\
0 & 0 & 0 & \frac{r}{r-t}
\end{pmatrix},
\end{align}
and is extended to the tensor product of many two-dimensional vector spaces in the same way with the 
$L$-operator.
The $R$-matrix satisfies the Yang-Baxter relation
\begin{align}
R_{\mu \nu}(r,t)R_{\mu \gamma}(r,s)R_{\nu \gamma}(t,s)
=R_{\nu \gamma}(t,s)R_{\mu \gamma}(r,s)R_{\mu \nu}(r,t),
\end{align}
acting on $V_\mu \otimes V_\nu \otimes V_\gamma$.

We introduce the monodromy matrix as the product of $L$-operators acting on $V_\mu \otimes V_1 \otimes \cdots \otimes V_L$
\begin{align}
T_\mu(r)=\prod_{j=1}^L L_{\mu j}(r)=\begin{pmatrix}
A(r) & B(r) \\
C(r) & D(r) \\
\end{pmatrix}_\mu.
\end{align}
The $A$- $B$- $C$- $D$-operators are
matrix elements of the monodromy matrix
with respect to the auxiliary space $V_\mu$.
The intertwining relation \eqref{RLL} implies the commutativity of the transfer matrix
\begin{align}
\tau(r)= A(r)+D(r),
\end{align}
i.e. from \eqref{RLL} it follows that
\begin{align}
[\tau(r),\tau(t)]=0.
\end{align}
Moreover, the transfer matrix yields (the $\gamma$-deformation of)
the Markov matrix $\mathcal{M}$ of TASEP
\begin{align}
-\frac{\partial}{\partial r} \mathrm{log} (r^{L/2} \tau(r))|_{r=1}
=\sum_{j=1}^L \Bigg\{
\ee^{-\gamma/2}
\sigma_j^+ \sigma_{j+1}^- + \frac{1}{4} (\sigma_j^z \sigma_{j+1}^z-1)
\Bigg\}=:\mathcal{M}.
\label{deformedMarkov}
\end{align}
Some of the explicit
elements of the intertwining relation \eqref{RLL} are the following:
\begin{align}
A(r)B(t)&=\frac{r}{r-t}B(t)A(r)+\frac{r^{1/2}t^{1/2} }{t-r}B(r)A(t), \label{ABcomm} \\
D(r)B(t)&=\frac{t}{t-r}B(t)D(r)+\frac{r^{1/2}t^{1/2}}{r-t}B(r)D(t), \label{DBcomm} \\
[B(r),B(t)]&=0.
\end{align}
Using these commutation relations and the action of $A(r)$ and $D(r)$ on the vacuum state
 $|\Omega \rangle:=|0 \rangle_1 \otimes \cdots \otimes |0 \rangle_L$
\begin{align}
A(r)|\Omega \rangle=\ee^{L \gamma/2} r^{-L/2} |\Omega \rangle, \ \ \
D(r)|\Omega \rangle=\ee^{-L \gamma/2} (r^{-1/2}-r^{1/2})^{L}|\Omega \rangle,
\end{align}
it is a standard procedure to show that the product of $B$-operators acting on the vacuum state
\begin{align}
|\psi(\{ z \}) \rangle=\prod_{j=1}^N B(z_j) |\Omega \rangle,
\end{align}
becomes an eigenstate of the transfer matrix $\tau(r)$
\begin{align}
\tau(r)|\psi(\{ z \}) \rangle&=\Theta(r,\{ z \}) |\psi(\{ z \}) \rangle, \label{eigenstaterel} \\
\Theta(r,\{ z \})&=\ee^{L \gamma/2} r^{-L/2} \prod_{j=1}^N \frac{r}{r-z_j}+\ee^{-L \gamma/2} (r^{-1/2}-r^{1/2})^L \prod_{j=1}^N \frac{-z_j}{r-z_j},
\end{align}
provided the spectral parameters $\{ z \}$ satisfy the Bethe ansatz equations (Eq.\eqref{eq:Bethe_in_one_line})
\begin{align}
\frac{z_k^N}{(1-z_k)^L}=(-1)^{N+1} \ee^{L \gamma} \prod_{j=1}^N z_j, \ \ \ k=1,\dots,N,
\label{BAEinreview}
\end{align}
and $z_j \neq z_k$ for $j \neq k$.
The eigenvalue of the (deformed) Markov matrix $\mathcal{M}$ is given in terms of the spectral parameters as
\begin{align}
\mathcal{M}(\{ z \})=-\frac{\partial}{\partial r} \log 
(r^{L/2} \Theta(r,\{ z \}) )|_{r=1}
=\sum_{j=1}^N \frac{z_j}{1-z_j}.
\label{deformedeigenvalue}
\end{align}
We can also show that the product of $C$-operators acting on the dual vacuum state $\langle \Omega|:=
{}_1 \langle 0| \otimes \cdots \otimes {}_L \langle 0|$
\begin{align}
\langle \psi(\{ t \})=\langle \Omega| \prod_{j=1}^N C(z_j),
\end{align}
becomes the dual eigenvector of $\tau(r)$
\begin{align}
\langle \psi(\{ z \})| \tau(r)=
\langle \psi(\{ z \})| \Theta(r,\{ z \}),
\end{align}
provided 
the spectral parameters $\{ z \}$ satisfy
the Bethe ansatz equations \eqref{BAEinreview} 
and $z_j \neq z_k$ for $j \neq k$.

Note that the commutativity for the $B$- and $C$-operators
$
[B(t),B(r)]=[C(t),C(r)]=0
$
follow from the intertwining relation \eqref{RLL}, and the ordering of the spectral parameters
for the $B$- and $C$-operators in $| \psi(\{ z \})\rangle $ and $\langle \psi(\{ z \})|$ does not matter.

Let us make some comments on the linear independence
on the Bethe vectors.
Take the operator $\displaystyle \tau_0:=  \lim_{r \to \infty} r^{N-L/2} \tau(r)$.
Then from \eqref{eigenstaterel} we have
\begin{align}
\tau_0|\psi(\{ z \}) \rangle=(-1)^{N+L} \ee^{-L \gamma/2} \prod_{j=1}^N z_j |\psi(\{ z \}) \rangle.
\label{limiteigen}
\end{align}

Consider two distinct Bethe roots $\{ z \}$ and$\{ z^\prime \}$.
If $\displaystyle \prod_{j=1}^N  z_j \neq \prod_{j=1}^N z_j^\prime$,
from \eqref{limiteigen}
we note $|\psi(\{ z \}) \rangle$ and $|\psi(\{ z^\prime \}) \rangle$ are eigenvectors of $\tau_0$
with different eigenvalues.

If $\displaystyle \prod_{j=1}^N  z_j= \prod_{j=1}^N z_j^\prime=v$,
since
\begin{align}
\Theta(r,\{z \})=
(\ee^{L \gamma/2} r^N-\ee^{-L \gamma/2} (-1)^N (1-r)^L v)
\prod_{j=1}^N \frac{1}{r-z_j},
\end{align}
and $\displaystyle
\prod_{j=1}^N (r-z_j)  \neq \prod_{j=1}^N (r-z_j^\prime) 
$,
we have $\Theta(r,\{z \}) \neq \Theta(r,\{z^\prime \})$.
Hence the eigenvalues of $\tau(r)$ are different for
Bethe vectors with distinct Bethe roots.
Together with the completeness of the Bethe ansatz equations,
we conclude the following for
$\gamma \neq 0$.
\begin{prop} \label{completenessvectorone}
For $\gamma \neq 0$,
if all Bethe roots $\{ z  \}$ are distinct, the Bethe vectors $|\psi(\{ z \}) \rangle$ corresponding to all of the Bethe roots
form the $\binom{L}{N}$-dimensional vector space.
\end{prop}

For no fugacity ($\gamma=0$) case, we need a little bit of care.
Note that the discussion of deriving
the Bethe ansatz equations by the algebraic Bethe ansatz
works when $z_j \neq z_k$ for $j \neq k$.
However, it is a fundamental fact that for the periodic
TASEP, the (unnormalized) steady state vector
\begin{align}
|S_N \rangle=\sum_{1 \le x_1<x_2<\cdots<x_N \le L} \prod_{j=1}^N \sigma_{x_j}^-  | \Omega \rangle,
\end{align}
gives the eigenvector corresponding to the zero eigenvalue of the Markov matrix, and the solution $z_1=\dots=z_N=0$ of the Bethe ansatz equations corresponds to the steady state.
Note that there exists the following relation \cite[(17)]{Bo}
\begin{align}
\lim_{z_j \to 0} \prod_{j=1}^N z_j^{(M-1)/2} B(z_j)|\Omega \rangle=|S_N \rangle, \label{limittosteady}
\end{align}
and we can obtain a generalization of \eqref{limittosteady} including fugacity
which is given in \cite[(4.35)]{MS} ($\alpha$ corresponds
to $\ee^{-\gamma}$ in this paper).
However, the vector obtained in this way does not correspond to the eigenvector
of the deformed Markov matrix \eqref{deformedMarkov}.
If it is an eigenvector, from \eqref{deformedeigenvalue}, the eigenvalue should be zero, but acting \eqref{deformedMarkov} on the vector does not give the zero vector unless $\gamma=0$.
Note that the algebraic Bethe ansatz basically works when $z_j \neq z_k$ for $j \neq k$ (the commutation relations \eqref{ABcomm} and \eqref{DBcomm} are used for deriving the Bethe ansatz equations, and singularities may emerge from the right hand side of them when the spectral parameters are not distinct, which means that we cannot apply the argument, at least naively),
and the fact that the vector corresponding to $z_1=\dots=z_N=0$ is not an eigenvector in general (when $\gamma \neq 0$) does not contradict this.

The fact that the steady state vector $|S_N \rangle$ is the eigenvector corresponding to the zero eigenvalue of the Markov matrix when $\gamma=0$, together with the completeness of the Bethe ansatz equations, implies the following.
\begin{prop} \label{completenessvectortwo}
For $\gamma = 0$, if all Bethe roots $\{ z  \}$ are distinct, the steady state vector $|S_N \rangle$ together with the Bethe vectors $|\psi(\{ z \}) \rangle$ corresponding to the rest of the Bethe roots form the $\binom{L}{N}$-dimensional vector space.
\end{prop}

\begin{rem}\label{rem:note}
We emphasize that there are two levels of completeness under consideration. 
In the first part of the paper, we established completeness at the level of equations in Section 2 (Theorem \ref{completenessequation}), namely that the Bethe equations admit $\binom{L}{N}$ solutions counted with multiplicities. 
We then address completeness at the level of eigenvectors in this section, which are Propositions \ref{completenessvectorone}, \ref{completenessvectortwo}: each solution of the Bethe equations gives rise to an eigenvector of the transfer matrix.

In the algebraic Bethe ansatz framework, as long as the Bethe roots are pairwise distinct ($z_i \neq z_j$ for $i \neq j$), the off-shell Bethe vector evaluated at a solution of the Bethe equations yields an eigenvector of the transfer matrix. 
Moreover, whenever two solutions $\{z\}$ and $\{z'\}$ do not coincide as sets, the corresponding eigenvalues are different, so the associated Bethe vectors are linearly independent. 
Therefore, under the assumptions above, which are stated in the Propositions and provided that the Bethe vector does not vanish, we can construct $\binom{L}{N}$ distinct eigenvectors, which implies completeness also at the level of eigenvectors.

A subtle point concerns the possibility that a Bethe vector becomes the zero vector. 
As is well known, each component of the Bethe vector
$\langle \Omega| \prod_{j=1}^N \sigma_{x_j}^+ \prod_{j=1}^N B(z_j)| \Omega \rangle$ is essentially given by a determinant $\displaystyle \det_{1 \le j,k \le N}(z_j^k(1-z_j)^{-x_k})$ (see around Eq.~(29) of \cite{ProlhacSciPost} under the same convention adopted in this paper, and \cite[(4.1)]{MS} for the precise relation
for example) and special values of the roots require some care. 
If one of the Bethe roots vanishes, $z_i=0$ for some $i$, then the Riemann surface formulation implies that in fact $z_j=0$ for all $j$, which can be shown as follows.
The defining equation for the singular curve $C_0$ implies that the values for $v$ are finite, and then from the equation in the middle of \eqref{eq:Bethe_original} we note $z_j$ are finite for all $j$.
The same equation also implies $v=0$ if $z_i=0$ for some $i$, and inserting into the remaining equations of \eqref{eq:Bethe_original} and from the fact that $z_j$s are finite, we conclude $z_j=0$ for all $j$.
The treatment for this case has been explained.

On the other hand, the case $z_i=1$ does not actually occur: from the equations in the top and the bottom of \eqref{eq:Bethe_original} and that the values for $v$ are finite, we conclude that $z_j \neq 1$ for all $j$. 
\end{rem}

Let us now turn to the introduction and evaluation of the free energy.
The norm of the on-shell Bethe vector is the inner product between the on-shell Bethe vector and its dual
\begin{align}
\langle \psi(\{ z \})|\psi(\{ z \} \rangle=\langle \Omega| \prod_{j=1}^N C(z_j) \prod_{j=1}^N B(z_j) | \Omega \rangle.
\end{align}

Changing variables from $u_j$ to $z_j=\alpha^{-1} u_j^{-2}$ and replacing $\alpha$ by $\ee^{-\gamma}$ in \cite[Cor. 3.7]{MS} (see \cite[(30), (31)]{MSS} for the case $\alpha=1$, \cite[(28)]{Bo} for a determinant representation, and also \cite[page 41]{ProlhacSciPost}), we have the following.

\begin{prop}
If the set of variables $\{ z \}$ satisfies the Bethe ansatz equations \eqref{BAEinreview} and $\{ z \} \neq \{0,\dots,0 \}$,
we have
\begin{align}
\langle \psi(\{ z \})|\psi(\{ z \} \rangle
=&\ee^{\gamma(M-(N-1)^2)}
\prod_{j=1}^N z_j^{N-L-1} \prod_{\substack{\ell,n=1 \\ \ell \neq n}}^N \frac{1}{z_n-z_\ell}
\prod_{j=1}^N \frac{N+(L-N)z_j}{1-z_j}
\Bigg(
1-\sum_{j=1}^N \frac{1-z_j}{N+(L-N)z_j}
\Bigg).
\label{onshellnorm}
\end{align}
\end{prop}
This factorization occurs for the five-vertex model, which does not occur for the more general six-vertex model
corresponding to the XXZ quantum spin chain \cite{Korepin}.
See also \cite{Slavnov,KBI,Slavnovscalarproduct}. 

\begin{figure}[htbp]
\centering
\includegraphics[width=8truecm]{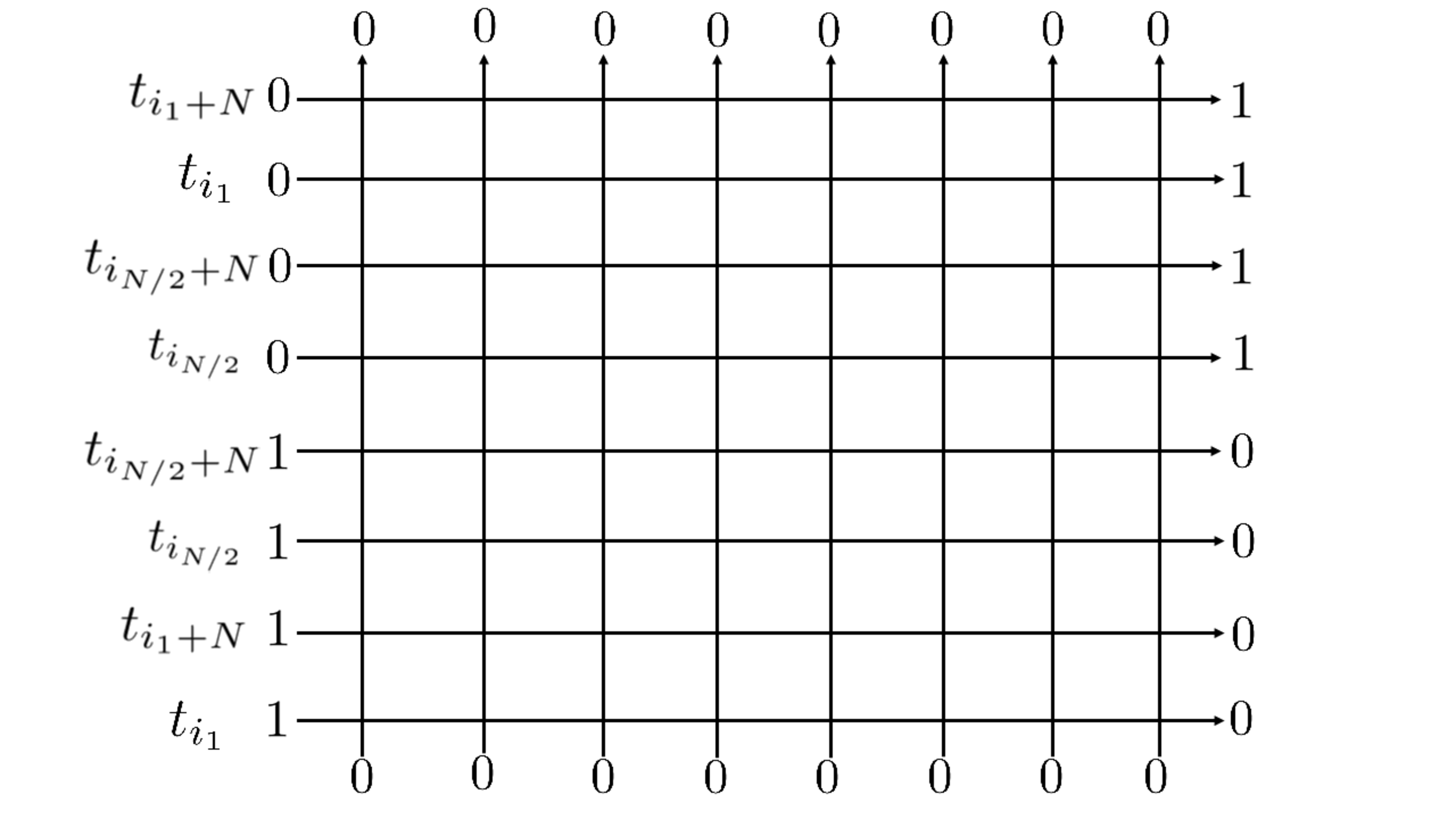}
\caption{
The on-shell Bethe vector norm \eqref{onshellnorm}, which can be regarded as the partition function of the five vertex model under a scalar product boundary condition.
For each row, we insert into the spectral variable of the $L$-operators the same value of $t$ denoted on the left.
} \label{figureforonshellnorm}
\end{figure}

The on-shell Bethe vector norm can also be regarded as the partition function of the five vertex model under scalar product boundary conditions (cf. \cite{BuPr}). 
See Figure \ref{figureforonshellnorm} for the specialization which we consider below.
Note that what are inserted into the spectral variables are the Bethe roots for `excited states'.
This type of insertion is different from the ones which we usually consider, for example, in \cite{BuPr,KoZJ,BF}, where the thermodynamic limit and asymptotics were investigated in detail.
An example of a type similar to what we consider here was investigated in \cite{deGKo}, for example.

For the rest of this section, we consider the case with no fugacity $\gamma=0$.
We evaluate the norm of the on-shell Bethe vector for the Golinelli-Mallick type Bethe roots
when $L=2N$, $N$: even.
Recall that the Bethe roots corresponding to $v=1$ are taking $N/2$ tuples of values out of $N$-tuples $(t_k,t_{N+k})$ ($k=1,\dots,N$) in total, each of which are solutions to
\begin{align}
t^2-(2+(-1)^{-1/N} \omega_N^{1-k})t+1=t^2-(2+\omega_N^{1/2-k})t+1=0.
\end{align}
By definition, $(t_k,t_{N+k})$ satisfy
$t_k+t_{N+k}=
2+\omega_N^{1/2-k}, \  t_k t_{N+k}=1$.
We take $(t_k,t_{N+k})$ for
$k=i_1,i_2,\dots,i_{N/2}$.
According to this notation, we rewrite \eqref{onshellnorm} as
(we replace $\{ z \}=\{ z_1,\dots,z_N \}$ by $(t_k,t_{N+k})$,
$k=i_1,i_2,\dots,i_{N/2}$)
\begin{align}
&\prod_{k=i_1,\dots,i_{N/2}} (t_kt_{N+k})^{-N-1} 
\prod_{k=i_1,\dots,i_{N/2}} \frac{1}{(t_k-t_{N+k})(t_{N+k}-t_k)} \nonumber \\
&\times \prod_{
\substack{j,k=i_1,\dots,i_{N/2} \\ j \neq k} }
\frac{1}{
(t_j-t_k)(t_j-t_{N+k})(t_{N+j}-t_k)(t_{N+j}-t_{N+k})
}
\nonumber \\
&\times \prod_{k=i_1,\dots,i_{N/2}} N^2 \frac{(1+t_k)(1+t_{N+k})}{(1-t_k)(1-t_{N+k})}
\Bigg(1-\frac{1}{N} \sum_{k=i_1,\dots,i_{N/2}}
\Bigg(
\frac{1-t_k}{1+t_k}+\frac{1-t_{N+k}}{1+t_{N+k}}
\Bigg)
\Bigg).
\end{align}
Using $t_k t_{N+k}=1$,
we can easily check
\begin{align}
\prod_{k=i_1,\dots,i_{N/2}} (t_kt_{N+k})^{-N-1}=1-\frac{1}{N} \sum_{k=i_1,\dots,i_{N/2}}
\Bigg(
\frac{1-t_k}{1+t_k}+\frac{1-t_{N+k}}{1+t_{N+k}} \Bigg)
=1.
\label{computepartone}
\end{align}
Using
$t_k+t_{N+k}=
2+\omega_N^{1/2-k}, \  t_k t_{N+k}=1$, we get
\begin{align}
\frac{(1+t_k)(1+t_{N+k})}{(1-t_k)(1-t_{N+k})}
=\frac{2+t_k+t_{N+k}}{2-t_k-t_{N+k}}
=\frac{4+\omega_N^{1/2-k}}{-\omega_N^{1/2-k}},
\end{align}
and
\begin{align}
&\frac{1}{(t_k-t_{N+k})(t_{N+k}-t_k)}=\frac{1}{2-t_k^2-t_{N+k}^2}
=\frac{1}{2-(t_k+t_{N+k})^2+2t_kt_{N+k}} \nonumber \\
=&\frac{1}{4-(2+\omega_N^{1/2-k} )^2}=\frac{1}{-4 \omega_N^{1/2-k}-\omega_N^{1-2k}}
=\frac{1}{-\omega_N^{1/2-k} (4+\omega_N^{1/2-k}) },
\end{align}
from which we find
\begin{align}
&\prod_{k=i_1,\dots,i_{N/2}} \frac{1}{(t_k-t_{N+k})(t_{N+k}-t_k)}
 \prod_{k=i_1,\dots,i_{N/2}} N^2 \frac{(1+t_k)(1+t_{N+k})}{(1-t_k)(1-t_{N+k})} \nonumber \\
=&N^{N} \prod_{k=i_1,\dots,i_{N/2}} \omega_N^{2k-1}.
\label{computeparttwo}
\end{align}
We can also evaluate the following product of four factors as 
\begin{align}
&
(t_j-t_k)(t_j-t_{N+k})(t_{N+j}-t_k)(t_{N+j}-t_{N+k}) \nonumber \\
=&(t_j^2-(2+\omega_N^{1/2-k})t_j+1)(t_{N+j}^2-(2+\omega_N^{1/2-k})t_{N+j}+1) \nonumber \\
=&
(t_j t_{N+j})^2+1+(2+\omega_N^{1/2-k})^2 t_j t_{N+j}+t_j^2+t_{N+j}^2
-(2+\omega_N^{1/2-k})(t_j+t_{N+j}) \nonumber \\
&-(2+\omega_N^{1/2-k})t_j t_{N+j} (t_j+t_{N+j}) \nonumber \\
=&2+(2+\omega_N^{1/2-k})^2+(t_j+t_{N+j})^2-2t_j t_{N+j}
-2(2+\omega_N^{1/2-k})(2+\omega_N^{1/2-j}) \nonumber \\
=&(2+\omega_N^{1/2-k})^2+(2+\omega_N^{1/2-j})^2
-2(2+\omega_N^{1/2-k})(2+\omega_N^{1/2-j}) \nonumber \\
=&(
2+\omega_N^{1/2-k}-2-\omega_N^{1/2-j}
)^2=(\omega_N^{1/2-k}-\omega_N^{1/2-j})^2,
\end{align}
to get
\begin{align}
& \prod_{
\substack{j,k=i_1,\dots,i_{N/2} \\ j \neq k} }
\frac{1}{
(t_j-t_k)(t_j-t_{N+k})(t_{N+j}-t_k)(t_{N+j}-t_{N+k})
}
\nonumber \\
=& \prod_{
\substack{j,k=i_1,\dots,i_{N/2} \\ j \neq k} }
\frac{1}{
(\omega_N^{1/2-k}-\omega_N^{1/2-j})^2
}. \label{computepartthree}
\end{align}
Combining
\eqref{computepartone}, \eqref{computeparttwo} and \eqref{computepartthree}, we get the following evaluation.
\begin{prop}
For $L=2N$ and $t_k+t_{N+k}=
2+\omega_N^{1/2-k}, \  t_k t_{N+k}=1$, $k=i_1,i_2,\dots,i_{N/2}$, we have
\begin{align}
&\langle \Omega| \prod_{j=i_1,\dots,i_{N/2}} (C(t_{j}) C(t_{j+N}))
\prod_{j=i_1,\dots,i_{N/2}} (B(t_{j}) B(t_{j+N}))|\Omega \rangle \nonumber \\
=&
N^{N} \prod_{k=i_1,\dots,i_{N/2}}  \omega_N^{2k-1}  
 \prod_{
\substack{j,k=i_1,\dots,i_{N/2} \\ j \neq k} }
\frac{1}{
(\omega_N^{1/2-k}-\omega_N^{1/2-j})^2
}. \label{evaluationnorm}
\end{align}
\end{prop}

$\langle \Omega| \prod_{j=i_1,\dots,i_{N/2}} (C(t_{j}) C(t_{j+N}))
\prod_{j=i_1,\dots,i_{N/2}} (B(t_{j}) B(t_{j+N}))|\Omega \rangle$ is a partition function of the five-vertex model on $2N \times 2N$-grid.
Note that the weights of the five-vertex model become complex in general.
Let us take
$i_j=j, \ j=1,\dots,N/2$.
We note the on-shell Bethe vector norm \\
$\langle \Omega| \prod_{j=1,\dots,N/2} (C(t_{j}) C(t_{j+N}))
\prod_{j=1,\dots,N/2} (B(t_{j}) B(t_{j+N}))|\Omega \rangle$ becomes real positive when
$N=4$ $(\mathrm{mod} \ 2)$ and real negative when $N=2$ $(\mathrm{mod} \ 2)$,
which can be checked by rewriting \eqref{evaluationnorm}
as
\begin{align}
&
N^{N} \prod_{k=i_1,\dots,i_{N/2}}  \omega_N^{2k-1}  
 \prod_{
\substack{j,k=1,\dots,N/2 \\ j \neq k} }
\frac{1}{ \omega_N^{1-2k}
(1-\omega_N^{k-j})^2
} \nonumber \\
=&N^N (-1)^{N^2/4} \prod_{1 \le j <k \le N/2}
\frac{1}{ \{ (1-\omega_N^{k-j})(1-\omega_N^{j-k}) \}^2
}.
\end{align}

For $N=4$ $(\mathrm{mod} \ 2)$ case, we introduce the
free energy (per site) as
\begin{align}
\mathcal{F}
=\lim_{N \to \infty} \frac{1}{4N^2} \mathrm{Log} 
\langle \Omega| \prod_{j=1,\dots,N/2} (C(t_{j}) C(t_{j+N}))
\prod_{j=1,\dots,N/2} (B(t_{j}) B(t_{j+N}))|\Omega \rangle
. \label{insidefreeenergy}
\end{align}
We evaluate $\mathcal{F}$ as follows.
From \eqref{evaluationnorm}, neglecting the factor $N^N$ (as the contribution of this factor vanishes in the limit we consider) and rewriting, we note the equation in
\eqref{insidefreeenergy} before taking the limit can be expressed as
\begin{align}
-\frac{1}{8} \frac{1}{(N/2)^2}
\sum_{\substack{j,k=1 \\ j \neq k }}^{N/2}
\mathrm{Log}
(
1-\ee^{\pi \sqrt{-1} (k-j)/(N/2)}
).
\end{align}
We view this as a Riemann sum and take the $N \to \infty$ limit to get the following sum
of double integrals
\begin{align}
\mathcal{F}
=-\frac{1}{8} 
\Bigg(
\int_{0}^1 \int_{y}^1 \mathrm{Log}(1-\ee^{\pi \sqrt{-1} (x-y)}) dx dy
+\int_{0}^1 \int_{0}^y \mathrm{Log}
(1-\ee^{\pi \sqrt{-1} (x-y)}) dx dy \Bigg).
\label{freeenergydoubleintegral}
\end{align}

To evaluate the double integrals, we use properties of the polylogarithm function defined as
\begin{align}
\mathrm{Li}_s(z)=\sum_{k=1}^\infty \frac{z^k}{k^s},
\end{align}
for $|z| <1$ and analytically continued to $|z| \ge 1$.
As a final result of the evaluation, the Riemann zeta function
\begin{align}
\zeta(s)=\sum_{k=1}^\infty \frac{1}{k^s},
\end{align}
appears. 
Note the relation $\mathrm{Li}_s(1)=\zeta(s)$.

Using $\mathrm{Li}_1(z)=-\mathrm{Log}(1-z)$ and the integral formula
\begin{align}
\mathrm{Li}_{s+1}(z)=\int_0^z \frac{\mathrm{Li}_s(t)}{t} dt,
\end{align}
and the evaluation $\mathrm{Li}_2(1)=\pi^2/6$, we have
\begin{align}
\int_{y}^1 \mathrm{Log}(1-\ee^{\pi \sqrt{-1} (x-y)}) dx=-\frac{\sqrt{-1} \pi}{6}+\frac{\sqrt{-1}}{\pi} \mathrm{Li}_2(-\ee^{-\sqrt{-1} \pi y}).
\label{firsteval}
\end{align}

Next, we note that the integration of the real part of $\mathrm{Li}_2(-\ee^{-\sqrt{-1} \pi y})$ over $y$ from 0 to 1 becomes zero, and we only need to take care of the imaginary part, i.e. consider the integration of $\sqrt{-1} \sum_{n=1}^\infty \frac{\sin (\pi(1-y)n)}{n^2}$.
We find
\begin{align}
\int_0^1 \mathrm{Li}_2(-\ee^{-\sqrt{-1} \pi y}) dy
=\frac{\sqrt{-1}}{\pi} \sum_{n=1}^\infty \frac{1-(-1)^n}{n^3}
=\frac{\sqrt{-1}}{\pi} (\zeta(3)-\mathrm{Li}_3(-1))=\frac{ 7 \sqrt{-1} }{4 \pi} \zeta(3).
\label{secondeval}
\end{align}
Here we used the relation $\mathrm{Li}_s(-1)=(2^{1-s}-1)\zeta(s)$.
From \eqref{firsteval} and \eqref{secondeval}, we get
\begin{align}
\int_{0}^1 \int_{y}^1 \mathrm{Log}(1-\ee^{\pi \sqrt{-1} (x-y)}) dx dy=
-\frac{\sqrt{-1} \pi}{6}-\frac{7}{4 \pi^2} \zeta(3). \label{evaldoubleintone}
\end{align}
In the same way, one has
\begin{align}
\int_{0}^1 \int_{0}^y \mathrm{Log}(1-\ee^{\pi \sqrt{-1} (x-y)}) dx dy=
\frac{\sqrt{-1} \pi}{6}-\frac{7}{4 \pi^2} \zeta(3).
\label{evaldoubleinttwo}
\end{align}

Combining
\eqref{freeenergydoubleintegral},
\eqref{evaldoubleintone} and
\eqref{evaldoubleinttwo}, we get the following result.
\begin{prop}
The free energy $\mathcal{F}$ is given by
\begin{align}
\mathcal{F}
=\frac{7 \zeta(3)}{16 \pi^2}.
\end{align}
\end{prop}
Numerically, 
$\mathcal{F} \sim 0.05328$,
which is smaller than the free energies of the six-vertex model for a variety of boundary conditions \cite{TRK}.
It seems that one of the reasons is that we treat the five-vertex model, which has less freedom for configurations than the six-vertex model in principle.
Another reason is that using complex weights for local configurations, and the partition function and the free energy become real means that those quantities are smaller than the values expected when using real positive weights.

\section{Other applications}
We discuss several other topics in this section.

\subsection{Particle-hole duality}

The TASEP possesses the particle-hole duality.
For the periodic boundary case we consider, systems with $N$ particles with $L$ sites can be regarded as $L-N$ particles with $L$ sites by exchanging the role of particles and holes.
Recall the Bethe ansatz equations (Eq.\eqref{eq:Bethe_in_one_line}) for $N$ particles with $L$ sites are
\begin{align} 
(1-z_i)^L=(-1)^{N+1} \ee^{-L \gamma} \prod_{j=1}^N \frac{z_i}{z_j}, \ \ \ i=1,\dots,N.
\end{align}
Let us denote the Bethe equations for $L-N$ particles with $L$ sites as
\begin{align}
(1-\tilde{z}_i)^L=(-1)^{L-N+1} \ee^{-L \gamma} \prod_{j=1}^{L-N} \frac{\tilde{z}_i}{\tilde{z}_j}, \ \ \ i=1,\dots,L-N.
\end{align}

The Bethe solution $\{ z_i \}$ for $N$ particles with $L$ sites and its corresponding solution $\{ \tilde{z}_i \}$ for $L-N$ particles with $L$ sites  are related by
\begin{align}
(z-1)^L+(-1)^{L-N} \ee^{-L \gamma} \prod_{i=1}^N z_i^{-1} z^N=\prod_{i=1}^{L-N} (z-\tilde{z}_i^{-1}) \prod_{i=1}^N (z-z_i).
\label{particleholerelation}
\end{align}
This can be obtained by performing change of variables and taking the $q=0$ (TASEP) limit of the corresponding equation for the ASEP case (see \cite[(44)]{ProlhacSciPost}, \cite[(II.101)]{Lazthesis} for example), which is derived by adapting the technique for the XXZ spin chain in \cite{PS}.

Also the relation $\prod_{i=1}^N z_i=\prod_{i=1}^{L-N} \tilde{z}_i$ holds.
From our point of view, this corresponds to the fact that the equation obtained from $C_0$ by setting $w=(-1)^{N+1} \ee^{L \gamma } v$ for $N$ particles with $L$ sites and that for $L-N$ particles with $L$ sites are given by the same equation and the solutions are the same.
Note also that \eqref{particleholerelation} can be rewritten as
\begin{align}
(z^{-1}-1)^L+(-1)^N \ee^{-L \gamma} \prod_{i=1}^{L-N} \tilde{z}_i^{-1} z^{N-L}=\prod_{i=1}^{L-N} (z^{-1}-\tilde{z}_i) \prod_{i=1}^N (z^{-1}-z_i^{-1}).
\end{align}

We can check the Bethe roots  $\{ z_i \}$ and $\{ \tilde{z}_i \}$ connected by \eqref{particleholerelation} give the same eigenvalue of the deformed Markov matrix for $N$ particles and $L-N$ particles as follows.
Taking the logarithm of both sides and differentiating with respect to $z$, we get
\begin{align}
&
\Bigg( L(z-1)^{L-1}+(-1)^{L-N} \ee^{-L \gamma} N \prod_{i=1}^N z_i^{-1} z^{N-1} \Bigg) 
\Bigg(
(z-1)^L+(-1)^{L-N} \ee^{-L \gamma} \prod_{i=1}^N z_i^{-1} z^N
\Bigg)^{-1} \nonumber \\
=&\sum_{i=1}^{L-N} \frac{1}{z-\tilde{z}_i^{-1}}+\sum_{i=1}^N \frac{1}{z-z_i}.
\end{align}
Setting $z=1$, one has
\begin{align}
N=\sum_{i=1}^{L-N} \frac{1}{1-\tilde{z}_i^{-1}}+\sum_{i=1}^N \frac{1}{1-z_i},
\end{align}
and rearranging gives
\begin{align}
\sum_{i=1}^{L-N} \frac{\tilde{z}_i}{1-\tilde{z}_i}=\sum_{i=1}^N \frac{z_i}{1-z_i}.
\end{align}

\subsection{Vanishing of special overlap}

One can generalize the vanishing of a special type of overlap for $L=2N$ \cite{MSS} to the case when the fugacity parameter is turned on.
First, we derive a factorized expression for a special case of the overlap.
Setting $M = 2N$ , $x_j = 2j$, $j = 1,\dots,N$ and changing variables from $u_j$ to $z_j =\alpha^{-1} u_j^{-2}$ and replacing $\alpha$ by $\ee^{-\gamma}$ in \cite[(4.2)]{MS},
we have
\begin{align}
\langle \psi(\{ z \})|0,1,0,1,\dots,0,1 \rangle
=\ee^{N \gamma/2}
\frac{\prod_{j=1}^N (1-z_j)^{2N} z_j^{-N-1/2} }{ \prod_{1 \le j < k \le N} (z_k-z_j) }
\det_{1 \le j,k \le N} \Bigg(
\Bigg(
\frac{z_j}{(1-z_j)^2}
\Bigg)^k
\Bigg). \label{specialforonehalf}
\end{align}
Using the factorization of the Vandermonde determinant 
$\displaystyle \det_{1 \le j,k \le N} (u_j^{k-1})=\prod_{1 \le j < k \le N} (u_k-u_j)$, we have
\begin{align}
\det_{1 \le j,k \le N} \Bigg(
\Bigg(
\frac{z_j}{(1-z_j)^2}
\Bigg)^k
\Bigg)&=\prod_{j=1}^N \frac{z_j}{(1-z_j)^2} 
\det_{1 \le j,k \le N} \Bigg(
\Bigg(
\frac{z_j}{(1-z_j)^2}
\Bigg)^{k-1}
\Bigg) \nonumber \\
&=\prod_{j=1}^N \frac{z_j}{(1-z_j)^2}
\prod_{1 \le j < k \le N}
\Bigg(
\frac{z_k}{(1-z_k)^2}-\frac{z_j}{(1-z_j)^2}
\Bigg) \nonumber \\
&=\prod_{j=1}^N \frac{z_j}{(1-z_j)^{2N}}
\prod_{1 \le j < k \le N} (z_k-z_j)
(1-z_j z_k). \label{applyvandermondeonehalf}
\end{align}
From \eqref{specialforonehalf} and \eqref{applyvandermondeonehalf}, we have
\begin{align}
\langle \psi(\{ z \})|0,1,0,1,\dots,0,1 \rangle
=\ee^{N \gamma/2}
\prod_{j=1}^N z_j^{-N+1/2}
\prod_{1 \le j < k \le N}
(1-z_j z_k). \label{factorizationonehalf}
\end{align}

From the factor
$\displaystyle \prod_{1 \le j < k \le N} (1-z_j z_k)$
in \eqref{factorizationonehalf}, the following is obvious.
\begin{lemma}
For $L=2N$, if the set of Bethe roots $\{ z \}$ contains a Golinelli-Mallick type tuple such that
$z_j z_k=1$, $z_j+z_k=2+\omega_N^{1-m} w^{-1/N}$ for some integer $m$, then we have
\begin{align}
\langle \psi(\{ z \})|0,1,0,1,\dots,0,1 \rangle=0.
\end{align}
\end{lemma}
This vanishing implies that not all of the Bethe roots contribute to the dynamics of the TASEP for the alternating initial condition.

We can show a similar vanishing for the case of 1/3-filling $L=3N$.
First, we derive a factorized expression for a special case of the overlap.
Setting $M=3N$, $x_j=3j$, $j=1,\dots,N$ and
changing variables from $u_j$ to $z_j=\alpha^{-1} u_j^{-2}$ and replacing $\alpha$ by $\ee^{-\gamma}$ in \cite[(4.2)]{MS}, we have 
\begin{align}
\langle \psi(\{ z \})|0,0,1,0,0,1,\dots,0,0,1 \rangle
=\ee^{N \gamma} \frac{\prod_{j=1}^N (1-z_j)^{3N} z_j^{-(3N+1)/2} }{ \prod_{1 \le j < k \le N} (z_k-z_j) }
\det_{1 \le j,k \le N} \Bigg(
\Bigg(
\frac{z_j}{(1-z_j)^3}
\Bigg)^k
\Bigg). \label{specialforonethird}
\end{align}
Using the factorization of the Vandermonde determinant, we have
\begin{align}
\det_{1 \le j,k \le N} \Bigg(
\Bigg(
\frac{z_j}{(1-z_j)^3}
\Bigg)^k
\Bigg)&=\prod_{j=1}^N \frac{z_j}{(1-z_j)^3} 
\det_{1 \le j,k \le N} \Bigg(
\Bigg(
\frac{z_j}{(1-z_j)^3}
\Bigg)^{k-1}
\Bigg) \nonumber \\
&=\prod_{j=1}^N \frac{z_j}{(1-z_j)^3}
\prod_{1 \le j < k \le N}
\Bigg(
\frac{z_k}{(1-z_k)^3}-\frac{z_j}{(1-z_j)^3}
\Bigg) \nonumber \\
&=\prod_{j=1}^N \frac{z_j}{(1-z_j)^{3N}}
\prod_{1 \le j < k \le N} (z_k-z_j)
(z_j z_k^2+z_j^2 z_k-3z_j z_k+1). \label{applyvandermonde}
\end{align}
From \eqref{specialforonethird} and \eqref{applyvandermonde}, we have
\begin{align}
\langle \psi(\{ z \})|0,0,1,0,0,1,\dots,0,0,1 \rangle
=\ee^{N \gamma} \prod_{j=1}^N z_j^{(1-3N)/2}
\prod_{1 \le j < k \le N}
(z_j z_k^2+z_j^2 z_k-3z_j z_k+1). \label{factorizationonethird}
\end{align}

From \eqref{factorizationonethird}, we can show the following vanishing.
\begin{prop}
For $L=3N$, if the set of Bethe roots $\{ z \}$ contains a Golinelli-Mallick type tuple such that
$z_j z_k z_\ell=1$, $z_j+z_k+z_\ell=3$, $z_j z_k+z_j z_\ell+z_k z_\ell=3+\omega_N^{1-m} w^{-1/N}$ for some integer $m$, then we have
\begin{align}
\langle \psi(\{ z \})|0,0,1,0,0,1,\dots,0,0,1 \rangle=0. \label{onethirdvanishing}
\end{align}
\end{prop}

\begin{proof}
We take the product of three factors from \eqref{factorizationonethird}
and rewrite using elementary polynomials of $z_j,z_k,z_\ell$ to get
\begin{align}
&
(z_j z_{k}^2+z_j^2 z_{k}-3z_j z_{k}+1)
(z_j z_{\ell}^2+z_j^2 z_{\ell}-3z_j z_{\ell}+1) 
(z_{k} z_{\ell}^2+z_{k}^2 z_{\ell}-3z_{k} z_{\ell}+1) 
\nonumber \\
=&1-3e_2+e_1 e_2-3e_3+9e_1 e_3-6e_1^2 e_3+e_1^3 e_3+6e_2 e_3-2e_1 e_2 e_3 \nonumber \\
&-24 e_3^2+18 e_1 e_3^2-3e_1^2 e_3^2-3 e_2 e_3^2+e_1 e_2 e_3^2-e_3^3, \label{onethirdfactors}
\end{align}
where $e_n=e_n(z_j,z_k,z_\ell)$, $n=1,2,3$.
Inserting $e_1(z_j,z_{k},z_{\ell})=3$, $e_3(z_j,z_{k},z_{\ell})=1$, we note the right hand side of \eqref{onethirdfactors}
vanishes and hence \eqref{onethirdvanishing} follows.

\end{proof}

\eqref{onethirdvanishing} implies that for the TASEP of 1/3-filling with the initial condition that the distances between neighboring particles are all equal, fewer Bethe roots are contributing to the dynamics.
We expect to have a similar vanishing of overlap for $1/\ell$-filling $(\ell=4,5,\dots)$.

\section*{Acknowledgment}

K.M.~thanks Kazumitsu Sakai and Jun Sato for previous collaborations, and Chikashi Arita for useful discussions on the TASEP Bethe ansatz equations. S.I.~is supported by Grant-in-Aid for Scientific Research, 22K03239 and 23K03056 from JSPS. K.M.~is supported by Grant-in-Aid for Scientific Research 21K03176 from JSPS.


\appendix

\section{Computation of symmetric polynomials}\label{sec:appA}

\subsection{$\Omega$-symmetric functions}


Given \( L \) indeterminates \( t_1, \dots, t_L \), 
the $\Omega$-elementary symmetric polynomials \( E_1, E_2, \dots, E_{\binom{L}{N}} \) and the $\Omega$-power sums \( P_1, P_2, \dots \)
are given by
\[
\prod_{1 \leq i_1 < \dots < i_N \leq L}(1 + t_{i_1} \cdots t_{i_N} T)
= 1 + E_1 T + E_2 T^2 + \cdots + E_{\binom{L}{N}} T^{\binom{L}{N}}
\]
and
\begin{equation}\label{eq:def_of_P}
P_n := \sum_{1 \leq i_1 < \dots < i_N \leq L} (t_{i_1} \cdots t_{i_N})^n.
\end{equation}

Then, $E_n$ and $P_n$ satisfy the same algebraic relations as $e_n$ and $p_n$.
\begin{lemma}[Newton's formula]\label{lemma:Newton}
We have
\[
\begin{gathered}
P_n=P_{n-1}E_1-P_{n-2}E_2+\cdots+(-1)^{n}P_1E_{n-1}+(-1)^{n+1}nE_n.    
\end{gathered}
\]
\end{lemma}


From \eqref{eq:def_of_P}, we immediately see that \( P_1 = e_N\).
In terms of \textit{monomial symmetric polynomials}, it is also written as $P_1=m_{(1^N)}$.
Since \( P_n \) is obtained from \( P_1 \) by the substitution map \( t_i \mapsto t_i^n \), which sends $m_{(1^a)}$ to $m_{(n^a)}$, we obtain
\begin{equation}\label{eq:mult_p}
P_n = m_{(n^N)}.
\end{equation}
Lemma \ref{lemma:Newton} and equation \eqref{eq:mult_p} allow us to compute $E_n$ step by step.
As a consequence, we obtain the determinantal formula
\[
E_i = \frac{1}{i!}
\begin{vmatrix}
m_{(1^N)} & 1\\
m_{(2^N)} & m_{(1^N)} & 2\\
\vdots & \vdots & \ddots & \ddots\\
m_{((i-1)^N)} & m_{((i-2)^N)} & \cdots & m_{(1^N)} & i - 1\\
m_{(i^N)} & m_{((i-1)^N)} & \cdots & m_{(2^N)} & m_{(1^N)}
\end{vmatrix}.
\]
We also note that the monomial symmetric polynomial $m_{(a^N)}=e_a|_{t_i\mapsto t_i^N}$ has the recursive equation
\[
p_{nN}=
p_{(n-1)N}\cdot m_{(1^N)}
-
p_{(n-2)N}\cdot m_{(2^N)}
+\cdots +(-1)^np_{N}\cdot m_{((n-1)^N)}
+(-1)^{n+1}n\cdot m_{(n^N)}
\]
and the determinantal formula
\begin{equation}\label{eq:det_formula_m}
m_{(a^N)}=e_N|_{t_i\mapsto t_i^a}=\frac{1}{N!}
\begin{vmatrix}
p_a & 1\\
p_{2a} & p_a & 2\\
\vdots & \vdots & \ddots & \ddots\\
p_{(N-1)a} & p_{(N-2)a} & \cdots & p_a & N - 1\\
p_{Na} & p_{(N-1)a} & \cdots & p_{2a} & p_{a}
\end{vmatrix}.    
\end{equation}

\begin{example}
When \( (L,N) = (4,2) \), we have $P_n=m_{(n,n)}$.
It follows from Lemma \ref{lemma:Newton} that
\[
\begin{gathered}
E_1=m_{1,1},\quad
E_2=m_{2,1,1}+3m_{1^4},\quad
E_3=m_{3,1,1,1}+m_{2,2,2}+2m_{2,2,1,1}\\
E_4=m_{3,2,2,1}+3m_{2^4},\quad
E_5=m_{3,3,2,2},\quad
E_6=m_{3,3,3,3}.
\end{gathered}
\]
Expanding the monomial symmetric polynomials as polynomials in $e_1,e_2,\dots$, we obtain
\[
\begin{gathered}
E_1=e_2,\quad E_2=e_1 e_3 - e_4,\quad 
E_3=e_1^2 e_4 - 2e_2 e_4 + e_3^2,\\
E_4=e_1e_3e_4-e_4^2,\quad
E_5=e_2e_4^2,\quad E_6=e_4^3.
\end{gathered}
\]
\end{example}

\subsection{$\mathcal{O}$-symmetric polynomials}\label{eq:appC}

An example of an $\mathcal{O}$-symmetric polynomial that is not contained in $\ZZ[w^{-1}]$ is given as follows.
When $(L,N)=(9,3)$, the number of connected components of $X$ is $\mathcal{N}=4$. (See Example \ref{ex:num_of_orbit_93}.)
These four components correspond to $[3,0,0]$, $[2,1,0]$, $[2,0,1]$, and $[1,1,1]$.
From $w=\frac{t^3}{(1-t)^9}$, we obtain
$$
t^3 - 3t^2 + (3 + \omega_3^{1-k} w^{-1/3})t - 1 = 0,\qquad (k=1,2,3),
$$
and
\[
t_k+t_{k+3}+t_{k+6}=3,\quad
t_{k,k+3}+t_{k,k+6}+t_{k+3.k+6}=3+\omega_3^{k-1}w^{-1/3},\quad
t_{k,k+3,k+6}=1.
\]
Let $p^{(k)}_n=t^n_{k}+t^n_{k+3}+t^n_{k+6}$.
Then, by direct calculations, we have
$p^{(k)}_1=3$,
$p^{(k)}_2=3-2\omega_3^{1-k}w^{-1/3}$,
$p^{(k)}_3=3-9\omega_3^{1-k}w^{-1/3}$,
and
$p^{(k)}_4=3-24\omega_3^{1-k}w^{-1/3}+2\omega_3^{2-2k}w^{-2/3}$.
One can calculate $\mathcal{O}$-symmetric polynomials using these polynomials.
For example, the $\mathcal{O}_{210}$-power sums are given by
\[
\begin{aligned}
P_n
&=\sum_{k=1}^3(t^n_{k,k+3}+t^n_{k,k+6}+t^n_{k+3,k+6})(t^n_{k+1}+t^n_{k+4}+t^n_{k+7})    
=\frac{1}{2}\sum_{k=1}^3
\{(p_{n}^{(k)})^2-p_{2n}^{(k)}\}p^{(k+1)}_n.
\end{aligned}
\]
Then, we have $P_1=27$, $P_2=27-6\omega_3^{-1}w^{-1},\textit{etc}$.
This $P_2$ is an example of an $\mathcal{O}$-power sum that is not contained in $\ZZ[w^{-1}]$.


\section{Numerical experiments}\label{sec:appB}

\subsection{$(L,N)=(6,3)$ case}
The defining equation of $C_0$ is as follows:
\begin{equation}\label{eq:def_poly_63}
\begin{aligned}
&(v^6 - 6v^5 + 15v^4 - (20+w^{-1})v^3 + 15v^2 - 6v + 1)^2\\
&\times    (v^8 - (8+w^{-1})v^7 + (28 - 12 w^{-1})v^6
- (56 + 7w^{-1})v^5
+ (70 + 40w^{-1} + w^{-2})v^4\\&
- (56 + 7 w^{-1})v^3
+ (28 - 12w^{-1})v^2
- (8 + w^{-1})v + 1)=0.
\end{aligned}
\end{equation}
The first factor corresponds to $[2,1,0]$ and $[2,0,1]$.
The second factor corresponds to $[1,1,1]$.

\subsection{$(L,N)=(8,4)$ case}

The defining equation of $C_0$ is as follows:
{\footnotesize
\begin{equation}\label{eq:def_poly_84}
\begin{aligned}
&(v-1)^6  \\
\times& (v^8 -8v^7 + (28-w^{-1})v^6 - (56 + 4w^{-1})v^5 + (70-6w^{-1})v^4 - (56+4w^{-1})v^3 + (28-w^{-1})v^2 - 8v + 1
)^2  \\
\times&(v^{16} - 16 v^{15} + (120 + 2 w^{-1})v^{14} - (560 -16w^{-1})v^{13} + (1820 - 132 w^{-1} + w^{-2})v^{12} - (4368 -272 w^{-1})v^{11}  \\
&+ (8008 - 2  w^{-1}  + 4  w^{-2})v^{10} -(11440+800 w^{-1} )v^9 + ( 12870 +1288  w^{-1}  + 6 w^{-2})v^8 -(11440+800 w^{-1} )v^7 \\
&+( 8008  -2 w^{-1} + 4  w^{-2})v^6 - ( 4368 -272  w^{-1})v^5 + (1820 -132  w^{-1}  +  w^{-2})v^4 \\
&-(560  +16  w^{-1})v^3+ ( 120 +2  w^{-1} )v^2 - 16 v + 1)^2  \\
\times&
(v^{16} - (16 - w^{-1})v^{15} + (120 + 52w^{-1})v^{14} -(560 - 67 w^{-1} + 2 w^{-2})v^{13} +(1820-1512 w^{-1} + 13 w^{-2})v^{12} \\
&-(4368 - 3633 w^{-1} + 226 w^{-2} - w^{-3})v^{11} +(8008 - 564 w^{-1} + 300 w^{-2} - 2 w^{-3})v^{10} \\
&-(11440 - 9845 w^{-1} + 676 w^{-2} - w^{-3})v^9 +(12870 + 16336 w^{-1} + 2574 w^{-2} + 4 w^{-3})v^8 \\
&-(11440 - 9845 w^{-1} + 676 w^{-2} -  w^{-3})v^7 +(8008 - 564 w^{-1} + 300 w^{-2} - 2 w^{-3})v^6 \\
&-(4368 - 3633 w^{-1} + 226 w^{-2} -  w^{-3})v^5 +(1820 -1512 w^{-1} + 13 w^{-2})v^4 -(560 - 67 w^{-1} + 2 w^{-2})v^3 \\
&+(120 + 52 w^{-1})v^2 -(16 -  w^{-1})v + 1)=0.
\end{aligned}
\end{equation}
}
The first factor corresponds to 
$[2,2,0,0]$ and
$[2,0,2,0]$,
the second to 
$[2,1,0,1]$,
the third to 
$[2,1,1,0]$ and
$[2,0,1,1]$, and
the fourth to
$[1,1,1,1]$.

\subsection{$(L,N)=(10,5)$ case}

Here we present the defining equation of $C_0$ after the substitution $w = v$ to avoid cumbersome expressions.
When $w=v$, the defining equation of $C_0$ factors as
\begin{align}
&f_1 f_2^2 f_3^2 f_4^6=0,
\end{align}
where
{\footnotesize
\begin{equation}
\begin{aligned}
f_1 
=&v ( v^{15}- 15 v^{14}+ 185 v^{13} - 1146 v^{12}+ 
   3994 v^{11} - 5728 v^{10} - 2155 v^9+ 8128 v^8\\
&- 10253 v^7- 
   4791 v^6+ 2267 v^5+ 5635 v^4+ 4262 v^3 - 2486 v^2+ 1459 v-126) \\
\times&
 (v^{16}- 17 v^{15}+ 55 v^{14}- 
   44 v^{13} + 854 v^{12}- 4590 v^{11}+ 16871 v^{10} 
- 44392 v^9 \\
&+   100017 v^8 - 116705 v^7+ 78637 v^6 - 29903 v^5
- 1518 v^4 - 68 v^3 + 
   9 v^2 + 22 v+2),
\end{aligned}
\end{equation}
}

{\footnotesize
\begin{equation}
\begin{aligned}
f_2
=\frac{1}{2}
\Bigg(
& 2 v^{40}- 
  80 v^{39}  + 1560 v^{38}  - 10 (1977 + \sqrt{5}) v^{37}
+ 
  15 (12171 + \sqrt{5}) v^{36}   \\
&+ 72 (-18173 + 35 \sqrt{5}) v^{35}
 + (7593106 - 
     35930 \sqrt{5}) v^{34}   \\
& + 
  4 (-9204963 + 59630 \sqrt{5}) v^{33}
 -   494 (-309287 + 1610 \sqrt{5}) v^{32}
 \\
& - 10 (55172277 + 10583 \sqrt{5}) v^{31}
 + (1751311821 + 
     16449515 \sqrt{5}) v^{30}   \\
& - 
  16 (306822061 + 6331855 \sqrt{5}) v^{29}
+ 
  2 (6081047431 + 190063695 \sqrt{5}) v^{28}   \\
&- 
  2 (13299458311 + 510594945 \sqrt{5}) v^{27}
 + (51251188141 + 
     2030715335 \sqrt{5}) v^{26}   \\
& - 
  12 (7236795556 + 239399295 \sqrt{5}) v^{25}
 + 
  5 (25862641743 + 444167195 \sqrt{5}) v^{24}  \\
&+ 
  20 (-8472997926 + 81572107 \sqrt{5}) v^{23}
 + (196465334245 - 
     9535819135 \sqrt{5}) v^{22}   \\
&+ 
  20 (-10199154351 + 1008294176 \sqrt{5}) v^{21} 
 + (193674439114 - 
     29782408770 \sqrt{5}) v^{20}  \\
&+ 
  60 (-2876401762 + 563723503 \sqrt{5}) v^{19}
 + (147663259320 - 
     29903370950 \sqrt{5}) v^{18}   \\
&+ 
  4 (-30678740223 + 4977812555 \sqrt{5}) v^{17}
 + (98365079806 - 
     8717157910 \sqrt{5}) v^{16}   \\
&+ (-74041526788 + 
     602602780 \sqrt{5}) v^{15}
 + (50650622192 + 
     3020431210 \sqrt{5}) v^{14}   \\
& - 
  10 (3075933945 + 327556037 \sqrt{5}) v^{13}+ 
  30 (543702269 + 71729211 \sqrt{5}) v^{12}   \\
& - 
  10 (739008265 + 104891721 \sqrt{5}) v^{11}  + 
  5 (557303357 + 79985265 \sqrt{5}) v^{10} 
 \\
& - 2 (430181323 + 57604835 \sqrt{5}) v^9+ (216416607 + 
     21302335 \sqrt{5}) v^8   \\
& - 
  2 (21720891 + 505115 \sqrt{5}) v^7+ (6818841 - 718315 \sqrt{5}) v^6 
+ 
  6 (-150401 + 41005 \sqrt{5}) v^5   \\
&+ (103191 - 33255 \sqrt{5}) v^4
+ 
  2 (-6301 + 995 \sqrt{5}) v^3 
+ 2 (694 + 5 \sqrt{5}) v^2
- 10 (9 + \sqrt{5}) v
+
2 
\Bigg),
\end{aligned}
\end{equation}
}

{\footnotesize
\begin{equation}
\begin{aligned}
f_3
=\frac{1}{2}
\Bigg(&2 v^{40}- 80 v^{39}+ 1560 v^{38} 
 + 
 10 (-1977 + \sqrt{5}) v^{37}
- 15 (-12171 + \sqrt{5}) v^{36}
 \\
& - 
 72 (18173 + 35 \sqrt{5}) v^{35}  + (7593106 + 35930 \sqrt{5}) v^{34}  \\
& 
- 
 4 (9204963 + 59630 \sqrt{5}) v^{33}
+ 
 494 (309287 + 1610 \sqrt{5}) v^{32}   \\
&  + 10 (-55172277 + 10583 \sqrt{5}) v^{31} + (1751311821 - 
    16449515 \sqrt{5}) v^{30}  \\
& + 
 16 (-306822061 + 6331855 \sqrt{5}) v^{29}+ (12162094862 - 
    380127390 \sqrt{5}) v^{28}   \\
& + 
 2 (-13299458311 + 510594945 \sqrt{5}) v^{27}+ (51251188141 - 
    2030715335 \sqrt{5}) v^{26}   \\
&+ 
 12 (-7236795556 + 239399295 \sqrt{5}) v^{25} 
 - 
 5 (-25862641743 + 444167195 \sqrt{5}) v^{24}  \\
&  - 
 20 (8472997926 + 81572107 \sqrt{5}) v^{23}+ 
 5 (39293066849 + 1907163827 \sqrt{5}) v^{22}  \\
&- 
 20 (10199154351 + 1008294176 \sqrt{5}) v^{21}+ 2 (96837219557 + 14891204385 \sqrt{5}) v^{20}   \\
&- 
 60 (2876401762 + 563723503 \sqrt{5}) v^{19} + 
 10 (14766325932 + 2990337095 \sqrt{5}) v^{18}   \\
& - 
 4 (30678740223 + 4977812555 \sqrt{5}) v^{17} 
 + (98365079806 + 
    8717157910 \sqrt{5}) v^{16} \\
&  - 
 4 (18510381697 + 150650695 \sqrt{5}) v^{15}+ (50650622192 - 
    3020431210 \sqrt{5}) v^{14}  \\
&+ 
 10 (-3075933945 + 327556037 \sqrt{5}) v^{13}
 - 
 30 (-543702269 + 71729211 \sqrt{5}) v^{12} 
 \\
&+ 
 10 (-739008265 + 104891721 \sqrt{5}) v^{11} 
 - 
 5 (-557303357 + 79985265 \sqrt{5}) v^{10}  \\
&+ 2 (-430181323 + 57604835 \sqrt{5}) v^9
 + (216416607 - 
    21302335 \sqrt{5}) v^8   \\
& + 
 2 (-21720891 + 505115 \sqrt{5}) v^7 + (6818841 + 718315 \sqrt{5}) v^6
 - 
 6 (150401 + 41005 \sqrt{5}) v^5   \\
&+ 3 (34397 + 11085 \sqrt{5}) v^4 - 
 2 (6301 + 995 \sqrt{5}) v^3- 2 (-694 + 5 \sqrt{5}) v^2
+ 10 (-9 + \sqrt{5}) v+
2 
\Bigg),
\end{aligned}
\end{equation}
}

\begin{align}
f_4 
=(v^5-5v^4+10v^3-11v^2+5v-1)(v^5-5v^4+10v^3-9v^2+5v-1).
\end{align}

$f_1$ corresponds to $[1,1,1,1,1]$, $f_2$ to $[2,1,1,1,0]$ and $[2,0,1,1,1]$,
$f_3$ to $[2,1,1,0,1]$ and $[2,1,0,1,1]$, $f_4$ to $[2,2,1,0,0]$, $[2,2,0,1,0]$, $[2,2,0,0,1]$, $[2,1,2,0,0]$,
$[2,0,2,1,0]$ and $[2,0,2,0,1]$.

Let us make a comment on the missing Bethe roots based on some ansatz on the distribution of the Bethe roots
by Golinelli-Mallick or Prolhac.
For the case of half-filling and with no fugacity parameter, the ansatz works perfectly up to $L=8$, $N=4$,
and when $L=10$, $N=5$, the Bethe roots whose corresponding quantum numbers given by
$(c(1), c(2), c(3), c(4), c(5)) = (2, 3, 4, 5, 6), (1, 2, 3, 4, 10)$ in Golinelli-Mallick's labeling \cite{GMone} are missing.
Comparing with the datum which we obtained,
we find the two values for $v$ which correspond to the missing Bethe roots are
two roots of the following equation which is a part of $f_1$
\begin{equation}
\begin{aligned}
 & v^{15}- 15 v^{14}+ 185 v^{13} - 1146 v^{12}+ 
   3994 v^{11} - 5728 v^{10} - 2155 v^9+ 8128 v^8  \\
&- 10253 v^7- 
   4791 v^6+ 2267 v^5+ 5635 v^4+ 4262 v^3 - 2486 v^2+ 1459 v-126=0.
\end{aligned}    
\end{equation}
The two numerical values are
$v=-1.0125-0.0436581 \sqrt{-1}$ and $v=-1.0125 + 0.0436581 \sqrt{-1}$.

For $v=-1.0125 - 0.0436581 \sqrt{-1}$, the Bethe roots are
\begin{equation}
\begin{aligned}
\{ t \} 
=&\{
0.503842 + 0.869555 \sqrt{-1}, 0.376385 + 0.382431 \sqrt{-1}, 0.38087 + 0.11074 \sqrt{-1},  \\
&2.43093 + 0.686883 \sqrt{-1}, 1.32373 + 1.32886 \sqrt{-1}
\}. \label{missingBethe}
\end{aligned}
\end{equation}
Note that all of the imaginary parts of $\{ t \}$ are positive.
Since the version of the variables we use for the Bethe roots are
the same with the one used by Prolhac rather than the version by Golinelli-Mallick,
we compare with \cite{ProlhacSciPost}.
The $L=10$, $N=5$ version of a figure like
\cite[Figure 17]{ProlhacSciPost} means that
for the case corresponding to quantum numbers $(c(1), c(2), c(3), c(4), c(5)) = (2, 3, 4, 5, 6)$,
an ansatz that four of the values of $\{ t \}$ are on the upper half plane,
and the remaining one value
is on the lower half plane, is imposed.
However,
for the Bethe roots \eqref{missingBethe} which we found, all values of $\{ t \}$ lie
on the upper half plane which implies the ansatz does not work for this case.

For $v=-1.0125+0.0436581 \sqrt{-1}$, the Bethe roots are
\begin{equation}
\begin{aligned}
\{ t \}
=&\{
0.503842 - 0.869555 \sqrt{-1}, 0.376385 - 0.382431 \sqrt{-1}, 0.38087 - 0.11074 \sqrt{-1},  \\
&2.43093 - 0.686883 \sqrt{-1}, 1.32373 - 1.32886 \sqrt{-1}
 \}. \label{missingBethetwo}
\end{aligned}
\end{equation}
Note that all imaginary parts of $\{ t \}$ are negative,
and \eqref{missingBethetwo} is complex conjugate to \eqref{missingBethe} as sets.

\subsection{$(L,N)=(9,3)$ case}

After substituting $w=v$, the defining equation of $C_0$ factors as
\begin{align}
g_1 g_2 g_3 g_4=0,
\end{align}
where
$g_1=(v-1)^3$ and
\begin{equation}
\begin{aligned}
g_2&=v
(
v^8-5v^7+31v^6-100v^5+223v^4-322v^3+271v^2-154v+28)  \\
&\times (
v^{18}-22v^{17}+210v^{16}-1089v^{15}+3307v^{14}-6017v^{13}+7176v^{12}-9178v^{11}+18275v^{10}
)  \\
&-30889v^9+34183v^8-25749v^7+12623v^6-2056v^5-294v^4+265v^3-11v^2-9v+3),
\end{aligned}
\end{equation}

{\footnotesize
\begin{equation}
\begin{aligned}
g_3=&v^{27}- 27 v^{26}+ 351 v^{25}- (2925 + 3 \omega) v^{24}
 + 17 547 v^{23}   \\
&-(80 616 - 318 \omega) v^{22}
- (-294 603 + 2496 \omega) v^{21}
 - (878 802 - 8511 \omega) v^{20}   \\
&-(-2 181 720 + 9201 \omega) v^{19}
 - (4 578 920 + 39 441 \omega) v^{18}
-(-8 232 429 - 211 302 \omega) v^{17}   \\
&- (12 813 462 + 522 114 \omega) v^{16} 
 -
(-17 377 049 - 816 555 \omega) v^{15}
 - (20 560 365 + 811 935 \omega) v^{14} 
  \\
& -(-21 155 793 - 358 215 \omega) v^{13}
 - (18 812 288 - 333 759 \omega) v^{12} 
-(-14 334 096 + 819 306 \omega)v^{11} 
 \\
& - (9 256 290 - 836 907 \omega)v^{10}
-(-4 998 854 +527 046\omega)v^9
- (2229447-205155\omega)v^8 
 \\
&-
(-816525 + 36138 \omega)v^7
-(247727 + 8658\omega)v^6
-(-63507-7875\omega)v^5  \\
&-(13881 + 2421\omega)v^4
-(-2513 -324\omega)v^3
-(348 -3\omega)v^2
- (-30 + 3\omega)v
-1,
\end{aligned}
\end{equation}

\begin{equation}
\begin{aligned}
g_4=
& v^{27}- 27 v^{26}+ 351 v^{25} + (-2928 + 3 \omega) v^{24}
+ 17 547 v^{23}
+(-80 298 - 318 \omega) v^{22}  \\
 &+ (292 107 + 2496 \omega) v^{21}
+ (-870 291 - 8511 \omega) v^{20} 
+(2 172 519 + 9201 \omega) v^{19}   \\
&
+ (-4 618 361 + 39 441 \omega) v^{18}
+
(8 443 731 - 211 302 \omega) v^{17}
+ (-13 335 576 + 522 114 \omega) v^{16}   \\
&+
(18 193 604 - 816 555 \omega) v^{15}
+ (-21 372 300 + 811 935 \omega) v^{14} 
 +(21 514 008 - 358 215 \omega) v^{13}   \\
&+ (-18 478 529 - 333 759 \omega) v^{12} 
+
(13 514 790 + 819 306 \omega) v^{11}
+ (-8 419 383 - 836 907 \omega) v^{10}   \\
&+
(4 471 808 + 527 046 \omega) v^9
+ (-2 024 292 - 205 155 \omega) v^8
+
(780 387 + 36 138 \omega) v^7    \\
&+ (-256 385 + 8658 \omega) v^6
+ (71 382 - 7875 \omega) v^5
+(-16 302 + 2421 \omega)v^4   \\
&+ (2837 - 324 \omega) v^3
+ (-345 - 3 \omega) v^2
+ (27 + 3 \omega) v
-1.
\end{aligned}
\end{equation}
}
Here, $\omega=\omega_3$.
$g_1$ corresponds to $[3,0,0]$,
$g_2$ to $[1,1,1]$,
$g_3$ to $[2,0,1]$,
$g_4$ to $[2,1,0]$.

\section{Proof of Equations \eqref{eq:explicit_form_m}, \eqref{eq:sum_of_m} and Lemma \ref{lemma:character_formula_2}}
\label{sec:appD}

Throughout, fix positive integers $L>N>0$.
Let $e=\mathrm{gcd}(L,N)$, $L=eL'$, and $N=eN'$.

\subsection{Equation \eqref{eq:explicit_form_m}}

Consider the set $Z(a_0,\dots,a_{L'})$ defined in \eqref{eq:def_of_Z(a,a)}.
For $\ell\mid e$, let $X_\ell$ be the subset of $Z(a_0,\dots,a_{L'})$ consisting of all fixed points of the action $\theta^\ell$.
Assume $e=\ell q$ for some integer $q$.
Then $X_{\ell}$ is non-empty if and only if $q\mid a_i$ for all $i$.
If $X_{\ell}\neq \emptyset$, let $a_i=q a_i'$.
Then, the cardinality of $X_{\ell}$ is given by
\[
\# X_{\ell}=\binom{\ell}{a_0',a_1',\dots,a_{L'}'}=
\binom{\ell}{\frac{a_0\ell}{e},\dots,\frac{a_{L'}\ell}{e}}.
\]

On the other hand, the number $m_\ell(a_0,\dots,a_{L'})$ of $\ZZ/e\ZZ$-orbits of length $\ell$ satisfies the equation
\[
\# X_{\ell}=\sum_{\ell'\mid \ell}m_{\ell}\cdot \ell.
\]
Solving this, we obtain the formula
\[
m_\ell\cdot \ell=\sum_{\ell'\mid \ell}\# X_{\ell'}\cdot \mu(\ell/\ell')=
\sum_{
\substack{
\ell'\mid\ell\\
e\mid a_i\ell'\ (\forall i)
}
}\binom{\ell'}{\frac{a_0\ell'}{e},\dots,\frac{a_{L'}\ell'}{e}}\mu(\ell/\ell'),
\]
where $\mu(n)$ is the M\"{o}bius function.
This implies \eqref{eq:explicit_form_m}.

\subsection{Equation \eqref{eq:sum_of_m}}

To show \eqref{eq:sum_of_m}, we use the formula
\begin{equation}\label{eq:eul_to_mob}
\varphi(n)=\sum_{d\mid n}\mu(n/d)\cdot d.
\end{equation}
Starting with the expression on the right-hand side of \eqref{eq:explicit_form_m}, we have
\begin{equation}
\begin{aligned}
\sum_{\ell\mid e}m_{\ell}(a_0,\dots,a_{L'})
&=
\sum_{
\substack{
\ell\mid e,\
\ell'\mid \ell\\
e\mid a_i\ell'\ (\forall i)
}
}\frac{1}{\ell}\binom{\ell'}{\frac{a_0\ell'}{e},\dots,
\frac{a_{L'}\ell'}{e}
}\cdot \mu(\ell/\ell')    \\
&=
\sum_{
\substack{
p\mid (e/\ell')\\
e\mid a_i\ell'\ (\forall i)
}
}\frac{p}{e}\binom{\ell'}{\frac{a_0\ell'}{e},\dots,
\frac{a_{L'}\ell'}{e}
}\cdot \mu(e/p\ell').
\end{aligned}
\end{equation}
By \eqref{eq:eul_to_mob}, this equals to
\[
\frac{1}{e}
\sum_{
\substack{
\ell'\mid e\\
e\mid a_i\ell'\ (\forall i)
}
}\binom{\ell'}{\frac{a_0\ell'}{e},\dots,
\frac{a_{L'}\ell'}{e}
}\cdot \varphi(e/\ell').
\]


\subsection{Lemma \ref{lemma:character_formula_2}}
Let $d=\mathrm{lcm}(N,L-N)$.
Recall that $\widehat{\sigma}:\Omega\to \Omega$ is of degree $d$.
For $1\leq k\leq d$, let $Y_k$ be the subset of $\Omega$ consisting of all fixed points of the action $\widehat{\sigma}^k$.

For an element $I$ of $\Omega$, we define the 01-sequence $x_I=x_1x_2\dots x_L\in \{0,1\}^L$ by putting $x_i=1$ if $i\in I$ and $x_i=0$ if $i\notin I$.
let $N_1:=\# (I\cap [1,N])$ and $N_2:=\# (I\cap [N+1,L])$.
Define $e_1:=\mathrm{gcd}(N,k)$, $e_2:=\mathrm{gcd}(L-N,k)$, $N=e_1q_1$, and $L-N=e_2q_2$.
We divide the 01-sequence $x_I$ into $(q_1+q_2)$ smaller parts as follows:
\[
\xymatrix@C=-1em@R=1em{
& \overbrace{0010\cdots 01}^L\ar[dl] \ar[dr] & \\
\overbrace{0010\cdots 11}^{\text{first }N\text{ digits}} \ar[d] & & 
\overbrace{111010\cdots 101}^{\text{last }L-N\text{ digits}} 
\ar[d] \\
\hbox{$
\begin{gathered}
00,\ 10,\ \dots,\ 11\\
(q_1 \text{ parts of size } e_1)
\end{gathered}
$}
&
&
\hbox{$
\begin{gathered}
111,\ 010,\ \dots,\ 101\\
(q_2 \text{ parts of size } e_2)
\end{gathered}
$}
}
\]
Then $I$ is contained in $Y_k$ if and only if 
\begin{itemize}
    \item all $q_1$ parts of size $e_1$ are identical, and
    \item all $q_2$ parts of size $e_2$ are identical.
\end{itemize}
In particular, if $Y_k\neq \emptyset$, the integers $q_1$ and $q_2$ must satisfy $q_1\mid N_1$ and $q_2\mid N_2$.
Assume $Y_k\neq \emptyset$, and let $N_1=q_1a_1$ and $N_2=q_2a_2$.
Then the cardinality of $Y_k$ is given by
\[
\# Y_k
=\sum_{N_1+N_2=N}\binom{e_1}{a_1}\binom{e_2}{a_2}
=\sum_{a_1\frac{N}{e_1}+a_2\frac{L-N}{e_2}=N}\binom{e_1}{a_1}\binom{e_2}{a_2}.
\]
This equation implies \eqref{subeq:to_prove_2}.

Similar arguments can be applied for proving \eqref{subeq:to_prove_3}.
Consider the action $\widehat{\tau}:\Omega\to \Omega$ of order is $L$.
Let $Z_k\subset \Omega$ be the subset consisting of all fixed points of the action $\widehat{\tau}^k$.
Define $e:=\mathrm{gcd}(L,k)$ and $L=eq$.
Then, $Z_k\neq \emptyset$ if and only if $q\mid N$.
Letting $N=qa$, we obtain 
\[
\# Z_k=\binom{e}{a}=\binom{\mathrm{gcd}(L,k)}{\mathrm{gcd}(L,k)\cdot N/L},
\]
which implies \eqref{subeq:to_prove_3}.

Equation \eqref{subeq:to_prove_4} follows from
\[
\begin{aligned}
\chi_{\mathbb{C}[\Omega]}(\widehat{\tau}^{-1} \cdot \widehat{\sigma})
&=\# \Omega -
\#\{I\subset \Omega\;;\;
I \text{ contains exactly one of } N \text{ or }L
\}
=\binom{L}{N}-2\binom{L-2}{N-1}.
\end{aligned}
\]





\begin{thebibliography}{99}
%
%
\bibitem{Bethe}
H. Bethe, Zur Theorie der Metalle, Z. Phys. 71, 205 (1931).

\bibitem{Baxterbook}
R.J. Baxter, Exactly solved models in statistical mechanics, Academic Press, Cambridge (1982).

\bibitem{Derrida}
B. Derrida, An exactly soluble non-equilibrium system: The asymmetric simple exclusion
 process, Phys. Rep. 301, 65 (1998).

\bibitem{Liggett}
T.M. Liggett, Stochastic interacting systems: Contact, voter and exclusion pro
cesses, Springer, Berlin (1999).

\bibitem{Ki}
A.N. Kirillov, Combinatorial identities and the completeness of states for Heisenberg
 magnet, J. Soviet Math. 30, 2298 (1985).

\bibitem{KR}
A.N. Kirillov and N.Yu. Reshetikhin, The Bethe ansatz and the combinatorics of Young tableaux, 
J. Soviet Math. 41, 925 (1988), 925.

\bibitem{LSA}
R.P. Langlands and Y. Saint-Aubin, Combinatorial aspects of the Bethe equations, CRM
 Proc. Lect. Notes 11, 231-301 (1997).

\bibitem{TV}
V. Tarasov and A. Varchenko, Completeness of Bethe vectors and difference equations with
 regular singular points, Int. Math. Res. Not. 13, 637 (1995).

\bibitem{Baxter}
R.J. Baxter, Completeness of the Bethe Ansatz for the six and eight-vertex models, J. Stat.
 Phys. 108, 1 (2002).

\bibitem{MTV}
E. Mukhin, V. Tarasov, and A. Varchenko, Bethe algebra of homogeneous 
XXX Heisenberg model has simple spectrum, Commun. Math. Phys. 288, 1 (2009).

\bibitem{Tarasov}
V. Tarasov, Completeness of the Bethe Ansatz for the periodic isotropic Heisenberg model,
Rev. Math. Phys. 30, 1840018 (2018)

\bibitem{BCPS}
A. Borodin, I. Corwin, L. Petrov, and T. Sasamoto,
 Spectral Theory for Interacting Particle Systems Solvable
 by Coordinate Bethe Ansatz,
Commun. Math. Phys. 339, 1167 (2015).
 

\bibitem{GSone}
L.-H. Gwa and H. Spohn, Bethe solution for the dynamical-scaling exponent of the noisy Burgers equation, Phys. Rev. A 46, 844 (1992).

\bibitem{GStwo}
L.-H. Gwa and H. Spohn, Six-vertex model, roughened surfaces, and an asymmetric spin
 Hamiltonian, Phys. Rev. Lett. 68, 725 (1992).

\bibitem{Kimone}
D. Kim, Bethe Ansatz solution for crossover scaling functions of the asymmetric XXZ chain and the Kardar-Parisi-Zhang-type growth model, Phys. Rev. E 52, 3512 (1995).
 
\bibitem{Kimtwo}
D. Kim, Asymmetric XXZ chain at the antiferromagnetic transition: Spectra and partition functions, J. Phys. A: Math. Gen. 30, 3817 (1997).

\bibitem{deGE}
J. de Gier and F. H. L. Essler, Bethe Ansatz solution of the asymmetric exclusion process with open boundaries, Phys. Rev. Lett. 95, 240601 (2005).






\bibitem{GMone}
O. Golinelli and K. Mallick, Bethe Ansatz calculation of the spectral gap of the asymmetric exclusion process, J. Phys. A: Math. Gen. 37, 3321 (2004).

\bibitem{GMtwo}
O. Golinelli and K. Mallick, Spectral gap of the totally asymmetric exclusion process at arbitrary filling, J. Phys. A: Math. Gen. 38, 1419 (2005).

\bibitem{GMthree}
O. Golinelli and K. Mallick, Hidden symmetries in the asymmetric exclusion process, J.
 Stat. Mech.: Theor. Exp. P12001 (2004).



\bibitem{GMspecdeg}
O. Golinelli and K. Mallick,
Spectral Degeneracies in the Totally Asymmetric Exclusion Process,
J. Stat. Phys. 120, 779 (2005).
 

\bibitem{GMfour}
O. Golinelli and K. Mallick, Family of commuting operators for the totally asymmetric exclusion process, J. Phys. A: Math. Theor. 40, 5795 (2007).







\bibitem{Prolhacone}
S. Prolhac, Spectrum of the totally asymmetric simple exclusion process on a periodic lattice-first excited states, J. Phys. A: Math. Theor. 47, 375001 (2014).

\bibitem{Prolhactwo}
S. Prolhac, Asymptotics for the norm of Bethe eigenstates in the periodic totally asymmetric exclusion process, J. Stat. Phys. 160, 926 (2015).

\bibitem{Prolhacthree}
S. Prolhac, Current fluctuations for totally asymmetric exclusion on the relaxation scale,
 J. Phys. A: Math. Theor. 48, 06FT02 (2015).

\bibitem{Prolhacfour}
S. Prolhac, Current fluctuations and large deviations for periodic TASEP on the relaxation scale, J. Stat. Mech.: Theory Exp. P11028 (2015).

\bibitem{Prolhacfive}
S. Prolhac, Finite-time fluctuations for the totally asymmetric exclusion process, Phys. Rev.
Lett. 116, 090601 (2016).

 

\bibitem{KPZ}
M. Kardar, G. Parisi and Y.-C. Zhang, Dynamic scaling of growing interfaces, Phys. Rev.
 Lett. 56, 889 (1986).

 \bibitem{Corwin}
I. Corwin, The Kardar-Parisi-Zhang equation and universality class, Random Matrices:
 Theory Appl. 01, 1130001 (2012).

\bibitem{Spohn}
H. Spohn, The Kardar-Parisi-Zhang equation: A statistical physics perspective, in
 Stochastic processes and random matrices, Oxford University Press, Oxford, UK, ISBN
 0198797311 (2017).

\bibitem{Takeuchi} 
K.A. Takeuchi, An appetizer to modern developments on the Kardar-Parisi Zhang universality class, Phys. A: Stat. Mech. Appl. 504, 77 (2018).


\bibitem{Johone}
K. Johansson, Shape ﬂuctuations and random matrices,
Comm. Math. Phys. 209, 437 (2000).

\bibitem{Johtwo}
K. Johansson,
Discrete polynuclear growth and determinantal processes,
Comm. Math. Phys. 242, 277  (2003).

\bibitem{BFPS}
A. Borodin, P.L. Ferrari, M. Pr\"ahofer, and T. Sasamoto, 
Fluctuation properties of the TASEP with periodic initial configuration, J. Stat. Phys. 129, 1055 (2007).

\bibitem{TW}
C.A. Tracy and H. Widom,
Asymptotics in ASEP with step initial condition,
Comm. Math. Phys. 290, 129 (2009).


\bibitem{SS}
T. Sasamoto and H. Spohn,
One-dimensional Kardar-Parisi-Zhang equation: an exact solution and its universality. 
Phys. Rev. Lett. 104, 230602  (2010).

\bibitem{ACQ}
G. Amir, I. Corwin, and J. Quastel,
Probability distribution of the free energy of the continuum directed random polymer in 1+ 1 dimensions,
Communications on Pure and Applied Mathematics 64, 466 (2011).






\bibitem{ProlhacRSone}
S. Prolhac, Riemann surfaces for KPZ with periodic boundaries, SciPost Phys. 8, 008
 (2020).

\bibitem{ProlhacRStwo}
S. Prolhac, Riemann surface for TASEP with periodic boundaries, J. Phys. A: Math. Theor.
 53, 445003 (2020).

\bibitem{ProlhacRSthree}
S. Prolhac,
 From the Riemann surface of TASEP to ASEP
J. Phys. A: Math. Theor. 54, 395002 (2021).



\bibitem{ProlhacSciPost}
S. Prolhac,
KPZ fluctuations in finite volume,
SciPost Phys. Lect. Notes 81 (2024).








\bibitem{deGKW}
J. de Gier, R. Kenyon, and S.S. Watson,
Limit shapes for the asymmetric five vertex model,
Commun. Math. Phys. 385, 793 (2021).



\bibitem{NPH}
G. Nakerst, T. Prosen, and M. Haque,
Spectral boundary of the asymmetric simple exclusion process: Free fermions, Bethe ansatz, and random matrix theory
Phys. Rev. E 110, 014110 (2024).







\bibitem{BDS}
E. Brattain, N. Do, and A. Saenz, The completeness of the Bethe Ansatz for the periodic
 ASEP, arXiv:1511.03762.


\bibitem{ISN}
Y. Ishiguro, J. Sato, and K. Nishinari,
Asymmetry-induced delocalization transition in the integrable non-Hermitian spin chain,
Physical Review Research 5, 033102 (2023).


\bibitem{BLone}
J. Baik and Z. Liu, Fluctuations of TASEP on a ring in relaxation time scale. Comm. Pure Appl.
 Math., 71, 0747 (2018).

\bibitem{BLtwo}
J. Baik and Z. Liu, Multipoint distribution of periodic TASEP, J. Amer. Math. Soc. 32, 609
 (2019).


\bibitem{AKSS}
C. Arita, A. Kuniba, K. Sakai, and T. Sawabe, Spectrum of a multi-species
asymmetric simple exclusion process on a ring, J. Phys. A: Math. Theor. 42, 345002 (2009).
 


\bibitem{Bo}
N.M. Bogoliubov,
Determinantal Representation of the Time-Dependent Stationary Correlation Function for the Totally Asymmetric Simple Exclusion Model,
SIGMA 5 052 (2009).

\bibitem{MSS}
K. Motegi, K. Sakai, and J. Sato,
Long time asymptotics of the totally asymmetric simple exclusion process,
J. Phys. A: Math. Theor. 45, 465004 (2012).

\bibitem{MS}
K. Motegi and K. Sakai,
Vertex models, TASEP and Grothendieck polynomials,
J. Phys. A. 46, 355201 (2013).


\bibitem{Korepin}
V.E. Korepin,
Calculation of norms of Bethe wave functions,
Commun. Math. Phys. 86, 391 (1982).

\bibitem{Slavnov}
N.A. Slavnov,
Algebraic Bethe Ansatz and Correlation Functions,
An Advanced Course, World Scientific
(2022).

\bibitem{KBI}
V.E. Korepin, N.M. Bogoliubov, and A.G. Izergin,
Quantum Inverse Scattering Method and Correlation
 Functions, Cambridge University Press
(1993).

\bibitem{Slavnovscalarproduct}
N.A. Slavnov, Calculation of scalar products of wave functions and form factors in the framework of the algebraic Bethe Ansatz, Theor. Math. Phys. 79, 502 (1989).

\bibitem{BuPr}
I.N. Burenev and A.G. Pronko,
Thermodynamics of the five-vertex model with scalar-product boundary conditions,
Commun. Math. Phys. 405, 148 (2024).



\bibitem{deGKo}
J. de Gier and V.E. Korepin,
Six-vertex model with domain wall boundary conditions: variable inhomogeneities,
J. Phys. A: Math. Gen. 34, 8135 (2001).





\bibitem{KoZJ}
V.E. Korepin and P. Zinn-Justin,
Thermodynamic limit of the Six-Vertex Model with Domain Wall Boundary Conditions,
J. Phys. A 33, 7053 (2000).

\bibitem{BF}
P.M. Bleher and V.V. Fokin,
Exact Solution of the Six-Vertex Model with Domain Wall
Boundary Conditions. Disordered Phase,
Commun. Math. Phys. 268, 223 (2006).


\bibitem{TRK}
T.S. Tavares, G.A.P. Ribeiro, and V.E. Korepin,
The entropy of the six-vertex model with variety of different boundary conditions,
J. Stat. Mech. P06016 (2015).


\bibitem{Lazthesis}
A. Lazarescu,
Exact Large Deviations of the Current in the Asymmetric Simple Exclusion Process with Open Boundaries,
Doctor thesis (Universit\'e Pierre et Marie Curie),
arXiv:1311.7370.

\bibitem{PS}
G.P. Pronko and Yu.G. Stroganov,
Bethe equations `on the wrong side of the equator',
J. Phys. A: Math. Gen. 32, 2333 (1999).

\bibitem{WM}
H.~Weyl, and G.~R. MacLane. The concept of a Riemann surface. Courier Corporation, (2009).

\end{thebibliography}
\end{document}